\documentclass[a4paper,12pt]{article}
\pdfoutput=1
\usepackage{cite}
\usepackage{epsfig}
\usepackage{amssymb,amsmath}
\usepackage{graphicx}
\usepackage{float}
\usepackage{comment}
\usepackage{soul}
\usepackage{multirow}

\usepackage{hyperref}

\usepackage{xcolor,colortbl}

\definecolor{Gray}{gray}{0.95}
\definecolor{RGray}{gray}{0.85}
\definecolor{CGray}{gray}{0.92}

\newcolumntype{a}{>{\columncolor{Gray}}c}
\newcolumntype{b}{>{\columncolor{white}}c}

\newlength{\dinwidth}
\newlength{\dinmargin}
\newcommand{\lsim}{~{}_{\textstyle\sim}^{\textstyle <}~}

\def\be{\begin{equation}}
\def\ee{\end{equation}}
\def\beqn{\begin{eqnarray}}
\def\eeqn{\end{eqnarray}}

\def\ba{\begin{array}{c}}
\def\bat{\begin{array}{cc}}
\def\ea{\end{array}}
\def\bi{\begin{itemize}}
\def\ei{\end{itemize}}

\def\be{\begin{equation}}
\def\ee{\end{equation}}

\usepackage{color}

\begin{document}

\title{
\begin{flushright}\vbox{\normalsize \mbox{}\vskip -15cm
FTUV/14--1024 \\[-3pt] IFIC/14--70
}
\end{flushright}\vskip 25pt
{\bf A class of invisible axion models\\ with FCNCs at tree level}}
\bigskip

\author{Alejandro Celis$^{a\,}$\thanks{alejandro.celis@ific.uv.es}\;, Javier Fuentes-Mart\'{\i}n$^{a\,}$\thanks{javier.fuentes@ific.uv.es}\;, Hugo Ser\^odio$^{b\,}$\thanks{hserodio@kaist.ac.kr}\\
{$^a$\small \textit{IFIC, Universitat de Val\`encia -- CSIC, Apt. Correus 22085, E-46071 Val\`encia, Spain}}\\
{$^b$\small\textit{Department of Physics, Korea Advanced Institute of Science and Technology,}}\\
{\small\textit{335 Gwahak-ro, Yuseong-gu, Daejeon 305-701, Republic of Korea}}}
\date{}
\maketitle
\bigskip \bigskip
\vspace{-1.cm}

\begin{abstract}
\noindent  We build a class of invisible axion models with tree-level Flavor Changing Neutral Currents completely controlled by the fermion mixing matrices.  The scalar sector of these models contains three-Higgs doublets and a complex scalar gauge singlet, with the same fermionic content as in the Standard Model.   A horizontal Peccei-Quinn symmetry provides a solution to the strong CP problem and predicts the existence of a very light and weakly coupled pseudo-Goldstone boson, the invisible axion or familon.     A phenomenological analysis is performed taking into account familon searches in rare kaon and muon decays, astrophysical considerations and axion searches via axion-photon conversion.   Drastic differences are found in the axion properties of different models due to the strong hierarchy of the CKM matrix, making some of the models considered much more constrained than others.   We also obtain that a rich variety of these models avoid the domain wall problem.     A possible mechanism to protect the solution to the strong CP problem against gravitational effects is also discussed. 

\end{abstract}

\newpage

\tableofcontents

\section{Introduction}
The recent discovery by the ATLAS~\cite{Aad:2012tfa} and CMS~\cite{Chatrchyan:2012ufa} collaborations of a Higgs-like particle with a mass around 125 GeV represents one of the greatest achievements of physics in the last decades and constitutes an indisputable success of the Standard Model (SM) of particle physics. This discovery reinforces the SM as the theory of electroweak and strong interactions, no significant deviations from this framework have been observed so far at LHC, in precision experiments at flavor factories nor in electroweak precision test at LEP.   However, the SM presents several unanswered questions which might be a hint of physics beyond the SM. One of such open questions is the so-called strong CP problem~\cite{Cheng:1987gp}.

The strong CP problem is tightly related to the $\mathrm{U(1)_A}$ problem of Quantum Chromodynamics (QCD). In the limit of massless quarks the QCD Lagrangian shows a chiral $\mathrm{U(1)_A}$ symmetry. The fact that after chiral symmetry breaking its associated pseudo-Goldstone boson was not found experimentally proved that this symmetry should be broken or not realized in nature~\cite{Glashow:1970rp}. This led to an apparent contradiction between theory and experiment which was termed as the $\mathrm{U(1)_A}$ problem. The solution to this issue came from the realization by t'Hooft that non-perturbative QCD effects explicitly break this symmetry~\cite{'tHooft:1976up}. However, with the resolution of this problem a new problem arose. The explicit breaking of the $\mathrm{U(1)_A}$ in QCD leads to the presence of an extra term in the Lagrangian
\be\label{eq:LCPstrong}
\mathcal{L}_{\mbox{\scriptsize CP}}^{\mbox{\scriptsize strong}}=\theta_{\mbox{\scriptsize QCD}} \frac{g_s^2}{32\pi^{2}}\,G_{a,\,\mu\nu}\tilde{G}^{\mu\nu}_a\,,
\ee
where $g_s$ is the strong coupling constant and $G^{\mu\nu}_a$ and $\tilde{G}^{\mu\nu}_a$ are the QCD field-strength tensor and its dual tensor, respectively. This way the QCD vacuum angle, $\theta_{\mbox{\scriptsize QCD}}$, together with $g_s$ remain as the only free parameters of massless QCD. If along with QCD the electroweak (EW) sector is introduced, one should take into account that the quark masses are complex in general. To get the Lagrangian in the physical basis a chiral $\mathrm{U(1)_A}$ transformation should be performed. As a result, the QCD vacuum angle in Eq.~\eqref{eq:LCPstrong} is substituted in the full theory by $\bar{\theta}$ defined as
\begin{align}
\bar{\theta}=\theta_{\mbox{\scriptsize QCD}}+\text{Arg}\left(\text{Det}\, M\right)\,,
\end{align}
being $M$ the quark mass matrix. For $\bar{\theta}\neq0$, Eq.~\eqref{eq:LCPstrong} introduces a violation of P and T but not C and consequently a violation of CP. However, the present bound on neutron dipole moment, $\left|d_n\right|<2.9\times10^{-26}$ e cm~\cite{Baker:2006ts}, set a stringent bound on this angle $\left|\bar{\theta}\right|\lsim10^{-11}$~\cite{Baluni:1978rf}. The reason why this parameter, coming from the strong and the electroweak sectors, is so small is unknown and gives rise to the Strong CP problem.

An elegant solution to the Strong CP problem was given by Peccei and Quinn~\cite{Peccei:1977hh}. This solution, commonly referred as the Peccei-Quinn (PQ) mechanism, consists on the introduction of a global chiral $\mathrm{U(1)}_{\mbox{\scriptsize{PQ}}}$  symmetry with mixed anomalies with QCD. This symmetry effectively replaces the CP-violating phase by a CP-odd field, the so-called axion, which correspond to the pseudo-Goldstone boson resulting from the spontaneous breaking of the PQ symmetry~\cite{Weinberg:1977ma}. The implementation of the PQ mechanism requires the extension of the matter content of the SM. In its original formulation, the scalar sector is enlarged to a two-Higgs-doublet model (2HDM) with the PQ charges implementation enforcing Natural Flavor Conservation (NFC)~\cite{Glashow:1976nt}. This way the severe experimental constraints from Flavor Changing Neutral Currents (FCNCs)~\cite{Bona:2007vi} are avoided. In this model the axion has a mass of few hundreds keV and presents large couplings to matter~\cite{Weinberg:1977ma}. This formulation was soon ruled out by experimental data.

In order to satisfy the experimental constraints one needs to decouple the PQ symmetry breaking and the EW scales. This is achieved by the introduction of a gauge singlet field that acquire a vacuum expectation value (vev) that breaks the PQ symmetry at a scale much higher than the EW scale. This results in invisible axion models where the mass and the couplings of the axion are suppressed by the vev of the scalar singlet and, therefore, are naturally small. In this class of models the axion possesses several interesting features. For instance, the invisible axion is a promising candidate for cold dark matter~\cite{Preskill:1982cy}. Additionally, the type I seesaw mechanism~\cite{Minkowski:1977sc} can be naturally implemented in these models allowing for the possibility to explain the smallness of the active neutrino masses and providing a dynamical origin to the heavy seesaw scale~\cite{Chikashige:1980ui}. 

Two models stand as benchmark invisible axion models: the Dine--Fischler--Srednicki--Zhitnitsky (DFSZ)~\cite{Zhitnitsky:1980tq} and the Kim--Shifman--Vainshtein--Zakharov (KSVZ)~\cite{Kim:1979if} models.   In the KSVZ model one adds to the SM particle content a heavy color triplet and $\mathrm{SU(2)}_L$ singlet vector-like quark and a complex scalar gauge singlet.     The SM fields carry no PQ charge in the KSVZ model.      On the other hand, in the DFSZ model one introduces an additional Higgs doublet and a complex scalar gauge singlet while the PQ symmetry enforces NFC just like in the original PQ model.     In this article we consider models of the DFSZ type where one only enlarges the scalar sector (possibly adding also right-handed neutrinos). In a recent paper~\cite{Celis:2014iua}, the authors presented an invisible axion model where the PQ symmetry is not family universal but rather a horizontal symmetry. As the PQ symmetry cannot be used now to implement NFC, potentially dangerous tree-level FCNCs might be present. The approach followed in this case to avoid large flavor violating scalar couplings was to implement the flavored PQ symmetry in the same fashion as in the Branco-Grimus-Lavoura (BGL) model~\cite{Branco:1996bq}. This way FCNCs appear at tree-level but they are controlled by the Cabibbo--Kobayashi--Maskawa (CKM)~\cite{Cabibbo:1963yz} and the Pontecorvo-Maki-Nakagawa-Sakata (PMNS)~\cite{Pontecorvo:1957qd} matrices. This axion model is characterized by several interesting features. Among these, we stress the possibility of avoiding the domain wall problem~\cite{Zeldovich:1974uw,Sikivie:1982qv,Vilenkin:1984ib}, also the presence of Flavor Changing Axion Interactions (FCAI) can introduce experimental constraints stronger than the astrophysical ones in some cases~\cite{Wilczek:1982rv}.   Invisible axion models with a horizontal PQ symmetry have been built previously in Refs.~\cite{Davidson:1983tp,Hindmarsh:1997ac,Hindmarsh:1998ph,Ahn:2014gva} and in the context of horizontal gauge symmetries in Ref.~\cite{Berezhiani:1990wn}.

The present work is devoted to the extension and detailed analysis of the model presented in Ref.~\cite{Celis:2014iua}. In Sec.~\ref{sec:notation} we introduce the notation and briefly review the main aspects of the BGL model in the two Higgs scenario. Sec.~\ref{sec:proof} is dedicated to the determination of the required conditions for this symmetry to be a chiral PQ symmetry, therefore demanding it to be QCD anomalous. We show that this condition cannot be fulfilled in the two-Higgs-doublet BGL model and that an extension of the scalar sector is required. In Sec.~\ref{sec:sol} we present a three-Higgs-doublet implementation with a horizontal PQ symmetry enforcing a BGL-like suppression in the FCNCs, we refer to this as the three-Higgs flavored Peccei-Quinn (3HFPQ) model. A full study of the axion properties of the model is done in Sec.~\ref{sec:axion}. The domain wall problem and gravitational effects are also considered in this section. All the results shown in the previous sections are obtained in a specific implementation with FCNCs in the down-quark sector and where the top quark is singled out. Sec.~\ref{sec:modelvar} is intended to the study of all the possible models variations. In Sec.~\ref{sec:discus} we perform a phenomenological analysis of the axion in these models taking into account flavor experiments, astrophysical considerations and axion searches via axion-photon conversion. Some details concerning Higgs decoupling scenarios and possible new physics signatures related to the Higgs sector can also be found in this section. We summarize our results and conclude in Sec.~\ref{sec:concl}.

\section{Notation and the Branco-Grimus-Lavoura model}  \label{sec:notation}
In this section we introduce the notation used throughout the article and present the so-called Branco-Grimus-Lavoura (BGL) model~\cite{Branco:1996bq}. We consider a 2HDM with the Higgs doublets parametrized as
\be\label{eq:Phi}
\Phi_j=e^{i\alpha_j}
\begin{pmatrix}
\varphi^+_j\\
\frac{1}{\sqrt{2}}\left(v_j+\rho_j+i\eta_j\right)
\end{pmatrix}\quad(j=1,2)\,.
\ee
Here $v_{1} > 0$ and $v_2 > 0$ generate the quark masses. We also set $\alpha_1 = 0$ and $\alpha_2 \equiv \alpha$ without loss of generality.    Due to the presence of an additional Higgs doublet the Yukawa Lagrangian takes the general form
\be
-\mathcal{L}_{\mbox{\scriptsize{Y}}}=\overline{Q_L^0}\,\left[\Gamma_1\, \Phi_1 +\Gamma_2\, \Phi_2\right]d^0_R+\overline{Q_L^0}\,[\Delta_1\, \widetilde{\Phi}_1+\Delta_2\, \widetilde{\Phi}_2] u^0_R+\text{h.c.} \,,
\ee  
where $\tilde \Phi_j = i \sigma_2 \Phi_j^*$, $\sigma_2$ being the Pauli matrix.   In order to study some of the new phenomena present in this framework it is convenient to work in the Higgs basis, where the Goldstone bosons $G^+$ and $G^0$ are singled out and only one Higgs doublet acquires a non-vanishing vev. For that, we perform the following transformations
\be
\begin{pmatrix}
G^+\\
H^+
\end{pmatrix}=O_2
\begin{pmatrix}
\varphi_1^+\\
\varphi_2^+
\end{pmatrix}\,,\quad \begin{pmatrix}
G^0\\
I
\end{pmatrix}=O_2
\begin{pmatrix}
\eta_1\\
\eta_2
\end{pmatrix}\,,\quad \begin{pmatrix}
H^0\\
R
\end{pmatrix}=O_2
\begin{pmatrix}
\rho_1\\
\rho_2
\end{pmatrix}\,,
\ee
with
\be
O_2=\frac{1}{v}
\begin{pmatrix}
v_1&v_2\\
v_2&-v_1
\end{pmatrix}\, \quad \text{and}\quad v\equiv \sqrt{v_1^2+v_2^2}=\left(\sqrt{2}G_F\right)^{-1/2}\,.
\ee
Expanding the Yukawa Lagrangian in the Higgs basis one obtains
\begin{align}
\begin{split}
-\mathcal{L}_{\mbox{\scriptsize{Y}}}\;\supset\;&\,  \frac{1}{v} \Bigl\{   \overline{d_L^0}  \left[vM_d+M_d H^0+N_d^0 R+iN_d^0I\right]d_R^0+\overline{u_L^0} \left[vM_u+M_u H^0+N_u^0 R-iN_u^0I\right]u_R^0\\
&+ \sqrt{2}H^+  \left(\overline{u_L^0}  \,N_d^0 \,d_R^0-\overline{u_R^0}\, N_u^{0\dagger} \,d_L^0\right)+\text{h.c.} \Bigr\}  \,.
\end{split}
\end{align}
 The matrices $M_q$ and $N^0_q$ $(q=u,d)$ encode the flavor structure in the 2HDM, these are given by
\be
M_d=\frac{1}{\sqrt{2}}\left(v_1\Gamma_1+v_2e^{i\alpha}\Gamma_2\right)\,,\quad M_u=\frac{1}{\sqrt{2}}\left(v_1\Delta_1+v_2e^{-i\alpha}\Delta_2\right) \,,
\ee
and
\begin{align}
\begin{split}
N_d^0\;=\;&\frac{1}{\sqrt{2}}\left(v_2\Gamma_1-v_1e^{i\alpha}\Gamma_2\right)=\frac{v_2}{v_1}M_d-\frac{v_2}{\sqrt{2}}\left(\frac{v_2}{v_1}+\frac{v_1}{v_2}\right)e^{i\alpha}\Gamma_2\,,\\
N_u^0\;=\;&\frac{1}{\sqrt{2}}\left(v_2\Delta_1-v_1e^{-i\alpha}\Delta_2\right)=\frac{v_2}{v_1}M_u-\frac{v_2}{\sqrt{2}}\left(\frac{v_2}{v_1}+\frac{v_1}{v_2}\right)e^{-i\alpha }\Delta_2\,.
\end{split}
\end{align}
The quark mass matrices $M_{u,d}$ determine the Yukawa couplings of the scalar field $H^0$ while the matrices $N_{u,d}^0$ determine the Yukawa couplings of the scalar $R$ and the pseudoscalar $I$.   Note that the fields $\{H^0, R, I\}$ are not mass eigenstates in general, the physical neutral scalar bosons will correspond to a linear combination of these fields.   

The quark mass matrices can be diagonalized through the bi-unitary transformations:
\be\label{eq:udmassbasis}
u_{L,R}^0=U_{uL,R}\, u_{L,R}\,,\quad d_{L,R}^0=U_{dL,R} \, d_{L,R}\,,
\ee
chosen appropriately so that
\be
U^\dagger _{uL} M_u U_{uR}=D_u=\text{diag}\left(m_u,\, m_c,\, m_t\right)\,,\quad U^\dagger_{dL} M_d U_{dR}=D_d=\text{diag}\left(m_d,\, m_s,\, m_b\right)  \,.
\ee
These transformations guarantee diagonal quark couplings for $H^0$ but, in general,
\be
N_u=U_{uL}^\dagger N_u^0 U_{uR}\neq \text{diag}\,,\quad
N_d=U_{dL}^\dagger N_d^0 U_{dR}\neq \text{diag}\,,
\ee
so that $R$ and $I$ have flavor violating couplings at tree-level.   These are the sources of dangerous FCNCs at tree level in the 2HDM. The most common solution to this problem is the NFC condition. This scenario is nothing more than the requirement of simultaneous diagonalization of $M_{u,d}$ and $N_{u,d}^0$, or equivalently, the simultaneous diagonalization of the Yukawa textures in each sector. In the two Higgs doublet models the NFC condition can be implemented in two ways:
\begin{itemize}
\item Through a discrete or continuous symmetry which restricts the number of Yukawas in each sector to one~\cite{Glashow:1976nt}.
\item Using the alignment condition, where the Yukawa matrices in the same sector have the same flavor structure up to an overall factor~\cite{Pich:2009sp}. This can be seen as an effective theory of a larger model with the first condition imposed at the UV level~\cite{Bae:2010ai}. It can also be seen as a first order expansion in a minimal flavor violating scenario~\cite{Buras:2000dm,Botella:2009pq,Buras:2010mh}.
\end{itemize}
There exists, however, a different scenario where NFC is only imposed in one sector and FCNCs present in the other sector are under control, this is known as the BGL model~\cite{Branco:1996bq}. The model uses an abelian symmetry to impose the Yukawa textures
\beqn\label{eq:BGLTextures}
\begin{split}
\Gamma_1^{\mbox{\scriptsize{BGL}}}&=&
\begin{pmatrix}
\times&\times&\times\\
\times&\times&\times\\
0&0&0
\end{pmatrix}\,,\quad
\Gamma_2^{\mbox{\scriptsize{BGL}}}=
\begin{pmatrix}
0&0&0\\
0&0&0\\
\times&\times&\times
\end{pmatrix}\,, \\[0.3cm]
\Delta_1^{\mbox{\scriptsize{BGL}}}&=&
\begin{pmatrix}
\times&\times&0\\
\times&\times&0\\
0&0&0
\end{pmatrix} \,\,,\quad
\Delta_2^{\mbox{\scriptsize{BGL}}}=
\begin{pmatrix}
0&0&0\\
0&0&0\\
0&0&\times
\end{pmatrix} \,. 
\end{split}
\eeqn
This implementation is also known as the top-BGL, since it singles out the top quark. In this case the flavor matrices responsible for the FCNCs take the form
\begin{align}\label{eq:NBGL}
\begin{split}
\left(N_{d}\right)^{\mbox{\scriptsize{BGL}}}_{ij}\;=\;&\frac{v_2}{v_1}\left(D_d\right)_{ij}-\left(\frac{v_2}{v_1}+\frac{v_1}{v_2}\right)(V^\dagger)_{i3}(V)_{3j}(D_d)_{jj} \,, \\
\left(N_u\right)^{\mbox{\scriptsize{BGL}}}\;=\;& \frac{v_2}{v_1}\text{diag}(m_u,m_c,0) -\frac{v_1}{v_2}\text{diag}(0,0,m_t)  \,,
\end{split}
\end{align}
with $V=U^\dagger_{uL}U_{dL}$ the CKM quark mixing matrix. This simple implementation introduces no FCNC effects in the up-quark sector. In the down-quark sector, one has tree-level FCNCs, however, those are highly suppressed. We can see from Eq.~\eqref{eq:NBGL} that the second term of $N_d$ will introduce FCNCs which are suppressed by:
\begin{itemize} 
\item The down-type quark masses.
\item The off-diagonal CKM matrix elements.
\end{itemize}
This way, the model implements controllable FCNCs at tree level within the 2HDM. As shown in Refs.~\cite{Ferreira:2010ir,Serodio:2013gka} this implementation is unique, up to trivial permutations, in models with abelian symmetries. A detailed phenomenological study of the experimental constraints on this model was presented in Refs.~\cite{Botella:2014ska,Bhattacharyya:2013rya,Bhattacharyya:2014nja
}.  

Although the BGL model presents several unique features, it still suffers from a few problems. The first problem is present in the scalar potential of the model. While the abelian flavor symmetry used to get the desired textures can be implemented through a discrete group, the scalar sector will exhibit an accidental global $\mathrm{U(1)}$ symmetry leading to a Goldstone boson after spontaneous symmetry breaking~\cite{Branco:1996bq}.    Alternatives to eliminate the accidental global symmetry have been discussed in Ref.~\cite{Branco:1996bq}, adding soft breaking terms to the scalar potential or extending the scalar sector with gauge singlet fields could protect the model against the phenomenologically dangerous Goldstone modes.    On the other hand, the strong CP problem is not addressed in this scenario.  While there are no large contributions to electric dipole moments in the BGL model~\cite{Botella:2012ab}, this is based on the assumption of a vanishing or very small $\overline{\theta}$ term~\cite{Cheng:1987gp}.    

In this work we suggest that these apparent problems of the BGL model could be solved in an unified way if the required Yukawa textures are imposed by a global chiral $\mathrm{U(1)}_{\mbox{\scriptsize{PQ}}}$ symmetry, bringing also other advantages we will discuss in the following sections.   The model then provides a dynamical solution to the strong CP problem via the PQ mechanism while an axion appears in the spectrum, which could in principle account for the dark matter of the Universe.

\section{The anomalous condition for a BGL-like model} \label{sec:proof}

In this section we shall find the anomalous condition for the abelian continuous symmetry that imposes the BGL Yukawa textures, extending the analyses done in Ref.~\cite{Celis:2014iua}. We are then interested in finding the abelian generators under which the fields must transform, i.e.
\be
Q_L^0\rightarrow \mathcal{S}_L\,Q_L^0\,,\quad d_R^0\rightarrow \mathcal{S}_R^d\, d_R^0\,,\quad u_R^0\rightarrow \mathcal{S}_R^u \,u_R^0\,,
\ee
with
\begin{align}
\begin{split}
\mathcal{S}_L=&\text{diag}(e^{iX_{uL}\,\theta},e^{iX_{cL}\,\theta},e^{iX_{tL}\,\theta})\,,\quad\mathcal{S}_{R}^d=\text{diag}(e^{iX_{dR}\,\theta},e^{iX_{sR}\,\theta},e^{iX_{bR}\,\theta})\,,\\\quad\mathcal{S}_R^u=&\text{diag}(e^{iX_{uR}\,\theta},e^{iX_{cR}\,\theta},e^{iX_{tR}\,\theta})\,,\\
\end{split}
\end{align}
and
\be\label{eq:PhiTrans}
\Phi\rightarrow \mathcal{S}_\Phi\, \Phi\,,
\ee
with
\be
\mathcal{S}_\Phi=\text{diag}(e^{iX_{\Phi 1}\,\theta},e^{iX_{\Phi 2}\,\theta})\,.
\ee
These field transformations will induce the following constraints
\be
\mathcal{S}_L^\dagger \,\Gamma_k \,\mathcal{S}_{R}^{d}\,\left(\mathcal{S}_{\Phi}\right)_{kk}=\Gamma_k\,,\quad \mathcal{S}_L^\dagger \,\Delta_k \,\mathcal{S}_{R}^{u}\,\left(\mathcal{S}^\ast_{\Phi}\right)_{kk}=\Delta_k\,,
\ee
with $k=1, 2$. The Yukawa texture patterns are dictated by the way the fermion fields transform, the Higgs field transformation will only select one of the allowed textures~\cite{Botella:2009pq,Botella:2012ab,Ferreira:2010ir,Serodio:2013gka}. The best way to find these fermion transformations is to study the Hermitian combinations $\Gamma_k\Gamma_k^\dagger$ and $\Gamma_k^\dagger \Gamma_k$ (and similarly for $\Delta_k$). The symmetry constraints on these combinations give
\begin{align}
\begin{split}
&\mathcal{S}_{L}^\dagger\, \left\{\Gamma_k \Gamma_k^\dagger\,, \Delta_k \Delta_k^\dagger\right\}\,\mathcal{S}_L= \left\{\Gamma_k \Gamma_k^\dagger\,, \Delta_k \Delta_k^\dagger\right\}\,,
\quad\mathcal{S}_{R}^{(d,u)\dagger}\, \left\{\Gamma^\dagger_k \Gamma_k\,, \Delta^\dagger_k \Delta_k\right\}\,\mathcal{S}^{(d,u)}_R= \left\{\Gamma^\dagger_k \Gamma_k\,, \Delta^\dagger_k \Delta_k\right\}\,.
\end{split}
\end{align}
The above equations are nothing more than the commutation of a diagonal matrix $\mathcal{S}_{L,R}$ with a Hermitian matrix. In order for these matrices to commute $\mathcal{S}_{L,R}$ must share the same eigenvectors as the Hermitian combination, or have degenerate eigenvalues for the non-shared eigenvectors. We then get three scenarios:
\begin{itemize} 
\item[i)] The matrix $\mathcal{S}_{L,R}$ has only one phase. The Hermitian combination has no restriction;

\item[ii)] The matrix $\mathcal{S}_{L,R}$ has two different phases. The Hermitian combination must be block diagonal, with the $2\times 2$ block in the same sector as the degeneracy in $\mathcal{S}_{L, R}$; 

\item[iii)] The matrix $\mathcal{S}_{L,R}$ has three different phases. The Hermitian combination must be diagonal.
\end{itemize}
  
From Eq.~\eqref{eq:BGLTextures} we see that the $\Delta_j$ Yukawa textures are block diagonal in the up-charm sector. The hermitian combinations $\Delta_k \Delta_k^\dagger$ and $\Delta_k^\dagger \Delta_k$ will also share the same form. This in turn implies that the symmetry generators for the left- and right-handed fields must belong to case $ii)$, i.e. the abelian generators have only two different phases
\be\label{eq:chargesaBGL1}
\mathcal{S}_L=\text{diag}\left(1,1,e^{iX_{tL}\,\theta}\right)\,,\quad \mathcal{S}^u_R=\text{diag}\left(e^{iX_{uR}\,\theta},e^{iX_{uR}\,\theta},e^{iX_{tR}\,\theta}\right)\,,
\ee
where we set one of the charges to zero using a global phase transformation. Notice that the charges should satisfy the conditions
\be \label{CA1}
\text{Condition A:}\quad X_{tL}\neq0\, \qquad \text{and} \qquad X_{uR}\neq X_{tR} \,, 
\ee
in order to stay in the scenario $ii)$. When the left-handed quark doublet and the right-handed up-type quark transform under this symmetry, the phases appearing in the Yukawa term are
\begin{equation}
\Theta_u\;=\;\theta
\begin{pmatrix}
X_{uR}&X_{uR}&X_{tR}\\
X_{uR}&X_{uR}&X_{tR}\\
X_{uR}-X_{tL}&X_{uR}-X_{tL}&X_{tR}-X_{tL}
\end{pmatrix}\,,
\end{equation}
with the additional condition
\be\label{CA2}
\text{Condition B:}\quad X_{tL}\neq-(X_{uR}-X_{tR}) \,.
\ee 
To the matrix $\Theta_u$ we call the up-quark phase transformation matrix. The generators in Eq.~\eqref{eq:chargesaBGL1}, together with the conditions A and B are the complete and minimal set of required conditions in order to have available the BGL textures for the up sector. In order to pick the desired textures we now have to attribute the correct charges to the Higgs fields. Remembering that in the up-quark sector we have the $\tilde{\Phi}_i$ field coupling, we choose for the scalar fields
\be
\mathcal{S}^{\mbox{\scriptsize{up}}}_{\Phi}=\text{diag}\left(e^{iX_{uR}\,\theta},e^{i(X_{tR}-X_{tL})\,\theta}\right)\,.
\ee
This choice makes $\tilde{\Phi}_j$ associated with the $\Delta_j$ of Eq.~\eqref{eq:BGLTextures}. We can now build the phase transformation matrix for the down-quark sector. The left-handed transformation is the same, since it is shared by the two sectors. Concerning the right-handed generator of $\Gamma_j$ (see Eq.~\eqref{eq:BGLTextures}), it belongs to the case $i)$ and, therefore, has the form
\be\label{eq:chargesaBGL2}
\mathcal{S}_R^d=e^{iX_{dR}\theta} \, \mathbb{I}\,.
\ee
The down-quark phase transformation matrix is then given by
\begin{equation}
\Theta_d\;=\;\theta
\begin{pmatrix}
X_{dR}&X_{dR}&X_{dR}\\
X_{dR}&X_{dR}&X_{dR}\\
X_{dR}-X_{tL}&X_{dR}-X_{tL}&X_{dR}-X_{tL}
\end{pmatrix}\,.
\end{equation}
Eqs.~\eqref{eq:chargesaBGL1} and~\eqref{eq:chargesaBGL2} together with the first part of condition A are the minimal set of required conditions necessary to obtain the BGL textures in the down sector. In order to pick the desired textures we would need the scalar transformation
\be
\mathcal{S}_{\Phi}^{\mbox{\scriptsize{down}}}=\text{diag}\left(e^{-iX_{dR}\,\theta},e^{i(X_{tL}-X_{dR})\,\theta}\right) \,.
\ee

To make the BGL textures in the up and down sectors compatible without introducing additional textures that spoil the nice features of the BGL-type models we need to impose some extra charge conditions, we call them texture matching conditions. They guarantee that the only non-BGL textures present in the Yukawa sector are the null textures, 
\begin{equation}\label{eq:compCond}
X_{dR}\neq -X_{tR},\quad
X_{tL}\neq X_{uR}+X_{dR},\quad
X_{tL}\neq X_{tR}+X_{dR},\quad
X_{tL}\neq \frac{1}{2}\left(X_{uR}+X_{dR}\right).
\end{equation}
Finally, we have to require an additional texture matching condition indicating how the up and down sectors match. In the original BGL formulation it is crucial that the Higgs doublet coupling to the $\Gamma^{\mbox{\scriptsize{BGL}}}_1$ and $\Delta^{\mbox{\scriptsize{BGL}}}_1$ textures is the same, the other possible implementation would violate one of the texture matching conditions in Eq.~\ref{eq:compCond}. This automatically leads to $X_{dR}=-X_{uR}$.

Since we introduced a chiral symmetry we get the anomaly free condition for the PQ symmetry with the QCD currents
\be
2X_{tL}-\left(2X_{uR}+X_{tR}+3X_{dR}\right)=0\quad\Rightarrow\quad X_{tL}=X_{uR}+\frac{1}{2}X_{tR}+\frac{3}{2}X_{dR}\,.
\ee
This anomaly free condition makes both $\mathcal{S}_{\Phi}^{\mbox{\scriptsize{up}}}$ and $\mathcal{S}_{\Phi}^{\mbox{\scriptsize{down}}}$ equal for $X_{dR}=-X_{uR}$, making the BGL implementation consistent and anomaly free with two Higgs doublets. If we want this symmetry to be anomalous then $ X_{tL}\neq-1/2( X_{uR}-X_{tR})$ and the model must be extended. We shall pursue a possible anomalous implementation in the multi-Higgs framework, making the three-Higgs doublet model the minimal extension.  In the three Higgs implementation we can just join the scalar generators $\mathcal{S}_{\Phi}^{\mbox{\scriptsize{up}}}$ and $\mathcal{S}_{\Phi}^{\mbox{\scriptsize{down}}}$ into a single one
\be  \label{caseIeq}
\mathcal{S}_{\Phi}=\text{diag}\left(e^{iX_{uR}\,\theta},e^{i(X_{tR}-X_{tL})\,\theta},e^{i(X_{tL}-X_{dR})\,\theta}\right)\,.
\ee
In this three-Higgs doublet model implementation we get the following Yukawa textures
\be\label{eq:aBGL}
\begin{array}{ccc}
\Gamma_1=
\begin{pmatrix}
\times&\times&\times\\
\times&\times&\times\\
0&0&0
\end{pmatrix}\,,&\Gamma_2=0\,,&
\Gamma_3=
\begin{pmatrix}
0&0&0\\
0&0&0\\
\times&\times&\times
\end{pmatrix} \,,  \\[1.2cm]
\Delta_1=
\begin{pmatrix}
\times&\times&0\\
\times&\times&0\\
0&0&0
\end{pmatrix}\,,&
\qquad \Delta_2=
\begin{pmatrix}
0&0&0\\
0&0&0\\
0&0&\times
\end{pmatrix}\,,&\Delta_3=0\,,
\end{array}
\ee
with the charge constraints
\begin{align}\label{eq:compCondTI}
\begin{split}
\text{Texture Matching Conditions: }&\left\{
\begin{array}{l}
X_{uR}=-X_{dR},\\
X_{uR}\neq X_{tR},\\
X_{tL}\neq X_{tR}-X_{uR},
\end{array}\right.\\
\text{Anomaly condition: }&X_{tL}\neq -\frac{1}{2}\left(X_{uR}-X_{tR}\right).
\end{split}
\end{align}

Since we extended the Higgs sector, in principle it is no longer necessary to have $\Gamma^{\mbox{\scriptsize{BGL}}}_1$ and $\Delta^{\mbox{\scriptsize{BGL}}}_1$ coupling to the same Higgs doublet. We can have another three different implementations:
\begin{itemize}
\item $\Gamma^{\mbox{\scriptsize{BGL}}}_1$ with $\Delta^{\mbox{\scriptsize{BGL}}}_2$. This implies $-X_{dR}=X_{tR}-X_{tL}$;

\item $\Gamma^{\mbox{\scriptsize{BGL}}}_2$ with $\Delta^{\mbox{\scriptsize{BGL}}}_1$. This implies $X_{uR}=X_{tL}-X_{dR}$;

\item $\Gamma^{\mbox{\scriptsize{BGL}}}_2$ with $\Delta^{\mbox{\scriptsize{BGL}}}_2$. This implies $X_{tR}-X_{tL}=X_{tL}-X_{dR}$.
\end{itemize}
However, the first two implementations violate the charge restrictions in Eq.~\eqref{eq:compCond} whereas the third one is a safe implementation and gives
\be  \label{caseIIeq}
\mathcal{S}_{\Phi}=\text{diag}\left(e^{iX_{uR}\,\theta},e^{-iX_{dR}\,\theta},e^{i(X_{tL}-X_{dR})\,\theta}\right)\,,
\ee
so we get the following Yukawa textures implementation
\be\label{eq:aBGLII}
\begin{array}{ccc}
\Gamma_1=
0\,,&\Gamma_2=\begin{pmatrix}
\times&\times&\times\\
\times&\times&\times\\
0&0&0
\end{pmatrix}\,,&
\Gamma_3=
\begin{pmatrix}
0&0&0\\
0&0&0\\
\times&\times&\times
\end{pmatrix} \,, \\[1.2cm]
\Delta_1=
\begin{pmatrix}
\times&\times&0\\
\times&\times&0\\
0&0&0
\end{pmatrix}\,,&
\Delta_2=0
\,,&\Delta_3=
\begin{pmatrix}
0&0&0\\
0&0&0\\
0&0&\times
\end{pmatrix}\,.
\end{array}
\ee
For this symmetry to be anomalous and in order not to introduce additional textures that spoil the desired behavior of the model we need to guarantee, in analogy to the previous case, the following charge restrictions
\begin{align}
\begin{split}\label{eq:compCondTII}
\text{Texture Matching Conditions: }&\left\{
\begin{array}{l}
X_{tR}=2X_{tL}-X_{dR},\\
X_{tL}\neq X_{uR}+X_{dR},\\
X_{tL}\neq \frac{1}{2}\left(X_{uR}+X_{dR}\right),
\end{array}\right.\\
\text{Anomaly Condition: }&X_{uR}\neq -X_{dR}.
\end{split}
\end{align}

The BGL 2HDM model needs condition B in its anomaly free implementation. However, when extending it to a three Higgs scenario, with the possibility of null couplings, this condition no longer needs to be satisfied. Relaxing this condition by setting $X_{tR}=X_{uR}+X_{tL}$, we get a new type of texture in the up sector (the combination of $\Delta_1^{\mbox{\scriptsize{BGL}}}$ and $\Delta_2^{\mbox{\scriptsize{BGL}}}$). Three new possible implementations become available:
\begin{itemize}
\item New texture coupling to $\Gamma_1^{\mbox{\scriptsize{BGL}}}$. This implies $X_{dR}=-X_{uR}$;
\item New texture coupling to $\Gamma_2^{\mbox{\scriptsize{BGL}}}$. This implies $X_{dR}-X_{tL}=-X_{uR}$;
\item New texture coupling to a null texture.
\end{itemize}
The first two cases violate the charge conditions in Eq.~\eqref{eq:compCond}, this is the reason why there is no BGL 2HDM with this texture. However, the third possibility gives a safe implementation in a three Higgs scenario with the same scalar charge assignments as in the previous case (i.e. Eq.~\eqref{caseIIeq}). The Yukawa textures implementation is then given by
\be\label{eq:aBGLIII}
\begin{array}{ccc}
\Gamma_1=
0\,,&\Gamma_2=\begin{pmatrix}
\times&\times&\times\\
\times&\times&\times\\
0&0&0
\end{pmatrix}\,,&
\Gamma_3=
\begin{pmatrix}
0&0&0\\
0&0&0\\
\times&\times&\times
\end{pmatrix} \,, \\[1.2cm]
\Delta_1=
\begin{pmatrix}
\times&\times&0\\
\times&\times&0\\
0&0&\times
\end{pmatrix}\,,&
\Delta_2=0
\,,&\Delta_3=0\,.
\end{array}
\ee
For the symmetry to be anomalous and to guarantee that we introduce no additional Yukawa textures, the following charge conditions apply
\begin{align}
\begin{split}\label{eq:compCondTIII}
\text{Texture Matching Conditions: }&\left\{
\begin{array}{l} 
X_{uR}\neq-X_{dR},\\
X_{tL}\neq X_{uR}+X_{dR},\\
X_{tL}\neq -\left(X_{uR}+X_{dR}\right),\\
X_{tL}\neq \frac{1}{2}\left(X_{uR}+X_{dR}\right),
\end{array}\right.\\
\text{Anomaly Condition: }&X_{tL}\neq 3\left(X_{uR}+X_{dR}\right).
\end{split}
\end{align}

In conclusion, in this section we have shown that it is not possible to build an anomalous two-Higgs-doublet model \`a la BGL and we have found three different implementations of the PQ symmetry for the three-Higgs-doublet model, up to permutations in the family or in the up-down sectors. These three cases are built from the generators in Eqs.~\eqref{eq:chargesaBGL1} and \eqref{eq:chargesaBGL2}. They read as follows:
\begin{itemize}
 \item \textbf{Case I:} where the Yukawa textures are given by Eq.~\eqref{eq:aBGL}, satisfies conditions A and B, and also the texture matching and anomaly conditions in Eq.~\eqref{eq:compCondTI}. 

The charges associated with the Higgs fields are
 \be\label{eq:PhiCaseI}
 X_{\Phi 1}=X_{uR}\,,\quad X_{\Phi 2}=X_{tR}-X_{tL}\,,\quad X_{\Phi 3}=X_{tL}+X_{uR}\,.
 \ee

\item \textbf{Case II:} with the Yukawa textures shown in Eq.~\eqref{eq:aBGLII}, satisfies conditions A and B, and also the texture matching and anomaly conditions in Eq.~\eqref{eq:compCondTII}. 

The charges associated with the Higgs fields are
 \be\label{eq:PhiCaseII}
 X_{\Phi 1}=X_{uR}\,,\quad X_{\Phi 2}=-X_{dR}\,,\quad X_{\Phi 3}=X_{tL}-X_{dR}\,.
 \ee
\item \textbf{Case III:} with the Yukawa textures shown in Eq.~\eqref{eq:aBGLIII}, satisfies the constraint $X_{tR} = X_{uR} + X_{tL}$, condition A, and also the texture matching and anomaly conditions in Eq.~\eqref{eq:compCondTIII}. 

The charges associated with the Higgs fields are the same as in case II.

\end{itemize}

\section{The three-Higgs-doublet class of anomalous models} \label{sec:sol}
In the previous section we have shown that the Yukawa textures in the BGL 2HDM cannot be imposed by a chiral PQ symmetry.    We also derived the necessary conditions to build three-Higgs doublet models with FCNC at tree-level completely determined by the fermion mixing matrices. In the latter scenario, we obtained all the possible Yukawa texture implementations imposed by a PQ symmetry and determined the restrictions to the PQ charges in each case.  We provide details about the quark Yukawa sector of these type of models in Sec.~\ref{ssec:quarks}.    The scalar potential of this class of models is discussed in detail in Sec.~\ref{ssec:scalar}.   The extension of the models considered to the leptonic sector is discussed in Sec.~\ref{ssec:leptonic}.

\subsection{The Yukawa quark sector \label{ssec:quarks}}
In similar a fashion to what was done in Sec.~\ref{sec:notation}, we shall build the relevant flavor matrix combinations that mediate the FCNCs.  The Yukawa Lagrangian in the three Higgs scenario is now written as 
\be
-\mathcal{L}_{\mbox{\scriptsize{Y}}}=\overline{Q_L^0}\,\left[\Gamma_1\, \Phi_1 +\Gamma_2\, \Phi_2+\Gamma_3\, \Phi_3\right]d^0_R+\overline{Q_L^0}\,[\Delta_1\, \widetilde{\Phi}_1+\Delta_2\, \widetilde{\Phi}_2+\Delta_3\, \widetilde{\Phi}_3] u^0_R+\text{h.c.}\,,
\ee  
where we just keep the same notation as in Eq.~\eqref{eq:Phi}, but for $j=1,2,3$ in this case.  We go once more to the Higgs basis, by preforming the following transformations
\be
\begin{pmatrix}  \label{eq:HiggsbasisG}
G^+\\
H^+\\
H^{\prime+}
\end{pmatrix}=O_3
\begin{pmatrix}
\varphi_1^+\\
\varphi_2^+\\
\varphi_3^+
\end{pmatrix}\,,\quad \begin{pmatrix}
G^0\\
I\\
I^\prime
\end{pmatrix}=O_3
\begin{pmatrix}
\eta_1\\
\eta_2\\
\eta_3
\end{pmatrix}\,,\quad \begin{pmatrix}
H^0\\
R\\
R^\prime
\end{pmatrix}=O_3
\begin{pmatrix}
\rho_1\\
\rho_2\\
\rho_3
\end{pmatrix}\,,
\ee
with 
\be
O_3=
\begin{pmatrix}
\dfrac{v_1}{v}&\dfrac{v_2}{v}&\dfrac{v_3}{v}\\
\dfrac{v_2}{v^\prime}&-\dfrac{v_1}{v^\prime}&0\\
\dfrac{v_1}{v^{\prime\prime}}&\dfrac{v_2}{v^{\prime\prime}}&-\dfrac{v^{\prime 2}}{v^{\prime\prime}v_3}
\end{pmatrix}\,,\quad v=\sqrt{v_1^2+v_2^2+v_3^2}\,,\quad v^\prime=\sqrt{v_1^2+v_2^2}\,,\quad v^{\prime\prime}=\frac{v^\prime v}{v_3}\,.
\ee
In the Higgs basis the mass and Yukawa interactions are given by
\begin{align}
\begin{split}
-\mathcal{L}_{\mbox{\scriptsize{Y}}}=&\,\overline{d_L^0}\left[M_d+\frac{1}{v}M_d H^0+\frac{1}{v^\prime}N_d^0 R+\frac{1}{v^{\prime\prime}}N_d^{\prime 0} R^\prime+i\frac{1}{v^\prime}N_d^0I+i\frac{1}{v^{\prime\prime}}N_d^{\prime0}I^\prime\right]d_R^0\\
&+\overline{u_L^0}\left[M_u+\frac{1}{v}M_u H^0+\frac{1}{v^\prime}N_u^0 R+\frac{1}{v^{\prime\prime}}N_u^{\prime0} R^\prime-i\frac{1}{v^\prime}N_u^0I-i\frac{1}{v^{\prime\prime}}N_u^{\prime0}I^\prime\right]u_R^0\\
&+\frac{\sqrt{2}}{v^\prime}H^+\left(\overline{u^0_L}N_d^0d^0_R-\overline{u^0_R}N_u^{0\dagger}d^0_L\right)
+\frac{\sqrt{2}}{v^{\prime\prime}}H^{\prime +}\left(\overline{u^0_L}N_d^{\prime 0}d^0_R-\overline{u^0_R}N_u^{\prime 0\dagger}d^0_L\right)+\text{h.c.} \,,
\end{split}
\end{align}
with the flavor matrices given by
\be
M_d=\frac{1}{\sqrt{2}}(v_1e^{i\alpha_1}\Gamma_1+v_2e^{i\alpha_2}\Gamma_2+v_3e^{i\alpha_3}\Gamma_3)\,,\quad
M_u=\frac{1}{\sqrt{2}}(v_1e^{-i\alpha_1}\Delta_1+v_2e^{-i\alpha_2}\Delta_2+v_3e^{-i\alpha_3}\Delta_3)\,,
\ee
and
\begin{align}\label{eq:NuNd3}
\begin{split}
N_d^0=&\frac{1}{\sqrt{2}}\left(v_2e^{i\alpha_1}\Gamma_1-v_1e^{i\alpha_2}\Gamma_2\right)\,,  \\
N_u^0=&\frac{1}{\sqrt{2}}\left(v_2e^{-i\alpha_1}\Delta_1-v_1e^{-i\alpha_2}\Delta_2\right)  \,, \\
N_d^{\prime0}=&\frac{1}{\sqrt{2}}\left(v_1e^{i\alpha_1}\Gamma_1+v_2e^{i\alpha_2}\Gamma_2-\frac{v'^2}{v_3}e^{i\alpha_3}\Gamma_3\right) \,,   \\
N_u^{\prime0}=&\frac{1}{\sqrt{2}}\left(v_1e^{-i\alpha_1}\Delta_1+v_2e^{-i\alpha_2}\Delta_2-\frac{v'^2}{v_3}e^{-i\alpha_3}\Delta_3\right) \,.
\end{split}
\end{align}
These last flavor matrix combinations are the ones mediating the FCNCs in our framework. We can now evaluate them for each of the three cases. In the basis where the quarks are mass eigenstates we get:
\begin{itemize}
\item \textbf{Case I:}
\be
\begin{array}{ll}
\left(N_d\right)_{ij}=\dfrac{v_2}{v_1}\left(D_d\right)_{ij}-\dfrac{v_2}{v_1}(V^\dagger)_{i3}(V)_{3j}(D_d)_{jj}\,,&N_u=-\dfrac{v_1}{v_2}\text{diag}(0,0,m_t)+\dfrac{v_2}{v_1}\text{diag}(m_u,m_c,0) \,, \\
\left(N_d^{\prime}\right)_{ij}=(D_d)_{ij}-\dfrac{v^2}{v^2_3}(V^\dagger)_{i3}(V)_{3j}(D_d)_{jj}\,, &N_u^{\prime}= D_u  \,.
\end{array}
\ee

\item \textbf{Case II:}
\be
\begin{array}{ll}
\left(N_d\right)_{ij}=-\dfrac{v_1}{v_2}\left(D_d\right)_{ij}+\dfrac{v_1}{v_2}(V^\dagger)_{i3}(V)_{3j}(D_d)_{jj}\,,&N_u=\dfrac{v_2}{v_1}\text{diag}(m_u,m_c,0) \,, \\
\left(N_d^{\prime}\right)_{ij}=(D_d)_{ij}-\dfrac{v^2}{v^2_3}(V^\dagger)_{i3}(V)_{3j}(D_d)_{jj}\,, &N_u^{\prime} = \text{diag}(m_u,m_c,0)-\dfrac{v'^2}{v_3^2}\text{diag}(0,0,m_t)  \,.
\end{array}
\ee

\item \textbf{Case III:}
\be
\begin{array}{ll}
\left(N_d\right)_{ij}=-\dfrac{v_1}{v_2}\left(D_d\right)_{ij}+\dfrac{v_1}{v_2}(V^\dagger)_{i3}(V)_{3j}(D_d)_{jj}\,,&N_u=\dfrac{v_2}{v_1}D_u \,, \\
\left(N_d^{\prime}\right)_{ij}=(D_d)_{ij}-\dfrac{v^2}{v^2_3}(V^\dagger)_{i3}(V)_{3j}(D_d)_{jj}\,, &N_u^{\prime} =D_u  \,.
\end{array}
\ee
\end{itemize}

As expected, in all cases the FCNCs will be mediated by quark masses and off-diagonal elements of the CKM quark mixing matrix. This is virtually the same type of suppression as the one obtained in the BGL 2HDM implementation. The difference lies in the vevs ratios that we get in front of each term. This actually contrasts with the anomaly free three Higgs BGL implementation~\cite{Botella:2009pq}. In that scenario the Yukawa textures, which differ from the 2HDM implementation, cannot give such a strong suppression to $|\Delta S|=2$ processes as compared to the original BGL implementation. One generally gets suppressions of the order of $(V_{cd}^\ast V_{cs})^2\sim\lambda^2$ ($\lambda \simeq 0.225$), requiring heavy neutral scalar fields. 
However, the fact that we kept the same Yukawa textures in passing from the two to the three Higgs implementation allows us to have suppressions of the type $(V_{td}^\ast V_{ts})^2\sim\lambda^{10}$ for $|\Delta S|=2$ processes, just like the original BGL scenario. 

\subsection{The scalar potential \label{ssec:scalar} }
Current experimental limits exclude axions coming from a PQ symmetry broken at the EW scale~\cite{Krauss:1986bq,Bardeen:1986yb,Feng:1997tn,Hindmarsh:1998ph}.  To obtain a viable axion model the PQ symmetry must be broken at a scale much higher than the EW scale.  The axion is then called invisible since its mass and couplings are suppressed by the large PQ symmetry breaking scale.   We can achieve this in a similar way as in the DFSZ and KSVZ invisible axion models, that is, by introducing a complex scalar singlet which acquires a very large vev $\langle0 |S| 0 \rangle = e^{i \alpha_{\mbox{\tiny PQ}}} v_{\mbox{\scriptsize PQ}}/\sqrt{2}$, with $v_{\mbox{\scriptsize PQ}}  \gg v$. The new complex field $S$ will have the following symmetry transformation
\be
S\rightarrow e^{iX_S\,\theta}S\,.
\ee
The introduction of the complex scalar singlet increases the number of independent charges in one unity. From the Yukawa sector alone, with fermion charges chosen in order for the symmetry to be anomalous, we are able to reduce the number of independent PQ charges to just three. In this way the number of independent charges increases to four. 

The scalar doublets transform as in Eq.~\eqref{eq:PhiTrans} with the charges $X_{\Phi i}$ expressed in terms of the three quark charges, their explicit form will depend on whether we are working in case I or II/III (as detailed in the previous section). We shall split the potential in two parts: the phase blind part $\left[V(\Phi,S)\right]_{\mbox{\scriptsize{blind}}}$, and the phase sensitive part $\left[V(\Phi,S)\right]_{\mbox{\scriptsize{sen}}}$, i.e.
\be
V(\Phi,S) \;=\;   \left[V(\Phi,S)\right]_{\mbox{\scriptsize{blind}}}  + \left[V(\Phi,S)\right]_{\mbox{\scriptsize{sen}}} \,.
\ee
The phase blind terms do not constrain the charge assignments, they are given by
\begin{align}   \label{VSblind}
\begin{split}
\left[V(\Phi,S)\right]_{\mbox{\scriptsize{blind}}}=&m_{i}^2\Phi_i^\dagger \Phi_i+\lambda_{ii,jj} \left(\Phi_i^\dagger\Phi_i\right)\left(\Phi_j^\dagger\Phi_j\right)+\lambda^\prime_{ij,ji} \left(\Phi_i^\dagger\Phi_j\right)\left(\Phi_j^\dagger\Phi_i\right)\\
&+m_S^2|S|^2+\lambda_S |S|^4+\lambda_{i}^{\Phi S}(\Phi_i^\dagger\Phi_i)|S|^2\,.
\end{split}
\end{align}
The parameters $\lambda_i^{\Phi S}$ and $\lambda_{ii,jj}$ run for all $i,j=1,2,3$, while the parameter $\lambda^\prime_{ij,ji}$ run for $i\neq j$. This part of the potential possesses a $\mathrm{U(1)}^4$ global symmetry. The role of the phase sensitive part is to introduce terms which break (explicitly) this symmetry down to $\mathrm{U(1)}_Y\times \mathrm{U(1)}_{\mbox{\scriptsize{PQ}}}$. With this symmetry we will have two complex phases to which the scalar potential will not be sensitive, one will be the neutral Goldstone boson and the other the axion. This will introduce two new additional constraints, reducing the number of independent charges down to two. 

We shall now present the possible phase sensitive terms that we may built and their constraints in terms of the PQ charges. We note that any term of the form $\Phi_i^\dagger\Phi_j$ (or any combination where this is the only phase sensitive part) implies the charge relation $X_{\Phi i}=X_{\Phi j}$, which is automatically excluded by the charge conditions, see Eqs.~\eqref{eq:PhiCaseI} and \eqref{eq:PhiCaseII}. Also, terms that are only sensitive to phases of one single field such as $S^k$, $\Phi_i^\dagger \Phi_i S^k$, etc. would imply a discrete phase, which is not allowed in our framework.

\begin{table}[ht]
\begin{center}
\doublerulesep 0.7pt \tabcolsep 0.07in
\begin{tabular}{cccc}
\hline \hline
Case&Phase sensitive&Constraint\\
\hline\hline
(1)&$\left(\Phi_1^\dagger \Phi_2\right)\left(\Phi_1^\dagger \Phi_3\right)$&$X_{\Phi 2}+X_{\Phi 3}-2X_{\Phi 1}=0$\\[0.1cm]  \rowcolor{Gray}
(2)&$\left(\Phi_2^\dagger \Phi_1\right)\left(\Phi_2^\dagger \Phi_3\right)$&$X_{\Phi 3}+X_{\Phi 1}-2X_{\Phi 2}=0$\\[0.1cm]
(3)&$\left(\Phi_3^\dagger \Phi_1\right)\left(\Phi_3^\dagger \Phi_2\right)$&$X_{\Phi 1}+X_{\Phi 2}-2X_{\Phi 3}=0$\\[0.1cm]  \rowcolor{Gray}
(4)&$\left(\Phi_1^\dagger \Phi_2\right)\{S,S^\ast\}^{k_1}$&$k_1X_S=\mp(X_{\Phi 2}-X_{\Phi 1})$\\[0.1cm]
(5)&$\left(\Phi_1^\dagger \Phi_3\right)\{S,S^\ast\}^{k_2}$&$k_2X_S=\mp (X_{\Phi 3}-X_{\Phi 1})$\\[0.1cm]  \rowcolor{Gray}
(6)&$\left(\Phi_2^\dagger \Phi_3\right)\{S,S^\ast\}^{k_3}$&$k_3X_S=\mp(X_{\Phi 3}-X_{\Phi 2})$\\
\hline\hline
\end{tabular}
\caption{  \it \small We consider $k_{i}=1,2$ due to renormalizability. The minus sign $(-)$ is associated with $S$ and the plus $(+)$ with the conjugated field $S^\ast$.}\label{tab:phase_sensitve}
\end{center}
\end{table}

In Table~\ref{tab:phase_sensitve} we present all the possible, renormalizable and gauge invariant, phase sensitive terms (up to hermitic conjugation).  We now have to check all the possible combinations of two terms from $(1)$ to $(6)$. Combining just the first three cases will lead to a constraint of the type $X_{\Phi i}=X_{\Phi j}$, which is excluded. When combining cases $(1)$ to $(3)$ with cases $(4)$ to $(6)$ all of these last three cases will be allowed simultaneously. After finding all the possible combinations and using the information about the explicit forms of $X_{\Phi i}$ in terms of the quark charges we get the following charge constraints:
\begin{itemize}
\item \textbf{Case I:} 
\be\label{eq:constraints_I}
X_{tL}=C_I(X_{uR}-X_{tR}) \,,\quad X_{S}= C_S^IX_{tL}\,.
\ee
\item \textbf{Case II/III:}
\be\label{eq:constraints_II}
X_{tL}=C_{II(III)}(X_{uR}+X_{dR}) \,,\quad X_{S}=C_S^{II(III)}X_{tL}\,.  
\ee
\end{itemize}
We must also have $C_I\neq 0,-1,-1/2$, $C_{II}\neq 0,1,1/2$ and $C_{III}\neq -1,0,\frac{1}{2},1,3$ (see Eqs.~\eqref{eq:compCondTI}, \eqref{eq:compCondTII} and \eqref{eq:compCondTIII}, respectively) in order to preserve the Yukawa textures and the symmetry to be anomalous. In Table~\ref{tab:aff} we present all possible values for $C_{I, II,III}$ and $C_{S}^{I, II,III}$ in each possible phase sensitive potential implementation.
\begin{table}[ht]
\begin{center}
\doublerulesep 0.7pt \tabcolsep 0.07in
\begin{tabular}{l|ll|cc|cc}
\hline \hline 
Term&\multicolumn{2}{|c|}{Combination}&$C_I$&$C_S^I$&$C_{II(III)}$&$C_S^{II(III)}$\\
\hline\hline
$T_1$&(4)+(5)&$k_1=1,\,k_2=2\,(S,S^\ast)$&$-2$&$1/2$&$3$&$1/3$\\  \rowcolor{Gray}
$T_2$&(4)+(5)&$k_1=2,\,k_2=1\,(S,S^\ast)$&$1$&$1$&$3/2$&$1/3$\\
$T_3$&(4)+(6)&$k_1=1,\,k_3=2\,(S,S)$&$-3/4$&$-1/3$&$-2$&$-1/2$\\ \rowcolor{Gray}
$T_4$&(4)+(6)&$k_1=2,\,k_3=1\,(S,S)$&$-3/5$&$-1/3$&$-1/2$&$-1$\\
$T_5$&(5)+(6)&$k_2=1,\,k_3=2\,(S,S^\ast)$&$-1/4$&$-1$&$2/3$&$1/2$\\  \rowcolor{Gray}
$T_6$&(5)+(6)&$k_2=2,\,k_3=1\,(S,S^\ast)$&$-2/5$&$-1/2$&$1/3$&$1$\\
$T_7$&(1)+(4)+(5)&$k_1=k_2=2\,(S,S^\ast)$&$-$&$-$&$2$&$1/4$\\  \rowcolor{Gray}
$T_8$&(2)+(4)+(6)&$k_1=k_3=2\,(S,S)$&$-2/3$&$-1/4$&$-1$&$-1/2$\\
$T_9$&(3)+(5)+(6)&$k_2=k_3=2\,(S,S^\ast)$&$-1/3$&$-1/2$&$-$&$-$\\  \rowcolor{Gray}
$T_{10}$&(1)+(4)+(5)+(6)&$k_1=1,\,k_2=1,\,k_3=2\,(S,S^\ast,S^\ast)$&$-$&$-$&$2$&$1/2$\\
$T_{11}$&(2)+(4)+(5)+(6)&$k_1=1,\,k_2=2,\,k_3=1\,(S,S,S)$&$-2/3$&$-1/2$&$-1$&$-1$\\   \rowcolor{Gray}
$T_{12}$&(3)+(4)+(5)+(6)&$k_1=2,\,k_2=1,\,k_3=1\,(S,S,S^\ast)$&$-1/3$&$-1$&$-$&$-$\\
\hline\hline
\end{tabular}
\caption{   \it \small Allowed values for the charge combinations $C_{I, II,III}$ and $C_{S}^{I, II,III}$.  Half of the possible values are not shown in the table as they can be trivially obtained by interchanging $S\leftrightarrow S^*$ in the above combinations, which amounts to a replacement $C_S^{I, II,III}\rightarrow -C_S^{I, II,III}$. The scenarios $T_1$, $T_8$ and $T_{11}$ are not possible in case III.  \label{tab:aff} }
\end{center}
\end{table}

At this point we have two free charges which we choose to be $X_{uR}$ and $X_S$, for all cases. We can normalize all charges to the scalar singlet charge, without loss of generality, just by setting the condition $X_S=1$. This allows the PQ quark charges to be written in terms of the values $C_{S}^{I,II,III}$, $C_{I,II,III}$ and one free charge, $X_{uR}$. They will now take the form:
\begin{itemize}
\item \textbf{Case I:}
\be\label{eq:chargesI}
X_{tL}= \frac{1}{C_S^I}\,,\quad X_{dR}=-X_{uR}\,,\quad X_{tR}=X_{uR} -\frac{1}{C_S^I C_I}\,.
\ee

\item \textbf{Case II:   }
\be \label{eq:chargesII}
X_{tL}= \frac{1}{C_S^{II}}\,,\quad X_{dR}=-X_{uR} +\frac{1}{C_S^{II} C_{II}}\,,\quad X_{tR}=X_{uR}-\frac{1-2C_{II}}{C_S^{II} C_{II}}\,.
\ee

\item \textbf{Case III:}
\be \label{eq:chargesIII}
X_{tL}=\frac{1}{C_S^{III}}\,,\quad X_{dR}=-X_{uR}+\frac{1}{C_S^{III} C_{III}}\,,\quad X_{tR}=X_{uR}+\frac{1}{C_S^{III}}\,.
\ee
\end{itemize} 

In this section we have found up to 12 distinct phase sensitive potential implementations, see Table~\ref{tab:aff}. For case I, $T_7$ and $T_{10}$ implementations are not compatible with the flavor PQ symmetry in the fermionic sector. In case II, the incompatible implementations are $T_9$ and $T_{12}$. Finally, case III has the same incompatible implementations as case II plus $T_1$, $T_8$ and $T_{11}$ implementations. As an illustrative example, let us choose the implementation $T_2$. The scalar potential would take the form
\be
V(\Phi,S)=  \left[V(\Phi,S)\right]_{\mbox{\scriptsize{blind}}}  +\left[\lambda(\Phi_1^\dagger\Phi_2)S^2+\mu (\Phi_1^\dagger\Phi_3)S^\ast+\text{h.c.}\right]\,,
\ee
with $\lambda$ dimensionless and $\mu$ with mass dimension. Under this particular potential implementation, and with our normalization, the PQ quark charges read  
\begin{itemize}
\item \textbf{Case I:}\quad $X_{tL}= 1\,,\quad X_{dR}=-X_{uR}\,,\quad X_{tR}=X_{uR} -1$.

\item \textbf{Case II:}\quad $X_{tL}= 3\,,\quad X_{dR}=-X_{uR} +2\,,\quad X_{tR}=X_{uR}+4$.

\item \textbf{Case III:}\quad $X_{tL}=3\,,\quad X_{dR}=-X_{uR}+2\,,\quad X_{tR}=X_{uR}+3$.
\end{itemize} 
While the scalar charges are: $X_{\Phi1}=X_{uR}\,,\quad X_{\Phi 2}=X_{uR}-2\,,\quad X_{\Phi3}=X_{uR}+1$. The fact that the scalar charges are the same for all the three cases should not be surprising. The scalar potential itself knows nothing about the distinct Yukawa implementations, that information enters only when we use the explicit expression of the scalar charges in terms of the quark ones. Therefore, the scalars charges will only depend on the the distinct potential implementations.

\subsection{The Yukawa leptonic sector  \label{ssec:leptonic}}
In this section we shall only be interested in the Yukawa couplings of the charged leptons and therefore we will say nothing on the Dirac or Majorana nature of the neutrinos. We will assume that the final neutrino mass matrix texture contains enough freedom, such that, in combination with the lepton mass matrix accommodates the full low-energy neutrino data. However, note that the neutrino Yukawa textures should satisfy some conditions such that the BGL quark and lepton textures are not spoiled through radiative corrections\cite{Botella:2011ne}.   In Ref.~\cite{Celis:2014iua} a particular model implementation has been presented were the neutrino sector has been worked out. However, since we are mostly interested in the axion properties in this class of models, we can just focus our attention to the charged lepton implementation.

The Yukawa leptonic Lagrangian will be of the form
\be
-\mathcal{L}^{\mbox{\scriptsize lep}}_{\mbox{\scriptsize Y}}=\overline{L_L^0}\,\left[\Pi_1\, \Phi_1 +\Pi_2\, \Phi_2+\Pi_3\, \Phi_3\right]l^0_R+\text{h.c.}
\ee
In a similar way as it happens in the quark Yukawa sector, it is convenient to rewrite the Yukawa lepton Lagrangian by rotating the Higgs doublets to the Higgs basis (see Eq.~\eqref{eq:HiggsbasisG}) and by diagonalizing the lepton mass matrices through the bi-unitary transformations
\be\label{eq:nuemassbasis}
\nu_{L}^0=U_{\nu L}\, \nu_{L}\,,\quad l_{L,R}^0=U_{eL,R} \, e_{L,R}\,.
\ee
The Yukawa Lagrangian now reads as
\begin{align}
\begin{split}
-\mathcal{L}^{\mbox{\scriptsize lep}}_{\mbox{\scriptsize Y}}=&\,\overline{e_L}\left[D_e+\frac{1}{v}D_e H^0+\frac{1}{v^\prime}N_e R+\frac{1}{v^{\prime\prime}}N_e^\prime R^\prime+i\frac{1}{v^\prime}N_eI+i\frac{1}{v^{\prime\prime}}N_e^{\prime}I^\prime\right]e_R\\
&+\frac{\sqrt{2}}{v^\prime}H^+\overline{\nu_L}N_ee_R+\frac{\sqrt{2}}{v^{\prime\prime}}H^{\prime +}\overline{\nu_L}N_e^{\prime}e_R+\text{h.c.}
\end{split}
\end{align}
where, as it happened with the quarks, $N_e$ and $N'_e$ will mediate the FCNCs. These flavor combinations will have the same expression, in the flavor basis, as $N_d$ and $N_d^\prime$ present in Eq.~\eqref{eq:NuNd3} with the replacement $\Gamma_i\rightarrow \Pi_i$.

Regarding the PQ symmetry transformations, the scalar field transformations are given in Eqs.~\eqref{eq:PhiCaseI} and \eqref{eq:PhiCaseII} for cases I and II/III respectively. We now need to determine the PQ charges of the leptonic fields. In general these will transform under the continuous symmetry as
\be
L^0_{L}\rightarrow \mathcal{S}_L^\ell\,L^0_L\,,\quad l_R^0\rightarrow \mathcal{S}_R^\ell l_R^0\,,
\ee
with
\be
\mathcal{S}_L^\ell=\text{diag}(e^{iX_{eL} \theta},e^{iX_{\mu L} \theta},e^{iX_{\tau L}  \theta })\,,\quad \mathcal{S}_R^\ell=\text{diag}(e^{iX_{e R} \theta },e^{iX_{\mu R} \theta },e^{iX_{\tau R} \theta })\,.
\ee
A global phase transformation allows us to set $X_{eL}=0$ without loss of generality, just as we did in the quark sector.

We could proceed with the symmetry implementation just like in the quark sector, however, we can also combine the BGL-like textures in the quark sector with NFC for the charged lepton such that we have several phenomenological models available. We shall then split these implementations into two classes:
\begin{itemize}
\item[(1)] With FCNCs in the charged lepton sector.

This is the extension to three Higgs doublets of the symmetry implementation in Ref.~\cite{Botella:2011ne}. In this case, in order to have the FCNCs under control we choose the implementation \textit{\`a la} BGL, i.e.
\be
\left\{\Pi_1,\, \Pi_2,\,\Pi_3\right\}\sim\left\{\Gamma^{\mbox{\scriptsize{BGL}}}_{1},\, \Gamma^{\mbox{\scriptsize{BGL}}}_{2},\, 0\right\}\,.
\ee
Just like in the quark sector, we need the other sector mass matrix (i.e. neutrino mass matrix) to be block diagonal, in order to have the PMNS mediating the FCNCs. The way to achieve this will depend on the Dirac or Majorana nature of neutrino and is out of the scope of this paper (see Ref.~\cite{Celis:2014iua} for more details). The symmetry implementation is just like the one in the quark sector, i.e. $X_{eL}=X_{\mu L}\equiv X_{l^\prime L}$ and $X_{eR}=X_{\mu R}=X_{\tau R}\equiv X_{lR}$. The constraints are
\be
X_{l^\prime L}-X_{lR}=X_{\Phi i}\,,\quad X_{\tau L}-X_{lR}=X_{\Phi j}\,.
\ee
The equivalent to conditions A and B in the quark sector also apply to the lepton charges. Since we have set $X_{e L}=0$, the charged lepton charges become completely defined by the known scalar charges, i.e.
\be
X_{\tau L}=X_{\Phi j}-X_{\Phi i}\,,\quad X_{lR}=-X_{\Phi i}\,.
\ee

\item[(2)] Without FCNCs in the charged lepton sector.

In this case there are six implementations possible, as it was shown in Ref.~\cite{Serodio:2013gka}. Using the information that all the charges of the scalar fields are different we get
\begin{itemize}
\item[(a)] 

\be
\left\{\Pi_1,\, \Pi_2,\,\Pi_3\right\}\sim\left\{\begin{pmatrix}
\times&\times&\times\\
\times&\times&\times\\
\times&\times&\times
\end{pmatrix}\,,\quad \begin{pmatrix}
&&\\
&&\\
&&
\end{pmatrix}\,,\quad  \begin{pmatrix}
&&\\
&&\\
&&
\end{pmatrix}\right\}\,.
\ee
In this scenario both left and right generators must be fully degenerate, i.e. $X_{eL}=X_{\mu L}=X_{\tau L}\equiv X_{lL}$ and $X_{eR}=X_{\mu R}=X_{\tau R}\equiv X_{lR}$. This implies the following constraint
\be
X_{lL}-X_{lR}=X_{\Phi i}\quad \left(\text{or}\quad X_{lR}=-X_{\Phi i}\right)\,.
\ee

\item[(b)]

\be
\left\{\Pi_1,\, \Pi_2,\,\Pi_3\right\}\sim\left\{
\begin{pmatrix}
\times&\times&0\\
\times&\times&0\\
0&0&\times
\end{pmatrix}\,,\quad  \begin{pmatrix}
&&\\
&&\\
&&
\end{pmatrix}\,,\quad  \begin{pmatrix}
&&\\
&&\\
&&
\end{pmatrix}\right\}\,.
\ee
In this scenario both left and right generators must be two-fold degenerate, i.e. $X_{eL}=X_{\mu L}\equiv X_{l^\prime L}$ and $X_{eR}=X_{\mu R}\equiv X_{l^\prime R}$. This implies the following constraints
\be
X_{l^\prime L}-X_{l^\prime R}=X_{\Phi i}\,,\quad X_{\tau L}-X_{\tau R}=X_{\Phi i}\,.
\ee

\item[(c)]

\be
\left\{\Pi_1,\, \Pi_2,\,\Pi_3\right\}\sim\left\{
\begin{pmatrix}
\times&\times&0\\
\times&\times&0\\
0&0&0
\end{pmatrix}\,,\quad \begin{pmatrix}
0&0&0\\
0&0&0\\
0&0&\times
\end{pmatrix}\,,\quad  \begin{pmatrix}
&&\\
&&\\
&&
\end{pmatrix}\right\}\,.
\ee
In this scenario the left and right generators have the same form as in the previous one. However, the constraints are
\be
X_{l^\prime L}-X_{l^\prime R}=X_{\Phi i}\,,\quad X_{\tau L}-X_{\tau R}=X_{\Phi j}\,.
\ee

\item[(d)]

\be
\left\{\Pi_1,\, \Pi_2,\,\Pi_3\right\}\sim\left\{
\begin{pmatrix}
\times&0&0\\
0&\times&0\\
0&0&\times
\end{pmatrix}\,,\quad \begin{pmatrix}
&&\\
&&\\
&&
\end{pmatrix}\,,\quad  \begin{pmatrix}
&&\\
&&\\
&&
\end{pmatrix}\right\}\,.
\ee
In this scenario the left and right generators must have no degeneracy. The constraint is given by
\be
X_{\alpha L}-X_{\alpha R}=X_{\Phi i}\quad (\alpha=e,\, \mu,\, \tau)\,.
\ee

\item[(e)]

\be
\left\{\Pi_1,\, \Pi_2,\,\Pi_3\right\}\sim\left\{
\begin{pmatrix}
\times&0&0\\
0&\times&0\\
0&0&0
\end{pmatrix}\,,\quad \begin{pmatrix}
0&0&0\\
0&0&0\\
0&0&\times
\end{pmatrix}\,,\quad  \begin{pmatrix}
&&\\
&&\\
&&
\end{pmatrix}\right\}\,.
\ee
In this scenario the left and right generators are the same as before. The constraints are given by
\be
X_{\alpha^\prime L}-X_{\alpha^\prime R}=X_{\Phi i}\,,\quad X_{\tau L}-X_{\tau R}=X_{\Phi j}\quad (\alpha^\prime=e,\, \mu)\,.
\ee

\item[(f)]

\be
\left\{\Pi_1,\, \Pi_2,\,\Pi_3\right\}\sim\left\{
\begin{pmatrix}
\times&0&0\\
0&0&0\\
0&0&0
\end{pmatrix}\,,\quad \begin{pmatrix}
0&0&0\\
0&\times&0\\
0&0&0
\end{pmatrix}\,,\quad  \begin{pmatrix}
0&0&0\\
0&0&0\\
0&0&\times
\end{pmatrix}\right\}\,.
\ee
In this scenario the left and right generators are the same as before. The constraints are given by
\be
X_{eL}-X_{e R}=X_{\Phi i}\,,\quad X_{\mu L}-X_{\mu R}=X_{\Phi j}\,,\quad X_{\tau L}-X_{\tau R}=X_{\Phi k}\,.
\ee
\end{itemize}
\end{itemize}

In general, we have only information on the difference between left- and right-handed charged lepton charges. The condition $X_{eL}=0$ allows us to have the charged lepton charges fully determined by the known scalar charges only in cases (1) and $(2a)$. For the other cases we would need to know the neutrino sector implementation. Nevertheless, as we shall see in the next section, the knowledge of the difference is enough to get most of the axion properties.

\section{Axion properties}   \label{sec:axion}
The anomalous $\mathrm{U(1)}_{\mbox{\scriptsize{PQ}}}$ symmetry of the class of models built in the previous sections is spontaneously broken by the vev of the singlet field $S$ at a very high scale, just like in the standard DFSZ and KSVZ models.   Non-perturbative QCD effects induce a potential for the axion field, allowing us to shift away the strong CP phase and also give a small mass to the axion~\cite{Weinberg:1977ma}, the physical one (denoted by $a$).  In the following we derive the most relevant axion properties for our model.  We start by writing the relevant Lagrangian for the physical axion
\be\label{eq:Laxion}
\mathcal{L}^{\mbox{\scriptsize{eff}}}_{\mbox{\scriptsize{axion}}}=\mathcal{L}_{SM}+\frac{1}{2}\partial_\mu a\,\partial^\mu a-\frac{1}{2}m_a^2\, a^2+\mathcal{L}_{a\gamma\gamma}
+\mathcal{L}_{a\bar{\psi}\psi}\,,
\ee
where $\mathcal{L}_{a\gamma\gamma}$ is the axion interaction to photons, we will shown in Sec.~\ref{subsec:aphoton}, and $\mathcal{L}_{a\bar{\psi}\psi}$ the axion interaction to matter, we will present in Sec.~\ref{subsec:amatter}.   The axion mass is given by~\cite{Weinberg:1977ma}:
\be
m_a = \frac{f_\pi m_\pi |C_{ag}|}{v_{\mbox{\scriptsize PQ}}} \left[\frac{z}{\left(1+z\right)\left(1+z+w\right) }\right]^{1/2}    \simeq 6~\text{meV} \times \left( \frac{  10^9~\text{GeV}  }{ v_{\mbox{\scriptsize PQ}}/|C_{ag}| }  \right) \,,
\ee
with $m_\pi \simeq 135$~MeV and $f_\pi \simeq 92$~MeV the pion mass and decay constant, respectively.   The parameters $z$ and $w$ denote the quark mass ratios $z=m_u/m_d \simeq 0.56$ and $w=m_u/m_s \simeq 0.029$.     The quantity $C_{ag}$ is determined by the chiral color anomaly of the current associated with the $\mathrm{U(1)}_{\mbox{\scriptsize{PQ}}}$ transformation~\cite{Adler:1969gk}, in our model it is given by
\be \label{chiralcolor}
C_{ag}\equiv\sum_{i=\text{colored}}X_{iR}-X_{iL}=  2X_{uR} + 3X_{dR} + X_{tR} - 2X_{tL}  \,.
\ee
This quantity turns out to be independent of the free charge and can be expressed as $C_{ag}^M$ (with $M=I,\, II,\, III$) and is given by
\be\label{eq:Cag}
\textbf{Case I:} \quad C^I_{ag}=-\frac{1+2C_I}{C_I C_S^I}\,,\quad \textbf{Case II:} \quad C^{II}_{ag}=\frac{2}{C_{II} C_S^{II}}\,,\quad \textbf{Case III:} \quad C^{III}_{ag}=\frac{3-C_{III}}{C_{III} C_S^{III}} \,.
\ee
The quantity $C^M_{ag}$ is therefore only dependent of the scalar implementation once the Yukawa textures are specified.

\subsection{Axion-photon coupling}\label{subsec:aphoton}
The axion two-photon interaction is described by the effective Lagrangian
\begin{align}   \label{aphotonco}
\begin{split}
\mathcal{L}_{a\gamma\gamma} =&   \frac{\alpha}{8\pi v_{\mbox{\scriptsize PQ}}} C_{ag} C_{a\gamma}^{\mbox{\footnotesize eff}} \,a \, F_{\mu\nu}  \widetilde{F}^{\mu\nu} \,,
 \end{split}
\end{align}
with $\alpha= e^2/4 \pi \simeq 1/137$, $F_{\mu \nu}$ is the electromagnetic field strength tensor and $\widetilde F_{\mu \nu}$ its dual tensor.  The effective factor $C_{a\gamma}^{\mbox{\footnotesize eff}}$ takes the form~\cite{Srednicki:1985xd}:
\be  \label{caf}
C_{a\gamma}^{\mbox{\footnotesize eff}} = \frac{C_{a\gamma}}{C_{ag}}-\frac{2}{3}\frac{4+z+w}{1+z+w},
\ee
where the second term is a model independent quantity which comes from the mixing of the axion with the $\pi^0$ and the $\eta$ while $C_{a\gamma}$ and $C_{ag}$ are model dependent quantities associated to the axial anomaly.  These are determined in terms of the fermion charges by
\begin{align}   \label{eq_phtc}
\begin{split}
C_{a\gamma}=&2\sum_{i=\text{charged}}(X_{iR}-X_{iL})Q_i^2\\
=&2\left[\frac{8}{3}X_{uR}+X_{dR}+\frac{4}{3}X_{tR}-\frac{5}{3}X_{tL}+\sum_{\alpha=e,\mu,\tau}\left(X_{\alpha R}-X_{\alpha L}\right)\right] \,,
\end{split}
\end{align}
while $C_{ag}$ was already introduced in Eq.~\eqref{chiralcolor}.    The quantity $C_{a\gamma}$ can be expressed as
\be \label{eq:Cagamma} 
C^{M}_{a\gamma}=\frac{2}{3}\frac{A_{M}}{C_{M}C_S^{M}}\,,
\ee
with $M=I,\, II,\, III$.  Here we have introduced the parameters $C_{I, II,III}$ and $C_{S}^{I, II,III}$ specified in Table~\ref{tab:aff} and a new combination of fermionic charges $A_{I, II,III}$ defined in Table~\ref{newcomb}. The charged lepton combinations are denoted by a vector $(i,j,k)$, which represents the Higgs doublet that is coupled to the left-handed leptons $(e,\mu,\tau)$. For example, the case $(1,1,3)$ tell us that $\Phi_1$ is coupled to $e_L$ and $\mu_L$ while $\Phi_3$ couples to $\tau_L$. This can correspond to the charged lepton Yukawa implementations (1), (2c) or (2e). Note also, that the case $(3,3,1)$ is not a relabeling of the scalar fields since we keep the quark sector unchanged and, therefore, we will get a distinct result. Once the choice on the Yukawa textures is made, the parameter $C^M_{a\gamma}$ will depend on the potential implementation and the way the charged leptons transform under the PQ symmetry.

\begin{table}[h]
\begin{center}
\begin{tabular}{r|ccc} 
Cases&$A_I$&$A_{II}$&$A_{III}$\\
\hline\hline
$(1,1,1)$&$-4-5C_I$&$-1+3C_{II}$&$3-C_{III}$\\  \rowcolor{Gray}
$(2,2,2)$&$5+4C_I$&$8+3C_{II}$&$12-C_{III}$\\
$(3,3,3)$&$-4-14C_I$&$8-6C_{II}$&$12-10C_{III}$\\  \rowcolor{Gray}
$(1,1,2)$&$-1-2C_I$&$2+3C_{II}$&$6-C_{III}$\\
$(1,1,3)$&$-4-8C_I$&$2$&$6-4C_{III}$\\  \rowcolor{Gray}
$(2,2,1)$&$2+C_I$&$5+3C_{II}$&$9-C_{III}$\\
$(2,2,3)$&$2-2C_I$&$8$&$12-4C_{III}$\\ \rowcolor{Gray}
$(3,3,1)$&$-4-11C_I$&$5-3C_{II}$&$9-7C_{III}$\\
$(3,3,2)$&$-1-8C_I$&$8-3C_{II}$&$12-7C_{III}$\\ \rowcolor{Gray}
$(1,2,3)$&$-1-5C_I$&$5$&$9-4C_{III}$\\
\hline\hline
\end{tabular}
\end{center}
\caption{ \it \small Charge combinations $A_I$, $A_{II}$ and $A_{III}$ entering in the description of the axion coupling to photons.    The numbers in the first column label the Higgs doublet
that is coupled to the left-handed charged leptons ($e, \mu, \tau$).  } \label{newcomb}      
\end{table}

\subsection{Axion couplings to matter}\label{subsec:amatter}
The interactions of the axion with fermions are described by
\begin{align}
\begin{split}
\mathcal{L}_{a\bar{\psi}\psi} =&   \frac{1}{2} \frac{\partial_\mu a}{v_{\mbox{\scriptsize PQ}}} \,\left[ \overline{e_\alpha}\gamma^\mu\, \left(\left(C^V_{a\ell}\right)_{\alpha\beta}+\gamma_5\left(C^A_{a\ell}\right)_{\alpha\beta}\right)\,e_\beta
+\, \overline{u_\alpha}\gamma^\mu\left(\left(C^V_{au}\right)_{\alpha\beta}+\gamma_5\, \left(C^A_{au}\right)_{\alpha\beta}\right)\,u_\beta\right.       \\ 
&\left.+ \overline{d_\alpha}\gamma^\mu\left(\left(C^V_{ad}\right)_{\alpha\beta}+\gamma_5\, \left(C^A_{ad}\right)_{\alpha\beta}\right)\,d_\beta-\eta\, C_{ag}\left(\overline{u}\gamma^\mu \gamma_5 u+z\overline{d}\gamma^\mu \gamma_5 d+w\overline{s}\gamma^\mu \gamma_5 s\right)\right],
 \end{split}
\end{align}
with $\eta=(1+z+w)^{-1}$. When calculating the axion couplings to matter one should redefine the axion current in such a way that it does not mix with the neutral Goldstone boson associated with the spontaneous symmetry breaking of the electroweak gauge symmetry. This redefinition results in a shift of the original scalar charges~\cite{Srednicki:1985xd},
\be\label{eq:orthoc}
X_{\Phi i}^\prime=X_{\Phi i}-Z\,,
\ee
which in terms of the fermion charges reads
\be\label{eq:Xshift}
X_{u,t\, R}^\prime=X_{u,t\, R}-Z\,,\quad X_{dR}^\prime=
X_{d R}+Z\,,\quad X_{tL}^\prime=X_{tL}\,,\quad X_{\ell L}^\prime=X_{\ell L}\,, \quad X_{\ell R}^\prime=X_{\ell R}+Z\,,
\ee
with $\ell=\left\{e,\mu,\tau\right\}$ and
\be
Z=\frac{1}{v^2}\sum_{i}v^2_i X_{\Phi i}\,.
\ee
The explicit expression for $Z$ will take the same form in cases II and III (they share the same Higgs charge assignments) but a different form in case I, i.e.
\be\label{eq:Z}
Z=\left\{
\begin{array}{ll}
X_{uR}-\dfrac{v_2^2\left(1+C_I\right) - v_3^2 C_I}{v^2C_S^I C_I}&\text{for case I,}\\
X_{uR}-\dfrac{v_2^2+v_3^2\left(1-C_{II(III)}\right)}{v^2C_S^{II(III)} C_{II(III)}}&\text{for case II and III}\,.
\end{array}\right.
\ee
We now define the shifted charge matrix as
\be
\mathcal{X}_X\equiv \frac{1}{i}\left.\frac{d\mathcal{S}^\prime_X}{d\theta} \right|_{\theta=0} \,,
\ee
which take the explicit form
\begin{align}\label{eq:chargesmatrix}
\begin{split}
\mathcal{X}_{uL}=&\mathcal{X}_{dL}=\text{diag}(0,\,0,\, X^\prime_{tL})\,,\quad
\mathcal{X}_{eL}=\text{diag}(X^\prime_{eL},\,X^\prime_{\mu L},\, X^\prime_{\tau L})\,,\\ 
\mathcal{X}_{uR}=&\text{diag}(X^\prime_{uR},\,X^\prime_{uR},\, X^\prime_{tR})\,,\quad \mathcal{X}_{dR}=X^\prime_{dR}\mathbb{I}\,,\quad \mathcal{X}_{eR}=\text{diag}(X^\prime_{eR},\,X^\prime_{\mu R},\, X^\prime_{\tau R})  \,.
\end{split}
\end{align}
These charge matrices determine the couplings of the axion to fermions in the flavor basis. By going to the mass basis the fermion fields are rotated through the unitary transformations in Eq.~\eqref{eq:udmassbasis} and in Eq.~\eqref{eq:nuemassbasis}. These transformations will change the charge matrix to 
\be
\widetilde{\mathcal{X}}_X=U^\dagger_X \mathcal{X}_X U_{X}\,.
\ee
In this new basis the quark charge matrices take the explicit form
\be\label{eq:quarkChi}
\widetilde{\mathcal{X}}_{uL}=\mathcal{X}_{uL}\,,\quad \widetilde{\mathcal{X}}_{uR}=\mathcal{X}_{uR}\,,\quad\widetilde{\mathcal{X}}_{dL}=
X_{tL}\begin{pmatrix}
|V_{td}|^2&V_{td}^\ast V_{ts}&V_{td}^\ast V_{tb}\\
V_{ts}^\ast V_{td}& |V_{ts}|^2&V_{ts}^\ast V_{tb}\\
V_{tb}^\ast V_{td}&V_{tb}^\ast V_{ts}&|V_{tb}|^2
\end{pmatrix}\,,\quad \widetilde{\mathcal{X}}_{dR}=\mathcal{X}_{dR}\,,
\ee
where we have used $X^{\prime}_{tL}=X_{tL}$. For the charged leptons we have in scenario $(1)$
\be\label{eq:leptonChi}
\widetilde{\mathcal{X}}_{eL}=
X_{\tau L}\begin{pmatrix}
|V_{\tau 1}|^2&V_{\tau 1}^\ast V_{\tau 2}&V_{ \tau 1}^\ast V_{\tau 3}\\
V_{\tau 2}^\ast V_{\tau 1}& |V_{\tau 2}|^2&V_{\tau 2}^\ast V_{\tau 3}\\
V_{\tau 3}^\ast V_{\tau 1}&V_{\tau 3}^\ast V_{\tau 2}&|V_{\tau 3}|^2
\end{pmatrix}\,,\quad \widetilde{\mathcal{X}}_{eR}=\mathcal{X}_{eR} \,,
\ee
where we have used $X^\prime_{\tau L}=X_{\tau L}$, in scenario $(2)$ we get
\be
\widetilde{\mathcal{X}}_{eL}=\mathcal{X}_{eL}\,,\quad \widetilde{\mathcal{X}}_{eR}=\mathcal{X}_{eR}\,.
\ee
The axion vector and axial couplings to matter are then given by
\be
C^{V,A}_{au}=\mathcal{X}_{uL}\pm\mathcal{X}_{uR}\,,\quad
C^{V,A}_{ad}= \widetilde{\mathcal{X}}_{dL}\pm\mathcal{X}_{dR}\,,\quad
C^{V,A}_{ae}= \widetilde{\mathcal{X}}_{eL}\pm\mathcal{X}_{eR}\,.
\ee
From the above equations we can see that the flavor changing axion interactions will be mediated by the off-diagonal elements of $\widetilde{\mathcal{X}}_{dL}$ (and $\widetilde{\mathcal{X}}_{eL}$ in case (1)). This is a common property of Goldstone bosons in flavor models, however it is an additional feature for the axion compared to the standard DFSZ and KSVZ scenarios.

Regarding the vectorial couplings, it is interesting to remark that the PQ symmetry implementation that we have used until now is defined up to a global vectorial phase transformation which allows us to remove some of the vectorial couplings. As a result, the Lagrangian will remain invariant if we redefine the PQ charges by performing the following transformation
\begin{align}
\mathcal{X}^V_X=\mathcal{X}_X+\alpha\;\mathbb{I}\,,
\end{align}
with $\mathcal{X}_X$ defined in Eq.~\eqref{eq:chargesmatrix} and $\alpha$ an arbitrary constant. In the DFSZ model this transformation is enough to remove all the vectorial couplings. However, this is not the case in the models we are presenting. For example, by setting $\alpha=-X^\prime_{uR}/2$ the transformed quark charges read as
\begin{align}
\begin{split}\label{eq:XV}
\mathcal{X}_{uL}^V=&\mathcal{X}_{dL}^V=\text{diag}(-X^\prime_{uR}/2,\,-X^\prime_{uR}/2,\, X^\prime_{tL}-X^\prime_{uR}/2)\,,\\
\mathcal{X}_{uR}^V=&\text{diag}(X^\prime_{uR}/2,\,X^\prime_{uR}/2,\, X^\prime_{tR}-X^\prime_{uR}/2)\,,\\
\mathcal{X}_{dR}^V=&\left(X^\prime_{dR}-X^\prime_{uR}/2\right)\mathbb{I}\,,
\end{split}
\end{align}
such that now $\left(C^{V}_{au}\right)_{11}=\left(C^{V}_{au}\right)_{22}=0$. Similarly, we could have set $\alpha=-\left(X_{tL}^{\prime}+X_{tR}^{\prime}\right)/2$ to make $\left(C^{V}_{au}\right)_{33}=0$ but there is no value of $\alpha$ that makes $C^{V}_{au}=0$ for the three families simultaneously. A similar procedure can be applied to the lepton sector.

Additionally, we can fix the value of the free PQ charge, $X_{uR}$, in order to remove extra vectorial couplings. For instance, setting $X_{uR}=X_{tR}+X_{tL}$ in Eq.~\eqref{eq:XV} would also make $\left(C^{V}_{au}\right)_{33}=0$ for $\alpha=-X^\prime_{uR}/2$. However, the condition $X_{uR}=X_{tR}+X_{tL}$ can only be satisfied in cases I and II while in case III it would violate the charge constraints in Eq.~\eqref{CA1}. The same happens for other values of $\alpha$ and thus one cannot use the freedom in $X_{uR}$ to remove vectorial couplings in case III.

Finally, note that the vectorial transformation and the freedom to fix the value of $X_{uR}$ only affect the diagonal vector couplings while the off-diagonal ones remain unchanged. In any case, it is simple to see that in the case of on-shell fermions the axion-fermion interaction is purely pseudoscalar for fermions of the same flavor 
\begin{align}
\begin{split}
\partial_\mu a \, \overline{\psi_\alpha}\,C_{\alpha\beta}^A \gamma^\mu\gamma_5 \,\psi_\beta=&i a\,\overline{\psi_\alpha}\,(m_\alpha+m_\beta)C_{\alpha\beta}^A\,\psi_\beta+\cdots\\
\partial_\mu a \, \overline{\psi_\alpha}\,C_{\alpha\beta}^V \gamma^\mu\,\psi_\beta=&i a\,\overline{\psi_\alpha}\,(m_\alpha-m_\beta)C_{\alpha\beta}^V\,\psi_\beta+\cdots
\end{split}
\end{align}
with $\psi$ representing a fermionic specie (up quarks, down quarks and charged leptons). Therefore, the nature of the axion interaction in the up quark sector is purely pseudoscalar for on-shell quarks. However, due to the presence of FCNCs in the quark sector, the axion interaction will no longer conserve flavor, reflecting a scalar (beside the pseudoscalar) nature of the axion field in models with FCNCs.

The axion axial couplings to light quarks are explicitly given by
\begin{align}\label{eq:guds}
\begin{split}
g_u&\equiv (C_{au}^A)_{11}=\left\{
\begin{array}{ll}
-\dfrac{v_2^2\left(1+C_I\right) - v_3^2 C_I}{v^2C_S^I C_I}&\text{for case I,}\\
-\dfrac{v_2^2+v_3^2\left(1-C_{II(III)}\right)}{v^2C_S^{II(III)} C_{II(III)}}&\text{for case II and III}\,,
\end{array}\right.\\
g_d&\equiv (C_{ad}^A)_{11}=-g_u+\left\{
\begin{array}{ll}
\dfrac{|V_{td}|^2}{C_{S}^{I}}&\text{for case I}\,,\\
\dfrac{|V_{td}|^2 C_{II(III)}-1}{C_{S}^{II(III)}C_{II(III)}}&\text{for case II and III}\,,
\end{array}
\right.\\
g_s&\equiv (C_{ad}^A)_{22}=g_d \quad (\text{with the replacement $V_{td}\rightarrow V_{ts}$}) \,.
\end{split}
\end{align}
Below the chiral symmetry breaking scale the axion nucleon interactions can be parametrized by
\be
\mathcal{L}_{\mbox{\scriptsize aN}}=\frac{1}{2}\frac{\partial_\mu a}{v_{\mbox{\scriptsize PQ}}}\,\overline{N}(g^0+g^{3} \sigma_3)\gamma^\mu\gamma_5\,N\,,
\ee
with $\sigma_3$ the Pauli matrix in the isospin space and $N=(p,\, n)^T$ the nucleon doublet. The isoscalar and isovector couplings are given in Refs.~\cite{Srednicki:1985xd,Hindmarsh:1997ac}. The couplings to protons and neutrons are given by the combinations
\begin{align}
\begin{split}
g_p\equiv g^0+g^3=&(g_u-2 \eta C_{ag})\Delta u+(g_d-2\eta  C_{ag} z)\Delta d+(g_s-2\eta  C_{ag} w)\Delta s\,,\\
g_n\equiv g^0-g^3=&(g_u- 2 \eta C_{ag})\Delta d+(g_d-2\eta  C_{ag} z)\Delta u+(g_s-2 \eta  C_{ag} w)\Delta s\,,
\end{split}
\end{align}
with $\Delta u=0.841\pm 0.020$, $\Delta d=-0.426\pm 0.020$ and $\Delta s=-0.085\pm 0.015$~\cite{Beringer:1900zz}.

The coupling to electrons in scenario $(1)$ is given by  
\be\label{eq:ge}
g_e\equiv\left(C_{ae}^A\right)_{11}=X_{\tau L}|V_{\tau 1}|^2-X_{eR}-Z\,.
\ee
The scenario $(2)$ can be obtained in the limit $|V_{\tau 1}|^2\rightarrow 0$. The explicit form of $g_e$ will depend on the scalar potential implementation and the vevs of the doublet fields.

\subsection{The domain wall problem}

During the evolution of the Universe the PQ symmetry gets broken in different ways. In the early Universe the PQ symmetry is spontaneously broken by the expectation value of the $S$ field. At this stage the potential has the mexican-hat shape, the angular part of the field becomes a Goldstone boson and the Lagrangian remains global phase invariant. As the Universe cools down non-perturbative instantonic effects at the QCD scale take place and the PQ symmetry gets explicitly broken by the QCD gluon anomaly $[\mathrm{SU(3)}_C]^2 \times \mathrm{U(1)}_{\mbox{\scriptsize{PQ}}}$~\cite{Weinberg:1977ma}. This is the PQ mechanism for the resolution of the strong CP problem.

However QCD instantons only break the symmetry down to a discrete $\mathcal{Z}_{N}$ subgroup. This can be easily seen from Eq.~\eqref{eq:LCPstrong} and the fact that the $\theta$ term is invariant under the shift $\theta\rightarrow \theta +2\pi k$. While before the QCD scale the shift $a/v_{\mbox{\scriptsize PQ}}\rightarrow a/v_{\mbox{\scriptsize PQ}}+\alpha$ was allowed for any $\alpha$, the presence of the axion coupling to gluons restricts the phase to the values $\alpha_k=2\pi k/|C_{ag}|$ (with $k=0,1,\ldots,|C_{ag}|-1$), which just reflects the original $\overline{\theta}$ periodicity. Therefore, the order $N$ of the discrete group is given by the color instantons effects to be $N=|C_{ag}|$.  As pointed out by Sikivie~\cite{Sikivie:1982qv}, these models will have $N_{\mbox{\scriptsize DW}}$ degenerate disconnected vacua. This in turn leads to an unwanted domain wall structure in the early universe~\cite{Zeldovich:1974uw,Sikivie:1982qv,Vilenkin:1984ib}. 

In general, the domain wall number, $N_{\mbox{\scriptsize DW}}$, coincides with the order of the discrete group, i.e. $N_{\mbox{\scriptsize DW}}=N=|C_{ag}|$.
Nonetheless, in some cases only a subgroup of $\mathcal{Z}_{N}$ acts non-trivially on the vacuum. To examine the vacuum structure one should analyze the gauge invariant order parameters of the theory. In this way the domain wall number will coincide with the dimension of the higher order subgroup of $\mathcal{Z}_{N}$ which acts non-trivially on at least one of the order parameters. For the models we are discussing, it suffices to notice that for the singlet condensate
\begin{align}\label{eq:S0}
\left<S\right>_k\rightarrow&\,\text{Exp}\left[\frac{2\pi k}{N/X_S}\right]\left<S\right>_0\,,
\end{align}
as we set $X_S=1$ the vacuum periodicity is $N$, i.e. all elements of the residual $\mathcal{Z}_{N}$ act non-trivially on the vacuum. As a result, we have $N_{\mbox{\scriptsize DW}}=|C_{ag}|$ for the class of models studied in this article.\footnote{Considering higher dimensional order parameters such as $\left<\Phi_i^\dagger \Phi_j\right>$ or $\left<\overline{q_{L\alpha}}q_{R\alpha}\right>$ would not change the periodicity of the full vacuum, since in our choice of normalization the residual discrete group always acts non trivially in $\left<S\right>$.}

Many axion models suffer from the domain wall problem. In particular, the well known DFSZ invisible axion model has a domain wall number $N_{\mbox{\scriptsize DW}}=2N_g$ or $N_{\mbox{\scriptsize DW}}=N_g$ depending on the scalar potential implementation, with $N_g$ the number of quark generations. In Table~\ref{DWtab} we present the values of the domain wall number for the different implementations of the models we are presenting. As we can see, while some of the implementations also suffer from the domain wall problem, others have $N_{\mbox{\scriptsize DW}}=1$, for which the resulting domain wall structure is harmless \cite{Barr:1986hs}.
\begin{table}[h]
\begin{center}
\begin{tabular}{c|cccccccccccc}
&$T_1$&$T_2$&$T_3$&$T_4$&$T_5$&$T_6$&$T_7$&$T_8$&$T_9$&$T_{10}$&$T_{11}$&$T_{12}$\\  
\hline\hline
$N_{\mbox{\scriptsize DW}}^I$&3&3&2&1&2&1&$-$&2&2&$-$&1&1\\   \rowcolor{Gray}
$N_{\mbox{\scriptsize DW}}^{II}$&2&4&2&4&6&6&4&4&$-$&2&2&$-$\\   
$N_{\mbox{\scriptsize DW}}^{III}$&$-$&3&5&7&7&8&2&$-$&$-$&1&$-$&$-$\\ 
\hline\hline
\end{tabular}
\caption{  \it \small Values for the domain wall number in each of the possible scenarios.}   \label{DWtab}
\end{center}
\end{table}

Even for $N_{\mbox{\scriptsize DW}}\neq1$, some solutions to the domain wall problem can be found in the literature.     It is possible to avoid the domain wall problem by assuming
that inflation has occurred after the PQ symmetry breaking.  In this case, one can derive limits on the inflationary
scale based on the observation of isocurvature fluctuations in the cosmic microwave background~\cite{Lyth:1989pb}.    Also, there have been several attempts to introduce an explicit breaking of the PQ symmetry that also breaks the $\mathcal{Z}_{N}$ discrete group in such a way that the PQ solution to the Strong CP problem is protected~\cite{Sikivie:1982qv,Holdom:1982ew}. This explicit breaking could come from gravity, giving rise to Planck scale suppressed operators~\cite{Abbott:1989jw,Banks:1989zw}. However, it was argued that this solution would give rise to long lived domain walls which introduce cosmological problems~\cite{Hiramatsu:2012sc}. Additionally, gravity violations of the PQ symmetry should be controlled in order not to spoil the PQ solution~\cite{Ghigna:1992iv, Holman:1992us, Kamionkowski:1992mf, Georgi:1981pu, Barr:1992qq, Kallosh:1995hi,Dobrescu:1996jp}. This issue will be considered in the next section.

\subsection{Protecting the axion against gravity \label{subsec:gravity}}

Until now we have discussed a model where an ad hoc PQ symmetry is imposed. However, as already mentioned, the presence of semi-classical gravitational effects can potentially violate global symmetries~\cite{Abbott:1989jw,Banks:1989zw}, spoiling the strong CP solution~\cite{Ghigna:1992iv, Holman:1992us, Kamionkowski:1992mf, Georgi:1981pu, Barr:1992qq, Kallosh:1995hi,Dobrescu:1996jp}. 

In the absence of gravity, the axion potential coming from the instantonic contributions can be written as~\cite{Peccei:1977hh}
\be
V_{\mbox{\scriptsize{{axion}}}}\simeq-\Lambda_{\mbox{\scriptsize{QCD}}}^4\,\cos\frac{a_{\text{phys}}}{v_{\mbox{\scriptsize PQ}}}\,,
\ee
which has a minimum at $\left<a_{\text{phys}}\right>=\bar{\theta}=0$ and where the estimated axion mass is $m_a\simeq \Lambda_{\mbox{\scriptsize{QCD}}}^2/v_{\mbox{\scriptsize PQ}}$. On the other hand, when including gravitational effects, the axion potential will change. For example in our invisible axion model, we should expect higher dimensional PQ violating terms of the type
\be
\frac{1}{M_{\mbox{\scriptsize{Pl}}}^{n-2}}\Phi_i^\dagger\Phi_j S^n\,,\quad \frac{1}{M_{\mbox{\scriptsize{Pl}}}^{n-4}} S^n\,,\cdots\,
\ee
with the Planck scale denoted by $M_{\mbox{\scriptsize{Pl}}}$. Let us consider, for simplicity, the second term in the above equation. By introducing this term in the Lagrangian the axion potential gets modified and takes the form~\cite{Barr:1992qq}
\be
\tilde{V}_{\mbox{\scriptsize{axion}}}\simeq-\Lambda_{\mbox{\scriptsize{QCD}}}\,\cos\frac{a_{\text{phys}}}{v_{\mbox{\scriptsize PQ}}}-\frac{c \,v_{\mbox{\scriptsize PQ}}^n}{M_{\mbox{\scriptsize{Pl}}}^{n-4}}\,\cos\left[\frac{a_{\text{phys}}}{v_{\mbox{\scriptsize PQ}}}+\delta\right]\,.
\ee 
The parameter $c$ is just a coupling constant and $\delta$ a CP violating phase. The problem with this new axion potential is that the minimum is no longer at $\left<a_{\text{phys}}\right>=0$, but rather at
\be
\bar{\theta}=\frac{\left<a_{\text{phys}}\right>}{v_{\mbox{\scriptsize PQ}}}\simeq c\,\sin\delta\frac{v^n_{\mbox{\scriptsize PQ}}}{M_{\mbox{\scriptsize{Pl}}}^{n-4}\Lambda_{\mbox{\scriptsize{QCD}}}^4}\,,
\ee
which in general will be far from zero. The axion mass will also be affected by these gravitational effects, taking the form
\be
m_a^2\simeq \frac{\Lambda_{\mbox{\scriptsize{QCD}}}^4}{v_{\mbox{\scriptsize PQ}}^2}+c\frac{v_{\mbox{\scriptsize PQ}}^{n-2}}{M_{\mbox{\scriptsize{Pl}}}^{n-4}}\,.
\ee
Therefore, in this simple scenario we see that gravitational effects will in general spoil the strong CP solution coming from the PQ symmetry. 

Fortunately, over the years several solutions to this problem have emerged, which allows us to preserve the PQ solution of the strong CP problem. GUT motivated models~\cite{Georgi:1981pu,Barr:1992qq}, extra dimensional~\cite{Choi:2003wr} ones and even models having neutrinos playing a big role in gravity protection~\cite{Dvali:2013cpa} can be found in the literature. However, many of these solutions need a significant extension of the original PQ model.

In this section we will focus on the use of gauge discrete symmetries to protect the PQ solution against gravity~\cite{Banks:1991xj}.  This solution has the interesting feature that the low energy spectrum of the theory does not need to be extended. Gauge discrete symmetries, which arise through the spontaneous symmetry breaking of a gauge symmetry, are not broken by gravity and can provide natural suppression to the harmful gravitational effects. The idea proposed is to have a large discrete abelian symmetry $\mathcal{Z}_P$ forbidding, up to a given order, these unwanted terms~\cite{Babu:2002ic,Dias:2002gg}. For example, if the symmetry only allows terms of the form $S^m/M_{\mbox{\scriptsize{Pl}}}^{m-4}$ for $m\geq 13$, we will just get irrelevant contributions to the axion mass and its vev~\cite{Dias:2014osa}.

In what follows we will identify the PQ symmetry as an accidental global symmetry at low energies, associated with the spontaneous breaking of a gauge symmetry, $\mathrm{U(1)}_A$, at high energies down to a discrete subgroup. We shall follow Ref.~\cite{Babu:2002ic}, using a discrete gauge symmetry to stabilize the axion without enlarging the low energy particle content. To this end, we shall use the discrete version of the Green-Schwarz anomaly cancellation mechanism~\cite{Green:1984sg}.

From the effective theory point of view, since we have at low energies the $\mathrm{SU(3)}_C\times \mathrm{SU(2)}_L\times \mathrm{U(1)}_Y$ gauge group, there are several possible anomalies we must consider:
\begin{align}
\begin{array}{lll}
\mathcal{A}_1:\; [\mathrm{U(1)}_Y]^2\times \mathrm{U(1)}_A\,, & \mathcal{A}_2:\; [\mathrm{SU(2)}_L]^2\times \mathrm{U(1)}_A\,, & \mathcal{A}_3:\; [\mathrm{SU(3)}_C]^2\times \mathrm{U(1)}_A\,,\\ 
\mathcal{A}_A:\; [ \mathrm{U(1)}_A]^3\,, & \mathcal{A}_G:\; [\text{gravity}]^2\times \mathrm{U(1)}_A\,.
\end{array}
\end{align}
The Green-Schwarz anomaly cancellation conditions are then given by 
\be
\frac{\mathcal{A}_1}{k_1}=\frac{\mathcal{A}_2}{k_2}=\frac{\mathcal{A}_3}{k_3}=\frac{\mathcal{A}_A}{k_A}=\frac{\mathcal{A}_G}{12}=\delta_{\mbox{\scriptsize GS}}\,,
\ee
with $\delta_{\mbox{\scriptsize GS}}$ a constant that cannot be specified by the low energy theory and $k_{1,2,3,A}$ the levels of the Kac-Moody algebra~\cite{Kac:1967jr} which are integers for non-abelian groups. The equality involving the hypercharge currents, $\mathcal{A}_1$, give no useful constraints since the associated level $k_1$ is not an integer in general \cite{Banks:1991xj}. Similarly, the anomaly $\mathcal{A}_A$ can be canceled by the Green-Schwarz mechanism but with no useful constraints due to the arbitrariness in the normalization of $\mathrm{U(1)}_A$. Finally, the anomaly $\mathcal{A}_G$ give no useful constraint either.

When the $\mathrm{U(1)}_A$ is broken down to a $\mathcal{Z}_P$, the effective low energy theory will satisfy the discrete version of the Green-Schwarz cancellation condition~\cite{Banks:1991xj,Ibanez:1991hv}
\be\label{eq:GSdiscrete}
\frac{\mathcal{A}_3+m P/2}{k_3}=\frac{\mathcal{A}_2+m^\prime P/2}{k_2}\,,
\ee 
with $m$ and $m^\prime$ integers. The model under discussion is non-supersymmetric, nevertheless, the Green-Schwarz mechanism should still be available since the breaking of supersymmetry can happen at the scale much higher that the weak scale.

Our goal is to build a $\mathrm{U(1)}_A$ symmetry that contains a discrete subgroup capable of solving the strong CP problem. The $\mathrm{U(1)}_{\mbox{\scriptsize{PQ}}}$ group is anomalous and, therefore, capable of giving such a solution (as it was seen in the previous sections). However, $\mathrm{U(1)}_{\mbox{\scriptsize{PQ}}}$ cannot be identified with $\mathrm{U(1)}_A$ as the PQ symmetry alone is, in general, not enough to satisfy the Green-Schwarz anomaly conditions. Fortunately, the model also presents Baryon number conservation ($+1$ charge for quarks, $-1$ for anti-quarks), which is QCD anomaly free but it is anomalous under $\mathrm{SU(2)}_L$. We shall then try to see if the combination $\mathrm{U(1)}_{\mbox{\scriptsize{PQ}}}+\gamma \mathrm{U(1)}_B$ is suitable to be our axial symmetry. As the lepton charges depend on the specific representation in the neutrino sector, we will focus on the quark sector. The generalization to the lepton sector will be discussed at the end. From the $\mathrm{U(1)}_{\mbox{\scriptsize{PQ}}}$ charge assignments in Eqs.~\eqref{eq:chargesI}, \eqref{eq:chargesII} and~\eqref{eq:chargesIII} we can find the anomaly coefficients
\begin{align}\label{eq:A2A3}
\begin{split}
\mathcal{A}_2=&\frac{3}{2}\left(3\gamma+X_{tL}\right)=\frac{3}{2}\left(3\gamma+\frac{1}{C_S^{M}}\right)\,,\\ 
\mathcal{A}_3=&\frac{1}{2}\left(2X_{tL}-2X_{uR}-X_{tR}-3X_{dR}\right)=-\frac{C_{ag}^M}{2}\,,
\end{split}
\end{align}
with $M=I,\, II,\, III$. The factor $\gamma$ is then found to be
\be
\gamma=-\frac{1}{9}\left(\frac{k_2}{k_3}C_{ag}^M+\frac{3}{C_S^{M}}\right)\,.
\ee
Using the simplest realization of the Kac--Moody algebra, i.e. $k_2=k_3=1$, we get for each case
\begin{align}
\begin{split}
\textbf{Case I:}&\quad\gamma=\frac{1-C_I}{9C_S^IC_I}\,,\quad\textbf{Case II:}\quad\gamma=-\frac{2+3C_{II}}{9C_S^{II}C_{II}}\,,\quad\textbf{Case III:}\quad\gamma=-\frac{3+2C_{III}}{9C_S^{III}C_{III}}\,.\\
\end{split}
\end{align}
Normalizing the combination to have all charges integer number we then define the axial abelian symmetry as
\be
\mathrm{U(1)}_A=9 \mathrm{U(1)}_{\mbox{\scriptsize{PQ}}}+9\gamma \mathrm{U(1)}_B\,.
\ee
The charges under this new symmetry are given in Table~\ref{tab:U1A}. Finally, note that the inclusion of PQ lepton charges would modify $\mathcal{A}_2$ in the following way
\begin{align}
\mathcal{A}_2\to\mathcal{A}_2+\frac{1}{2}\left(X_{eL}+X_{\mu L}+X_{\tau L}\right)\,,
\end{align}
while $\mathcal{A}_3$ would remain unaltered. This accounts to a correction of $\gamma$ of the form
\begin{align}
\gamma\to\gamma-\frac{1}{9}\left(X_{eL}+X_{\mu L}+X_{\tau L}\right)\,,
\end{align}
which transforms the quark charges in Table~\ref{tab:U1A} to
\begin{align}
\begin{split}
Q_{Li}&\to Q_{Li}-\left(X_{eL}+X_{\mu L}+X_{\tau L}\right)\,,\\
u_{Ri}&\to u_{Ri}-\left(X_{eL}+X_{\mu L}+X_{\tau L}\right)\,,\\
d_{Ri}&\to d_{Ri}-\left(X_{eL}+X_{\mu L}+X_{\tau L}\right)\,,
\end{split}
\end{align}
and leaves the lepton and scalar charges unchanged.

\begin{table}[t]
\begin{center}
\begin{tabular}{c|lll}
$\mathrm{U(1)}_A$&Case I&Case II&Case III\\
\hline\hline
$Q_{L1,2}$&$\frac{1-C_I}{C_SC_I}$&$-\frac{2+3C_{II}}{C_S^{II}C_{II}}$&$-\frac{3+2C_{III}}{C_S^{III}C_{III}}$\\    \rowcolor{Gray}
$Q_{L3}$&$\frac{8C_I+1}{C_SC_I}$&$\frac{6C_{II}-2}{C_S^{II}C_{II}}$&$\frac{7C_{III}-3}{C_{S}^{III}C_{III}}$\\
$u_{R1,2}$&$x+\frac{1-C_I}{C_SC_I}$&$x-\frac{2+3C_{II}}{C_S^{II}C_{II}}$&$x-\frac{3+2C_{III}}{C_S^{III}C_{III}}$\\    \rowcolor{Gray}
$u_{R3}$&$x-\frac{C_1+8}{C_SC_I}$&$x+\frac{15C_{II}-11}{C_S^{II}C_{II}}$&$x+\frac{7C_{III}-3}{C_S^{III}C_{III}}$\\
$d_{R1,2,3}$&$-x+\frac{1-C_I}{C_SC_I}$&$-x+\frac{7-3C_{II}}{C_S^{II}C_{II}}$&$-x+\frac{6-2C_{III}}{C_S^{III}C_{III}}$\\    \rowcolor{Gray}
$\Phi_1$&$x$&$x$&$x$\\
$\Phi_2$&$x-\frac{9+9C_I}{C_S^IC_I}$&$x-\frac{9}{C_S^{II}C_{II}}$&$x-\frac{9}{C_S^{III}C_{III}}$\\     \rowcolor{Gray}
$\Phi_3$&$x+\frac{9}{C_S^I}$&$x+\frac{9C_{II}-9}{C_S^{II}C_{II}}$&$x+\frac{9C_{III}-9}{C_S^{III}C_{III}}$\\
$S$&9&9&9\\
\hline\hline
\end{tabular}
\caption{ \it \small Charge assignments under $\mathrm{U(1)}_A$,    $x=9X_{uR}$.}\label{tab:U1A}
\end{center}
\end{table}

In Table~\ref{tab:exT8} we present the axial symmetry and its discrete $\mathcal{Z}_{13}$ version in the $T_6$ scenario, for each case I, II and III. The discrete anomaly coefficients are $\mathcal{A}_3=-2$ and $\mathcal{A}_2=24$ for case I, $\mathcal{A}_3=-15/2$ and $\mathcal{A}_2=12$ for case II, and $\mathcal{A}_3=3$ and $\mathcal{A}_2=45/2$ for case III. In each case, by construction, the anomaly coefficients satisfy the discrete Green-Schwarz cancellation condition Eq.~\eqref{eq:GSdiscrete}. 

\begin{table}[h]
\begin{center}
\begin{tabular}{c|ccccccccc}
$T_6$&$Q_{L1,2}$&$Q_{L3}$&$u_{R1,2}$&$u_{R3}$
&$d_{R1,2,3}$&$\Phi_1$&$\Phi_2$&$\Phi_3$&$X_S$\\
\hline\hline
Case I:&&&&&&&&&\\    \rowcolor{Gray}
$\mathrm{U(1)}_A$&7&$-11$&$x+7$&$x-38$&$-x+7$&$x$&$x-27$&$x-18$&9\\
$\mathcal{Z}_{13}$&7&2&$x+7$&$x+1$&$-x+7$&$x$&$x+12$&$x+8$&9\\
\hline
Case II:&&&&&&&&&\\    \rowcolor{Gray}
$\mathrm{U(1)}_A$&$-9$&$0$&$x-9$&$x-18$&$-x+18$&$x$&$x-27$&$x-18$&9\\
$\mathcal{Z}_{13}$&4&0&$x+4$&$x+8$&$-x+5$&$x$&$x+12$&$x+8$&9\\
\hline
Case III:&&&&&&&&&\\    \rowcolor{Gray}
$\mathrm{U(1)}_A$&$-11$&$-2$&$x-11$&$x-2$&$-x+16$&$x$&$x-27$&$x-18$&9\\
$\mathcal{Z}_{13}$&2&11&$x+2$&$x+11$&$-x+3$&$x$&$x+12$&$x+8$&9\\
\hline\hline
\end{tabular}
\caption{ \it \small Particular example with the phase sensitive scalar potential $T_6$. \label{tab:exT8}}
\end{center}
\end{table}
In this example the phase sensitive terms are explicitly given by
\be\label{eq:T6terms}
T_6:\quad \Phi_1^\dagger\Phi_3S^2\,,\quad \Phi_2^\dagger\Phi_3 S^*\,.
\ee
Due to gravity effects we expect the most relevant $\mathrm{U(1)}_{\mbox{\scriptsize{PQ}}}$ breaking contributions to be of the type
\be
\left[\mathcal{O}_{4-d}\right]\times \frac{S^k}{M_{Pl}^{k-d}}:\quad \left\{
\begin{array}{ll}
d=4& \mathcal{O}_0\sim \text{const}\\
d=2&\mathcal{O}_2\sim |\Phi_i|^2,\, |S|^2,\,\cdots\\
d=1& \mathcal{O}_3\sim \Phi_2^\dagger\Phi_3 S^*\\
d=0&\mathcal{O}_4\sim |\Phi_i|^4,\, |S|^4,\,\cdots
\end{array}\right.
\ee
with $k$ integer. The largest contribution will be the one coming from the $\mathcal{O}_0$ operator. Due to the $Z_{13}$ symmetry this contribution will only take place for $k=13$, i.e. $S^{13}/M_{Pl}^9$. This operator will give a contribution to the axion mass squared of the order $v_{\mbox{\scriptsize{PQ}}}^{11}/M_{Pl}^9\sim \left[10^{-72},10^{-39}\right]\,\text{GeV}^2$ for PQ scales between $10^{9}$ to $10^{12}$ GeV. The contribution to the $\bar{\theta}$ will be between $10^{-54}$ to $10^{-15}$. These are extremely small contributions, making the model safe against large gravitational corrections.

In this section we have shown how we could avoid large contributions to the axion mass, as well as to the $\bar{\theta}$ parameter, just by invoking a discrete gauge symmetry. However, there are many more effective operators that will be induced by gravity than those presented above. Some of them could give contributions to the original Yukawa textures potentially spoiling the good behavior of the BGL-like textures. 

Let us choose case I as a particular scenario. From the $\mathcal{Z}_{13}$ charge assignments we have the Yukawa term $\overline{Q}_{L1}\,u_{R3}\,\tilde{\Phi}_2$ carrying a net charge $8$. This term is not allowed at the renormalizable level, but the gravity induced effects can introduce the $\mathcal{Z}_{13}$ invariant term $\overline{Q}_{L1}\,u_{R3}\,\tilde{\Phi}_2\,(S/M_{Pl})^2$. This term will contribute to the Yukawa textures once $S$ spontaneously breaks the PQ symmetry with a correction of the order $y\times v_{\mbox{\scriptsize PQ}}^2/M_{Pl}^2$, with $y$ the associated Yukawa coupling. For a PQ breaking scale of order $v_{\mbox{\scriptsize PQ}}\lsim\mathcal{O}\left(10^{15}\right)$~GeV this operator give a harmless contribution. However, for higher scales this term could give significant corrections to the BGL-like suppression when $y\sim\mathcal{O}(1)$. Nevertheless, even for a high PQ breaking scale, we could have $\mathcal{O}(y)\ll 1$ suppressing this additional contribution. This is not so strange taking into account that in our framework no information on the Yukawa hierarchy is given. We know that $\mathcal{O}(y_u,\,y_d,\,\cdots)\ll 1$ and in our framework this is imposed by hand. In a more complete model, these hierarchies could be made dynamical and there we should also take attention to these additional gravity induced terms.

\section{Model variations}\label{sec:modelvar}

The models presented in the main sections of the paper had FCNCs in the down-quark sector and the top quark was singled out. However, there are many other possible implementations that will still give the same minimal flavor violating scenario. These model variations can be found by performing any of the two operations:
\begin{itemize}
\item[(i)] Symmetric permutations in the flavor space;

\item[(ii)] Changing up and down right-handed generators.

\end{itemize}
We can apply these two operations to the models previously studied in order to get all possible model variants. 

\subsection{Type \texorpdfstring{$(i)$}{(i)} operation}
The permutations in flavor space will change the textures in the sector with no FCNCs, i.e the up sector if we apply this operation in the original formulation. The symmetry generators take now the form 
\be\label{eq:generatorsP}
\mathcal{S}_{L}\rightarrow P^T\mathcal{S}_LP\,,\quad \mathcal{S}_{R}^{u,d}\rightarrow P^T\mathcal{S}_R^{u,d}P\quad
\mathcal{S}^e_{L}\rightarrow P^{\prime T}\mathcal{S}_LP^\prime\,,\quad \mathcal{S}_{R}^{e}\rightarrow P^{\prime T}\mathcal{S}_R^{e}P^\prime\,,
\ee
with $P$ and $P^\prime$ $3\times3$ permutation matrices. The 2 by 2 block in the NFC sector will change structure, we get 
\begin{subequations}
\begin{equation}\label{eq:P23}
P,P^\prime=P_{23}\quad\longrightarrow\quad
\left\{
\begin{pmatrix}
\times&&\times\\
&&\\
\times&&\times
\end{pmatrix},\quad
\begin{pmatrix}
&&\\
&\times&\\
&&
\end{pmatrix}
,\,\quad
\begin{pmatrix}
\times&&\times\\
&\times&\\
\times&&\times
\end{pmatrix}
\right\}:\,
\text{Block }1-3
\end{equation}
\begin{equation}\label{eq:P13}
P,P^\prime=P_{13}\quad\longrightarrow\quad
\left\{
\begin{pmatrix}
&&\\
&\times&\times\\
&\times&\times
\end{pmatrix},\quad
\begin{pmatrix}
\times&&\\
&&\\
&&
\end{pmatrix}
,\,\quad
\begin{pmatrix}
\times&&\\
&\times&\times\\
&\times&\times
\end{pmatrix}
\right\}:\,
\text{Block }2-3
\end{equation}
\end{subequations}
Where $P_{ij}$ permutes the lines $i$ and $j$ (when applied on the left) and columns $i$ and $j$ (when applied on the right). Besides the textures the only changes due to $(i)$ are in the axion-matter couplings. The permutation matrices single out other flavors. Therefore, the action of the permutation matrices will change the CKM and PMNS elements entering in Eq.~\eqref{eq:quarkChi} and Eq.~\eqref{eq:leptonChi}, respectively.   We get the following redefinition
\begin{align}
\begin{split}
P,P^\prime&=P_{23}\quad\longrightarrow\quad t\rightarrow c, \tau\rightarrow\mu\\
P,P^\prime&=P_{13}\quad\longrightarrow\quad t\rightarrow u, \tau\rightarrow e
\end{split}
\end{align}
Consequently, the couplings $u$, $d$, $s$ and $e$ are appropriately changed.

\subsection{Type \texorpdfstring{$(ii)$}{(ii)} operation}
We change the sector where the FCNCs are present, that can be accounted with the following symmetry generators
\be\label{eq:generatorsii}
\mathcal{S}_L=\text{diag}(1,1,e^{iX_{tL}\theta})\,,\quad \mathcal{S}_R^u=e^{iX_{dR}\theta}\mathbb{I}\,,\quad \mathcal{S}_R^d=\text{diag}(e^{iX_{uR}\theta},e^{iX_{uR}\theta},e^{iX_{tR}\theta})\,.
\ee
We have switched the $\mathcal{S}_R^u$ and $\mathcal{S}_R^d$ generators, keeping the same labels for the charges. Thus, in this scenario the $\mathcal{S}_R^u$ is completely degenerate, but instead of labeling the charge $X_{uR}$ we keep it labeled as $X_{dR}$, just as in the previous case. By keeping the same label we can easily compare this new case with the previous scenario where the FCNCs where in the down sector. The Higgs charges, for the three cases, take the same form as in the original scenario (see Eqs.~\eqref{eq:PhiCaseI} and~\eqref{eq:PhiCaseII}) but with an overall minus sign. 
\begin{align}\label{eq:PhiCase2}
\begin{split}
\textbf{Case I:}&\quad X_{\Phi 1}=-X_{uR}\,,\quad X_{\Phi 2}=-(X_{tR}-X_{tL})\,,\quad X_{\Phi 3}=-(X_{tL}+X_{uR})\,.\\
\textbf{Case II:}&\quad X_{\Phi 1}=-X_{uR}\,,\quad X_{\Phi 2}=X_{dR}\,,\quad X_{\Phi 3}=-(X_{tL}-X_{dR})\,.\\
\textbf{Case III:}&\quad X_{\Phi 1}=-X_{uR}\,,\quad X_{\Phi_2}=X_{dR}\,,\quad X_{\Phi_3}=-(X_{tL}-X_{dR})\,.
\end{split}
\end{align}

The quark charges take the same form as in Eq.~\eqref{eq:chargesI},~\eqref{eq:chargesII} and~\eqref{eq:chargesIII} as long as in the scalar sector the role of the $S$ field is substituted by the $S^\ast$, keeping the $X_S=1$ normalization. This will account for the overall minus sign coming from the Higgs charges. While the left-handed quark charges have the same expression as in the original scenario, the right-handed ones switched sectors. The coupling to gluons will not change, however the coupling to photons changes since the up and down electric charges are different. It will be given by
\be
C_{a\gamma}=2\left[4X_{dR}+\frac{2}{3}X_{uR}+\frac{1}{3}X_{tR}-\frac{5}{3}X_{tL}+\sum_\alpha (X_{\alpha R}-X_{\alpha L})\right]\,.
\ee
This will have the same form as Eq.~\eqref{eq:Cagamma}, but now the coefficients take the form given in Table~\ref{newcomb2}

\begin{table}[h]
\begin{center}
\begin{tabular}{r|ccc}
Cases&$A_I$&$A_{II}$&$A_{III}$\\
\hline\hline
$(1,1,1)$&$-1-5C_I$&$11-3C_{II}$&$12-4C_{III}$\\     \rowcolor{Gray}
$(2,2,2)$&$-10-14C_I$&$2-3C_{II}$&$3-4C_{III}$\\
$(3,3,3)$&$-1+4C_I$&$2+6C_{II}$&$3+5C_{III}$\\  \rowcolor{Gray}
$(1,1,2)$&$-4-8C_I$&$8-3C_{II}$&$9-4C_{III}$\\
$(1,1,3)$&$-1-2C_I$&$8$&$9-C_{III}$\\  \rowcolor{Gray}
$(2,2,1)$&$-7-11C_I$&$5-3C_{II}$&$6-4C_{III}$\\
$(2,2,3)$&$-7-8C_I$&$2$&$3-C_{III}$\\  \rowcolor{Gray}
$(3,3,1)$&$-1+C_I$&$5+3C_{II}$&$6+2C_{III}$\\
$(3,3,2)$&$-4-2C_I$&$2+3C_{II}$&$3+2C_{III}$\\  \rowcolor{Gray}
$(1,2,3)$&$-4-5C_I$&$5$&$6-C_{III}$\\
\hline\hline
\end{tabular}
\end{center}
\caption{ \it \small Charge combinations $A_I$, $A_{II}$ and $A_{III}$ entering in the description of the axion coupling to photons.    The numbers in the first column label the Higgs doublet giving mass to the charged leptons ($e, \mu, \tau$); for example (1,1,2) stands for the case where $\Phi_1$ gives mass to $e$ and $\mu$ while $\Phi_2$ to $\tau$.   \label{newcomb2} }   
\end{table}

The axion coupling to matter will also change, since we have changed the FCNC sector. The shift we need to perform in order to account for the orthogonality between the axion and the Goldstone boson will get an overall minus sign with respect to Eq.~\eqref{eq:Z}. Because we changed sectors without changing the labels of the charges, the charge shifts will coincide with the ones in Eq.~\eqref{eq:Xshift}.  The shifted charge matrix takes the form
\begin{align}\label{eq:chargesmatrix2}
\begin{split}
\mathcal{X}_{uL}=&\mathcal{X}_{dL}=\text{diag}(0,\,0,\, X^\prime_{tL})\,,\quad
\mathcal{X}_{eL}=\text{diag}(X^\prime_{eL},\,X^\prime_{\mu L},\, X^\prime_{\tau L})\,,\\ 
\mathcal{X}_{uR}=&X^\prime_{dR}\mathbb{I}\,,\quad \mathcal{X}_{dR}=\text{diag}(X^\prime_{uR},\,X^\prime_{uR},\, X^\prime_{tR})\,,\quad \mathcal{X}_{eR}=\text{diag}(X^\prime_{eR},\,X^\prime_{\mu R},\, X^\prime_{\tau R})  \,.
\end{split}
\end{align}
In the mass basis they take the explicit form
\be\label{eq:quarkChi2}
\widetilde{\mathcal{X}}_{uL}=X_{tL}\begin{pmatrix}
|V_{ub}|^2&V_{ub} V^\ast_{cb}&V_{ub} V^\ast_{tb}\\
V_{cb} V^\ast_{ub}& |V_{cb}|^2&V_{ub} V_{tb}^\ast\\
V_{tb} V^\ast_{ub}&V_{tb} V^\ast_{cb}&|V_{tb}|^2
\end{pmatrix}\,,\quad \widetilde{\mathcal{X}}_{uR}=\mathcal{X}_{uR}\,,\quad\widetilde{\mathcal{X}}_{dL}=\mathcal{X}_{dL}
\,,\quad \widetilde{\mathcal{X}}_{dR}=\mathcal{X}_{dR}\,,
\ee
The axial couplings to the light quarks are now given by
\begin{align}\label{eq:guds2}
\begin{split}
g_d&\equiv (C_{ad}^A)_{11}=\left\{
\begin{array}{ll}
-\dfrac{v_2^2\left(1+C_I\right) - v_3^2 C_I}{v^2C_S^I C_I}&\text{for case I,}\\
-\dfrac{v_2^2+v_3^2\left(1-C_{II(III)}\right)}{v^2C_S^{II(III)} C_{II(III)}}&\text{for case II and III}\,,
\end{array}\right.\\
g_s&\equiv (C_{ad}^A)_{22}=g_d\,,\\
g_u&\equiv (C_{au}^A)_{11}=-g_{d}+\left\{
\begin{array}{ll}
\dfrac{|V_{ub}|^2}{C_{S}^{I}}&\text{for case I}\,,\\
\dfrac{|V_{ub}|^2 C_{II(III)}-1}{C_{S}^{II(III)}C_{II(III)}}&\text{for case II and III}\,.
\end{array}
\right.\
\end{split}
\end{align}

\subsection{Model variations dictionary}
In this section we present a compilation of the most significant changes resulting from applying the operations $(i)$ or $(ii)$ to the original formulation. These can be found in Table~\eqref{tab:modelvar}.  
\begin{table}[ht]
\begin{center}
\begin{tabular}{c||c|c|c|c||}
&\multirow{2}{*}{Original}&\multicolumn{2}{|c|}{Operation (i)}&\multirow{2}{*}{Operation (ii)}\\\cline{3-4}
&&$P_{23}$&$P_{13}$&\\
\hline\hline\vspace{-0.2cm}
Yukawa&\multirow{2}{*}{Eq.~\eqref{eq:aBGL},~\eqref{eq:aBGLII},~\eqref{eq:aBGLIII}}&\multirow{2}{*}{Eq.~\eqref{eq:P23}}&\multirow{2}{*}{Eq.~\eqref{eq:P13}}&\multirow{2}{*}{Original}\\
Textures&&&&\\\hline\vspace{-0.2cm}
Symmetry&\multirow{2}{*}{Eqs.~\eqref{eq:chargesaBGL1},~\eqref{eq:chargesaBGL2}}&\multicolumn{2}{|c|}{\multirow{2}{*}{Eq.~\eqref{eq:generatorsP}}}&
\multirow{2}{*}{Eq.~\eqref{eq:generatorsii}}\\
Generators&&\multicolumn{2}{|c|}{}&\\\hline\vspace{-0.2cm}
Scalar&\multicolumn{3}{|c|}{\multirow{2}{*}{Eqs.~\eqref{eq:PhiCaseI},~\eqref{eq:PhiCaseII}}}&\multirow{2}{*}{Eq.~\eqref{eq:PhiCase2}}\\
Charges&\multicolumn{3}{|c|}{}&\\\hline\vspace{-0.2cm}
Coupling to&\multicolumn{3}{|c|}{\multirow{2}{*}{Table~\ref{newcomb}}}&\multirow{2}{*}{Table~\ref{newcomb2}}\\
Photons&\multicolumn{3}{|c|}{}&\\\hline\vspace{-0.2cm}
Orthogonality&\multicolumn{3}{|c|}{\multirow{2}{*}{Eq.~\eqref{eq:Z}}}&\multirow{2}{*}{Overall minus sign}\\
Shift&\multicolumn{3}{|c|}{}&\\\hline\vspace{-0.2cm}
Mass Basis&\multirow{2}{*}{Eqs.~\eqref{eq:chargesmatrix},~\eqref{eq:quarkChi}}&$t\rightarrow c$&$t\rightarrow u$&\multirow{2}{*}{Eqs.~\eqref{eq:chargesmatrix2},~\eqref{eq:quarkChi2}}\\
Charges&&$\tau \rightarrow \mu$&$\tau\rightarrow e$&\\\hline\vspace{-0.2cm}
$u$, $d$, $s$&\multirow{2}{*}{Eqs.~\eqref{eq:guds}}&\multirow{2}{*}{$t\rightarrow c$}&\multirow{2}{*}{$t\rightarrow u$}&\multirow{2}{*}{Eqs.~\eqref{eq:guds2}}\\
couplings&&&&\\\hline\vspace{-0.2cm}
Electron&\multirow{2}{*}{Eqs.~\eqref{eq:ge}}&\multirow{2}{*}{$\tau\rightarrow \mu$}&\multirow{2}{*}{$\tau\rightarrow e$}&\multirow{2}{*}{Original}\\
coupling&&&&\\\hline\hline
\end{tabular}
\caption{\label{tab:modelvar}  \it \small  This table summarizes the changes we need to do in order to get all possible model variations. The firs column, i.e. \textit{Original}, presents the various relevant equations in the scenario where the down sector has FCNCs and the top is singled out. The second column, i.e. \textit{Operation (i)}, represents models with permutations in the flavor space of each sector. The last column, i.e. \textit{Operation (ii)}, represent models with quark sectors interchanged.}
\end{center}
\end{table}

\section{Discussion} \label{sec:discus}
In the previous sections we have characterized a class of invisible axion models with tree-level FCNCs. We have also detailed the most relevant properties of the axion in these models.   In this section we will analyze the different constraints on these models due to familon searches in kaon and muon decays, astrophysical considerations, as well as axion searches that rely on the axion-photon conversion mechanism.   We will separate the discussion as follows:  In Sec.~\ref{ssec:photon} we discuss the constraints that can be extracted from the axion-photon coupling.    In Sec.~\ref{ssec:matter} we consider constraints on the axion from its flavor diagonal couplings to matter (nucleons and electrons).    In Sec.~\ref{ssec:flavor} we discuss constraints derived from familon searches in rare kaon and muon decays.    Our main results regarding the constraints on the axion are summarized in Figs.~\ref{fig:constraintsO} and~\ref{fig:constraintsP}, the reader familiar with the axion phenomenology might prefer to skip directly to these figures.  Finally, in Sec.~\ref{ssec:higgs} we discuss the phenomenology of the Higgs sector within the frameworks considered, for this we feel it is important to discuss also the possible decoupling limits of the scalar sector.

\subsection{Constraining the axion coupling to photons  \label{ssec:photon}}

\begin{figure}[ht!]
\centering
\includegraphics[width=0.8\textwidth]{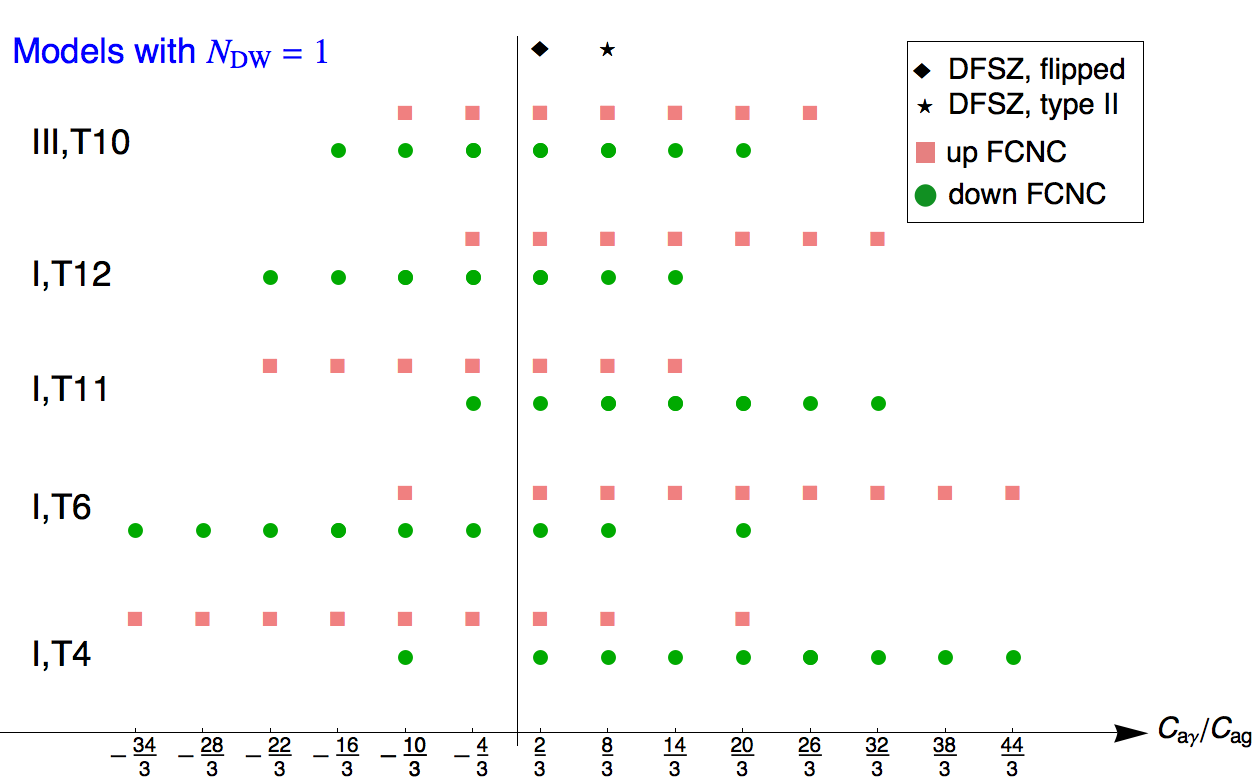}
\caption{\label{fig:varationCag} \it \small     Values of $C_{a\gamma}/C_{ag}$ in 3HFPQ models with $N_{\mbox{\scriptsize{DW}}}=1$.    Squares stand for models with FCNC in the up-quark sector while those with FCNC in the down-quark sector are denoted with circles. The corresponding values for $C_{a\gamma}/C_{ag}$ in the DFSZ models (type II and flipped) are also shown.  }
\end{figure}

The axion couples to photons through the dimension 5 operator in Eq.~\eqref{aphotonco}. The axion-photon coupling constant is defined by
\be
g_{a\gamma}=\frac{\alpha}{2\pi v_{\mbox{\scriptsize{PQ}}}}C_{ag}C_{a\gamma}^{\mbox{\footnotesize{eff}}}\,.
\ee
In various extensions of the SM, weakly coupled light pseudoscalar particles emerge naturally. However, the axion possesses (due to QCD effects) an inherent correlation between the photon coupling and its mass
\be
\left(m_a/1\,\text{eV}\right)\simeq 0.5\, \xi\, g_{10}\,,
\ee
where $g_{10}=|g_{a\gamma}|/(10^{-10}\,\text{GeV}^{-1})$ and $\xi=1/|C_{a\gamma}^{\text{eff}}|$. The dimensionless coefficient $\xi$ is in many axion models of order 1. In the well-known DFSZ (type II and flipped) and KSVZ models $\xi$ takes the approximate values $1.4\,(0.8)$ and $0.5$, respectively.%
\footnote{For the KSVZ model we have taken the benchmark of $X_{Q}^{em}=0$, with $X_{Q}^{em}$ denoting the electric charge of the exotic color triplet $Q$.   For the DFSZ model there are two possible implementations of NFC, the Higgs doublet coupling to $l_R$ can couple either to down-type quarks (type II) or to up-type quarks (flipped).} 
In our 3HFPQ scenario we can get, besides the standard values, an additional set of discrete values allowing us to cover a large range of axion-photon couplings.    
We present in Fig.~\ref{fig:varationCag} the values of the model dependent quantity $C_{a \gamma}/C_{ag}$ for those 3HFPQ models with $N_{\mbox{\scriptsize{DW}}}=1$, where the different values for each potential implementation represent distinct models for the charged leptons.

It is well known that the evolution of stars place strong constraints on the axion coupling to photons. A strong bound can be derived from globular-cluster stars~\cite{Raffelt:2006cw}. These are homogeneous gravitationally bound systems of stars formed around the same time, allowing for detailed tests of stellar-evolution theory.   The actual experimental bound gives $g_{10}\lsim 1$ for axion masses up to $30 \, \text{keV}$. Recently, the analysis of the evolution of massive stars lead to the bound $g_{10}\lsim 0.8$, based on the fact that Cepheid variable stars exist~\cite{Friedland:2012hj}. An even more stringent bound from an updated analysis of 39 Galactic Globular Clusters has been reported~\cite{Ayala:2014pea}, setting the limit $g_{10}\lsim 0.66$ at $95\%$~CL. 

Several helioscope and haloscopes experiments are currently involved in probing the $g_{a\gamma}$ coupling. The most powerful axion helioscope experiment is the CERN Axion Solar Telescope (CAST), which searches for solar axions via axion-photon conversion using a dipole magnet directed towards the sun.    The CAST experiment achieved the limit $g_{10} \lsim 0.88$ for $m_a\lsim0.02\, \text{eV}$, while slightly weaker bounds were obtained for heavier axions~\cite{Zioutas:2004hi}. Still the astrophysical bounds represent a slight improvement over the CAST results.  It is expected that the next generation of axion helioscope experiments, such as the International Axion Observatory (IAXO)~\cite{Irastorza:2011gs}, will provide better bounds on the axion-photon couplings in the future.   Microwave cavity haloscopes, including the Axion Dark Matter experiment (ADMX), exclude a window for the axion around a few $\mu$eV~\cite{DePanfilis:1987dk,Asztalos:2003px,Carosi:2013rla}.   These experiments search for cold dark matter axions in the local galactic dark matter halo. 
\subsection{Constraining the axion couplings to matter  \label{ssec:matter} }

We define the axion-electron coupling constant as $h_{aee}=|g_e| m_e/v_{\mbox{\scriptsize PQ}}$, with $g_e$ given by Eq.~\eqref{eq:ge}.    The axion-electron coupling is bounded from astrophysical sources.   In globular clusters, energy losses in red-giant stars due to axion emission would delay helium-ignition and make the red-giant branch extend to brighter stars.   Helium ignition arguments in red-giant branch stars place the following upper bound on the axion-electron coupling, $h_{aee} \lsim 3 \times 10^{-13}$~\cite{Raffelt:1994ry}.  A more restrictive bound comes from white-dwarf (WD) cooling due to axion losses~\cite{Raffelt:2006cw,Bertolami:2014wua}, 
\be
h_{aee}<1.3\times 10^{-13}\quad \Rightarrow\quad v_{\mbox{\scriptsize PQ}}  > |g_e| \times (4\times 10^9\,\text{GeV})\,.
\ee
The bounds that can be extracted on the PQ symmetry breaking scale, or alternatively, on the axion mass, are very model dependent for this observable.  The value of $g_{e}$ given by Eq.~\eqref{eq:ge} not only depends on the particular charge assignments of the model considered but also on the vevs of the Higgs doublets.  In some regions of the parameter space it is even possible to obtain $g_{e} \simeq 0$ so that WD cooling arguments would not place a strong bound on the axion mass.\footnote{For a very small axion-electron coupling one should also include the one loop contributions~\cite{Srednicki:1985xd}.}  Taking the benchmark point $|g_{e}|/N_{\mbox{\scriptsize{DW}}} =10^{-1}$ for example, implies the upper bound $m_a \lsim 15$~meV.  

Axion-nucleon interactions are constrained by the requirement that the neutrino signal of the supernova SN 1987A is not excessively shortened by axion losses~\cite{Keil:1996ju,Raffelt:2006cw}.  We find these constraints to be similar than those coming from the WD cooling arguments in general.  However, the SN 1987A limit involves many uncertainties which are not easy to quantify~\cite{Raffelt:2006cw}.  Once more, the bound extracted on the PQ scale from the SN 1987A will depend on the vevs of the Higgs doublets.     

The axion couplings to matter can also be tested in terrestrial laboratories, with promising prospects of probing unexplored regions of the axion parameter space.   Dark matter axions can cause transitions between atomic states that differ in energy by an amount equal to the axion mass.    By tuning the atomic states energy using the Zeeman effect it is possible in principle to detect axion dark matter candidates in the $10^{-4}$~eV mass range~\cite{Sikivie:2014lha}.     The axion can also be tested in dedicated laboratory experiments looking for oscillating nucleon electric dipole moments (EDMs)~\cite{Graham:2011qk,Stadnik:2013raa,Roberts:2014cga}, and, oscillating parity- and time reversal-violating effects in atoms and molecules~\cite{Stadnik:2013raa,Roberts:2014cga}.       The proposed Cosmic Axion Spin Precession Experiment (CASPEr) for example, could cover the entire range of axion dark matter masses $m_a \lesssim \mu$eV by looking for oscillating EDMs in a nuclear magnetic resonance solid-state experiment~\cite{Budker:2013hfa}.

\begin{figure}[ht!]
\centering
\includegraphics[width=0.8\textwidth]{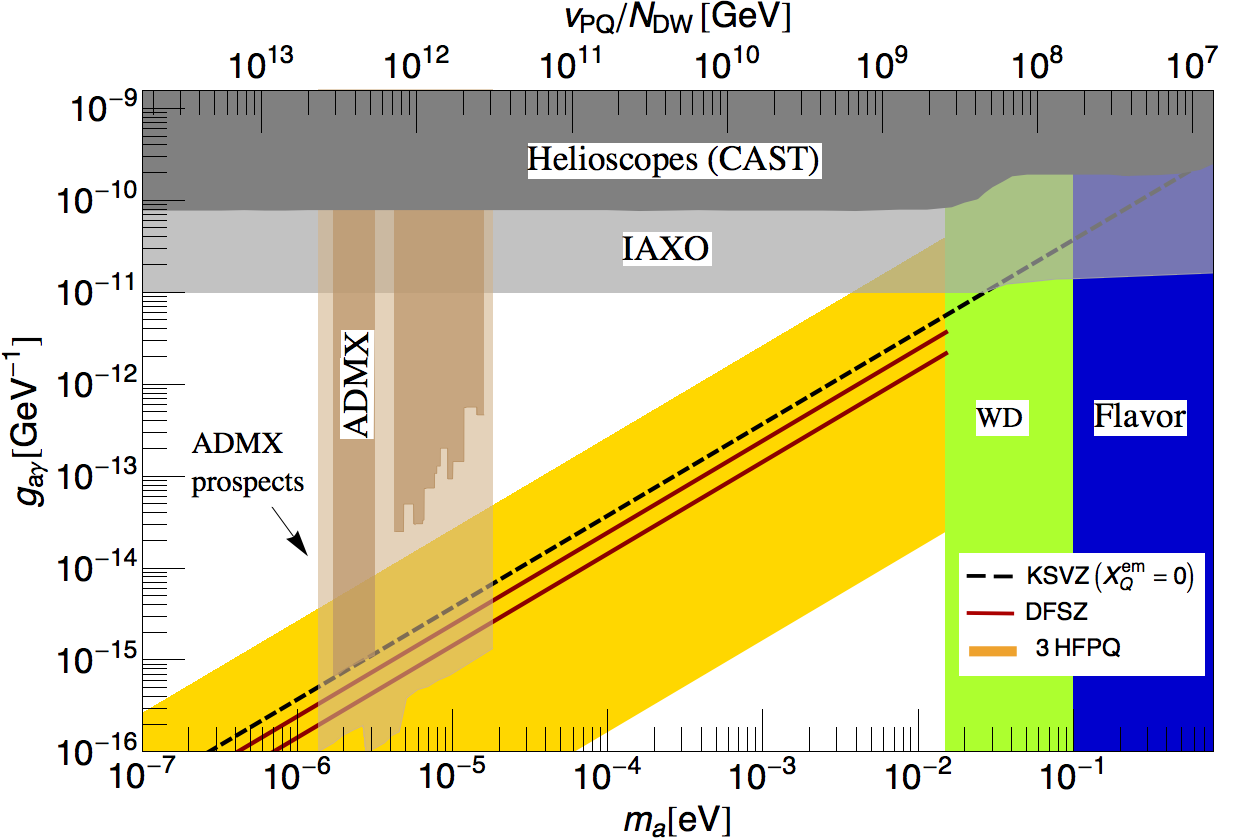}
\caption{\label{fig:constraintsO} \it \small    Constraints on the invisible axion of the 3HFPQ models where the top-quark is singled out.  Familon searches in flavor experiments, astrophysical considerations and axion-photon conversion experiments are taken into account.      The yellow wide band represents a scan over all possible 3HFPQ models considered.     Constraints from white-dwarfs (WD) cooling are shown taking the benchmark point $|g_{e}|/N_{\mbox{\scriptsize{DW}}}=10^{-1}$.     The dark blue band represents the most conservative upper bound on the axion mass from $\mu^+ \rightarrow e^+ a \gamma$.    Predictions for the KSVZ and DFSZ models (type II and flipped) are also shown.       }
\end{figure}

\begin{figure}[ht!]
\centering
\includegraphics[width=0.8\textwidth]{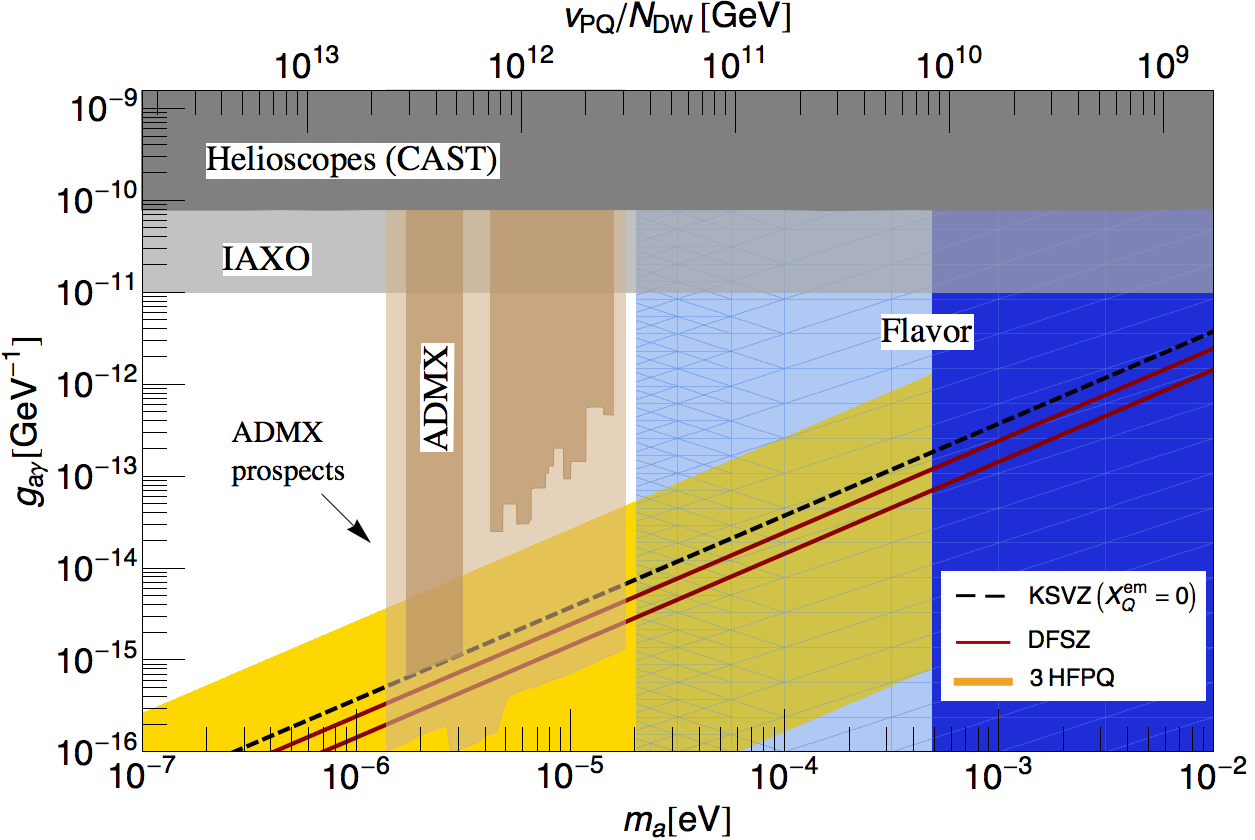}
\caption{\label{fig:constraintsP} \it \small   Constraints on the invisible axion of the 3HFPQ models where the up-quark and charm quarks are singled out.  Familon searches in flavor experiments, astrophysical considerations and axion-photon conversion experiments are taken into account.             The yellow wide band represents a scan over all possible 3HFPQ models considered.      The dark blue band corresponds to the most conservative upper bound on the axion mass extracted from $K^+ \rightarrow \pi^+ a$, the light blue band corresponds to the strongest upper bound from this process.   Predictions for the KSVZ and DFSZ models (type II and flipped) are also shown.    }
\end{figure}

\subsection{Constraining flavor changing axion interactions   \label{ssec:flavor} }

In our framework, and contrarily to what happens in the DFSZ and KSVZ models, the axion couples differently to different flavors and has flavor changing interactions at tree level. Pseudo-Goldstone bosons arising from the spontaneous breaking of a horizontal symmetry are known as familons~\cite{Reiss:1982sq,Wilczek:1982rv}. In our class of models this familon is the axion, and it will have non-diagonal interactions in the up-quark sector or in the down-quark sector depending on the model considered.   There are also flavor changing axion interactions in the charged lepton sector for some of the leptonic implementations considered.

The most stringent bounds on flavor changing axion interactions are extracted from flavor violating decays of kaon or muons into the axion and some other particle(s).    Flavor processes in which the axion enters with a double insertion of the axion coupling ($\mu \rightarrow e \gamma$, $\mu-e$ conversion in nuclei, $K^0- \bar K^0$ mixing, $B_s \rightarrow \mu^+ \mu^-$, among others) are very suppressed by an extra $v_{\mbox{\scriptsize{PQ}}}^{-1}$ factor at the amplitude level and do not put relevant bounds.

The leptonic decay $\mu^+\rightarrow e^+\, a$ can in principle be used to constrain charged lepton flavor violating interactions of the axion.    In~Ref.~\cite{Jodidio:1986mz} the authors reported the experimental bound $\mathrm{Br}(\mu^+ \rightarrow e^+ a) < 2.6 \times 10^{-6}$ at $90\%$~CL. This result however relies on the assumption that the positron is emitted isotropically to avoid large backgrounds from the ordinary muon decay $\mu^+ \rightarrow e^+ \nu_e \bar \nu_{\mu}$. This assumption would be valid if we only had vectorial couplings but not axial ones. In our scenarios, the lepton flavor violating axial and vectorial couplings are equal, so the assumptions behind the $\mu^+ \rightarrow e^+ a$ bound do not apply. In this case the best process to bound the charged lepton flavor violating axion couplings is the radiative decay $\mu^+\rightarrow e^+\, a \, \gamma$.   With this process it is possible to extract limits which are independent of the chirality properties of the axion couplings~\cite{Goldman:1987hy}. The most stringent experimental bound at the moment is $\mathrm{Br}(\mu^+\rightarrow e^+\,a\,\gamma)<1.1\times 10^{-9}$ at $90\%$~CL~\cite{Bolton:1988af}, obtained at the Los Alamos Meson Physics Facility (LAMPF) using the Crystal Box detector. From this process we can extract
\be\label{eq:vpqmu}
v_{\mbox{\scriptsize PQ}}>\left[\left|g_{\mu e}^V\right|^2+\left|g_{\mu e}^A\right|^2\right]^{1/2}\times\left(1.6\times 10^9\,\text{GeV}\right)=\left|g^V_{\mu e}\right|\times\left(2.3\times 10^9\,\text{GeV}\right)\,,
\ee
with the axion-lepton couplings $g^V_{\mu e}=g^A_{\mu e}=\left(C_{ae}^A\right)_{21}$ in the scenario with FCNCs in the charged lepton sector.   We obtain a robust bound from this process since the flavor changing couplings $g^{V,\,A}_{\mu e}$ are completely determined by elements of the PMNS lepton mixing matrix due to the underlying PQ symmetry.   The bound extracted on the PQ scale, or equivalently the axion mass, from $\mu^+ \rightarrow e^+ a \gamma$ does not vary much between all the models with FCNCs in the charged lepton sector.  The reason being that that the PMNS matrix is very anarchical, that is, $| V_{\tau 2}^* V_{\tau 1} | \sim  | V_{\mu 2}^*  V_{\mu 1} | \sim |  V_{e 2}^*  V_{e 1}  | $. Obviously, models without tree-level FCNCs in the charged lepton sector avoid the constraints coming from $\mu^+ \rightarrow e^+ a \gamma$.

Models with FCNCs in the up-quark sector do not receive strong constraints from flavor observables, all the relevant observables involve a double insertion of the axion couplings in this case.  On the other hand, models with FCNCs in the down-quark sector are strongly constrained by limits on $K^+ \rightarrow \pi^+ a$.   To the best of our knowledge, the strongest bound on $K^+ \rightarrow \pi^+ a $ decays has been set by the E787 Collaboration at Brookhaven National Laboratory, achieving $\mathrm{Br}(K^+ \rightarrow \pi^+ a ) < 4.5 \times 10^{-11}$ at $90\%$~CL~\cite{Adler:2002hy}.  The partial decay width for this process is given by
\be
\Gamma(K^+\rightarrow \pi^+\, a)=\frac{1}{64\pi}\frac{m_K^3}{v_{\mbox{\scriptsize PQ}}^2}\left|g^V_{sd}\right|^2\beta^3\left|F_1(0)\right|^2\,,
\ee
with $\beta=1-m_{\pi}^2/m_K^2$ and $g^V_{sd}=\left(C_{ad}^{V}\right)_{21}$. The relevant hadronic matrix element 
\be 
\left<\pi^+(p^\prime)\right|\overline{s}\gamma^\mu d\left|K^+(p)\right>=F_1(q^2)(p+p^\prime)^\mu \,,
\ee
can be extracted in the limit of exact $\mathrm{SU(3)}$ flavor symmetry. At the zero momentum transfer the form factor has the fixed normalization $F_1(0)=1$~\cite{Leutwyler:1984je}.   We have $\left<\pi^+(p^\prime)\right|\overline{s}\gamma^\mu \gamma_5d\left|K^+(p)\right>=0$ because $K^+$ and $\pi^+$ are pseudoscalar mesons.     From this result we can extract a lower bound on the PQ scale
\be\label{eq:vpqK}
v_{\mbox{\scriptsize PQ}}>\left|g_{sd}^V\right|\times\left(4.4\times 10^{11}\, \text{GeV}\right)\,.
\ee
Just like in the charged lepton sector, the coupling $g_{sd}^V$ is fixed in terms of elements of the CKM quark mixing matrix due to the underlying PQ symmetry.     A robust bound can then be extracted on the axion mass which is independent of the many free parameters of the model.  

Future improvements on the $\mu^+ \rightarrow e^+ a \gamma$ bounds are difficult to achieve with present facilities, see discussion in Ref.~\cite{Hirsch:2009ee}. 
On the other hand, improvements on the $K^+ \rightarrow \pi^+ a$ limits can be expected from the NA62 experiment at CERN~\cite{Fantechi:2014hqa}.

In Figs.~\ref{fig:constraintsO} and~\ref{fig:constraintsP} we summarize all the constraints discussed so far on the axion properties.  We do not show explicitly the limits from massive stars on $g_{a\gamma}$ though these are similar to that from CAST.  In Fig.~\ref{fig:constraintsO} we show constraints on models with FCNCs in the charged lepton sector and in the down-quark sector which select the top-quark.    The strongest bound from flavor observables arises in this case from $\mu^+ \rightarrow e^+ a \gamma$ because of the strong suppression factor $|V_{ts}^* V_{td}|$ entering in $K^+ \rightarrow \pi^+ a$ decays.   The wide yellow band represents the prediction scanning over all the 3HFPQ models of this type.  For this type of models astrophysical bounds from WD cooling put in general a stronger limit on the axion mass than flavor processes, the WD bound however depends strongly on the vevs of the Higgs doublets while the flavor limits do not.        This is precisely what occurs for the model analyzed in Ref.~\cite{Celis:2014iua}, which corresponds to a case I model with scalar potential implementation $T_{11}$ and leptonic implementation $(3,3,2)$.   Predictions for the KSVZ and DFSZ invisible axion models are also shown in Figs.~\ref{fig:constraintsO} and~\ref{fig:constraintsP}.  For the KSVZ model we assume that the exotic color triplet has no electric charge ($X_{Q}^{em}=0$).      In both Figures, the upper DFSZ line corresponds to the flipped scenario while the bottom one to the type II case. 

In Fig.~\ref{fig:constraintsP} we show the constraints on those models with FCNCs in the down-quark sector which select the up or charm quark.  The most relevant limit on the axion mass comes now from $K^+ \rightarrow \pi^+ a$ due to the value of the product of CKM matrix elements $|V_{ud}^* V_{us}|^2 \sim |V_{cd}^* V_{cs}|^2 \gg |V_{td}^* V_{ts}|^2$ lifting the decay rate.   For some models of this type the bound from kaon decays can be as strong as $m_a \lsim 2 \times 10^{-5}$~eV.  This is one of the main results of our work.   Among all the models with FCNCs in the down-quark sector, those which select the top quark are much less constrained because of the very effective CKM suppression entering in $K^+ \rightarrow \pi^+ a$ decays.

\subsection{Higgs physics   \label{ssec:higgs}}
The scalar sector of the model contains three complex Higgs doublets $\Phi_j$ ($j=1,2,3$) and a complex scalar gauge singlet $S$.  The scalar fields are then parametrized in terms of 14 real degrees of freedom (each doublet carrying 4 and the singlet 2).  Three degrees of freedom correspond to the usual Goldstone bosons $G^{\pm, 0}$ responsible of giving mass to the massive weak gauge bosons. These have already being isolated by going to the Higgs basis in Eq.~\eqref{eq:HiggsbasisG}.    Another degree of freedom corresponds to the axion which, up to corrections of order $\mathcal{O}\left(v/v_{\mbox{\scriptsize{PQ}}}\right)$, is given by the phase of the scalar gauge singlet. The other 10 degrees of freedom become physical scalar fields, leaving 2 electrically charged and 6 neutral physical scalars.  It is not our intent to present a detailed analysis of the Higgs phenomenology in this class of models. We will, nevertheless, say a few words on some of these aspects. However, before discussing the Higgs phenomenology it is necessary to have some basic grasp of the decoupling structure of the kind of models considered.  

\subsubsection{Decoupling in the scalar sector}
After the scalar fields acquire a vev, mixing among the scalars with the same charge is induced. Due to the large hierarchy between the vevs, i.e. $ v_{\mbox{\scriptsize{PQ}}}   \gg v$, the radial part of the gauge singlet acquires a large mass and we can treat the mixing as $\mathrm{SU(2)}_L$-conserving. The scalar potential of the $\mathrm{SU(2)}_L$ doublets is then given by
\begin{align}\label{eq:phipotential}
V(\Phi)=(
\Phi_i^\dagger
\mathcal{M}^2_{ij}
\Phi_j+\mathcal{O}(v^2)+\text{h.c.})+\text{quartic terms on $\Phi_i$}\,,
\end{align}
with the square mass matrix taking the form
\begin{align}\label{eq:massmatrix}
\mathcal{M}^2=v_{\mbox{\scriptsize PQ}}^2
\begin{pmatrix}
\dfrac{m_1^2}{v_{\mbox{\scriptsize PQ}}^2}+\lambda_1^{\Phi S} &\lambda_4 & \lambda_5\\
\lambda_4^* & \dfrac{m_2^2}{v_{\mbox{\scriptsize PQ}}^2}+\lambda_2^{\Phi S} & \lambda_6\\
\lambda_5^* & \lambda_6^* & \dfrac{m_3^2}{v_{\mbox{\scriptsize PQ}}^2}+\lambda_3^{\Phi S}
\end{pmatrix}\,.
\end{align}
The couplings $\lambda_i$ are associated with a phase sensitive terms $(i)$ of Table~\ref{tab:phase_sensitve}. In the case where a phase sensitive term has mass dimension, such as in cases $(4)$, $(5)$ and $(6)$ with $k_i=1$, we parametrize it as $\mu_i=v_{\mbox{\scriptsize PQ}}\lambda_i$. Additionally, note that for models $T_1$ to $T_9$ the PQ symmetry forces one of the couplings to be zero, that is $\lambda_k=0$ with $k=4,5$ or $6$, while for models $T_{10}$ to $T_{12}$ all the couplings are expected to be non-zero.

Because of the large value of the PQ symmetry breaking scale, the scalar sector of invisible axion models usually presents a decoupling scenario. This way only the axion and a SM-like Higgs remains at the EW scale while the others acquire a mass of order $v_{\mbox{\scriptsize{PQ}}}$. However, it is possible to achieve specific values for the parameters in order to avoid the decoupling limit in such a way that two (or three) Higgs doublets get masses around the EW scale. 

In what follows, we analyze the different decoupling limits and give a possible texture reproducing each scenario. In the textures we use the parameters $b,c\sim\mathcal{O}(1)$ and $\epsilon\sim\mathcal{O}\left(v^2/v_{\mbox{\scriptsize{PQ}}}^2\right)$. The last parameter, $\epsilon$, has been introduced in order to show how EW corrections coming from the terms we have neglected in Eq.~\eqref{eq:phipotential} can lift the zero eigenvalues to the EW scale. We also distinguish between the case where one of the $\lambda$-couplings is zero as in models $T_1$ to $T_9$ and the case where all the couplings are non-zero, corresponding to models $T_{10}$ to $T_{12}$. 
\begin{itemize}
 \item \textbf{One doublet at the electroweak scale.}\\
 This scenario is characterized by the presence of a SM-like Higgs at the EW scale, with the other two scalar doublets having masses at the PQ symmetry breaking scale. As a result, the infrared theory will correspond to the SM plus the axion (whose properties and couplings were discussed in Sec.~\ref{sec:axion}) supplemented by higher dimension operators suppressed by the PQ breaking scale which can be neglected.%
 \footnote{Additionally, one should take special care of higher dimension operators coming from gravitational effects as they give non-negligible contributions. For a detailed analysis see subsection~\ref{subsec:gravity}.}

 We list two textures generating this scenario
  \begin{align}
  \mathcal{M}^2=v_{\mbox{\scriptsize PQ}}^2
  \begin{pmatrix}
  \sqrt{2}\,b+\epsilon & b & b\\
  b & \sqrt{2}\,b & 0\\
  b & 0 & \sqrt{2}\,b
  \end{pmatrix}\,.
  \end{align}
  This texture can be implemented in models $T_1$ to $T_9$, even though the zero has been located in the position of $\lambda_5$. The same mass spectrum is generated by permuting the value of the parameters appropriately. On the other hand, for the models $T_{10}$ to $T_{12}$ one possible texture is given by
  \begin{align}
  \mathcal{M}^2=v_{\mbox{\scriptsize PQ}}^2
  \begin{pmatrix}
  b+\epsilon & b & c\\
  b & b+\epsilon & c\\
  c & c & c
  \end{pmatrix}\,,
  \end{align}
 where the constraint $b\neq c$ needs to be satisfied to have just one doublet at the EW scale.
  
 \item \textbf{Two doublets at the electroweak scale.}\\
 In this decoupling scenario we obtain a 2HDM with tree-level FCNCs controlled by the CKM and PMNS matrices, the quark masses and the ratio of the vevs of the two Higgs doublets whose masses are at the EW scale (in a similar fashion as in the BGL 2HDM). However, the values of the flavor changing scalar couplings cannot be determined in general as it will depend on the specific implementation of the scalar parameters in $\mathcal{M}^2$. In any case, as these couplings present the same structure as in the BGL models, similar constraints in the parameter space are expected. 
  
 For models $T_1$ to $T_9$ it is not possible to reproduce this scenario unless some of the parameters are ultraweak, i.e. of order $\mathcal{O}\left(v^2/v_{\mbox{\scriptsize{PQ}}}^2\right)$. In this case one possible texture is given by
  \begin{align}
  \mathcal{M}^2=v_{\mbox{\scriptsize PQ}}^2
  \begin{pmatrix}
  b+\epsilon & b & \epsilon\\
  b & b+\epsilon & 0\\
  \epsilon & 0 & \epsilon
  \end{pmatrix}\,,
  \end{align}
  which is only valid for models $T_1$, $T_2$ and $T_7$. The equivalent texture for models $T_3$ to $T_6$, $T_8$ and $T_9$ can be directly obtained from the previous texture by permuting the entries in the matrix. Finally, one texture reproducing this scenario in models $T_{10}$ to $T_{12}$ is
  \begin{align}
  \mathcal{M}^2=v_{\mbox{\scriptsize PQ}}^2
  \begin{pmatrix}
  b+\epsilon & b & b\\
  b & b+\epsilon & b\\
  b & b & b
  \end{pmatrix}\,.
  \end{align}

 \item \textbf{Three doublets at the electroweak scale.}\\
 Having the three doublets at the EW scale is only possible if we force all the parameters in Eq.~\eqref{eq:massmatrix} to be ultraweak, that is if every term in $\mathcal{M}^2$ take values around the EW scale. As we have discussed in Sec.~\ref{sec:sol}, this scenario gives rise to FCNCs which are suppressed by the CKM and the PMNS matrices with the explicit scalar flavor violating couplings depending on the model implementation.
\end{itemize}

This simple analysis of the possible decoupling scenarios is by no means a full and detailed study of the scalar spectrum. The textures above are just illustrative and many other textures with different degrees of tuning might be present for any of the three relevant decoupling scenarios.  Finally, it should be noted that the scalar sector in this class of models suffers from a fine tuning problem (commonly known as the hierarchy problem), just like most models where more than one scale is present in the theory. A solution for this problem is out of the scope of the present work. However, some promising directions have been pursued in the literature within the framework of invisible axion models~\cite{Allison:2014hna}. 

\subsubsection{Higgs phenomenology}
If there is a decoupling in the scalar sector where one Higgs doublet remains at the weak scale while the other scalar fields become very heavy, three degrees of freedom of this doublet correspond to the Goldstone bosons giving mass to the massive gauge bosons while the remaining degree of freedom corresponds to a SM-like Higgs boson.    The possibility of a richer decoupling structure in the scalar sector, with two or three Higgs doublets at the weak scale would give rise to potentially new physics signatures at flavor factories and collider experiments like the LHC.    The latter scenario would imply the existence of additional neutral Higgs boson (besides the $125$~GeV SM-like Higgs boson) and charged scalars with masses around the EW scale.    Neglecting mixing effects among the scalar fields, the phenomenology of these scalars would be basically the same than in the BGL 2HDMs analyzed in Refs.~\cite{Botella:2014ska,Bhattacharyya:2013rya,Bhattacharyya:2014nja}.  For example, dangerous $|\Delta S| =2$  contributions to $K^0- \bar K^0 $ mixing due to neutral scalars would be very suppressed in the top BGL models because the flavor changing couplings are proportional to  $|V_{ts}^* V_{td}|$, allowing the mass of these scalars to be at the weak scale~\cite{Botella:2014ska}.

A classification of flavor observables which receive important contributions in the BGL 2HDMs and a comprehensive phenomenological analysis of this models was presented in Ref.~\cite{Botella:2014ska}.    Additional neutral scalars with flavor changing couplings will enter at tree level in pseudoscalar meson leptonic decays $M^0 \rightarrow \ell^+ \ell^-$, neutral meson mixing $M^0 - \bar M^0$, as well as in lepton flavor violating transitions of the type: $\ell_1^- \rightarrow \ell_2^{-} \ell_3^{+} \ell_{4}^{-}$, $\tau \rightarrow \ell \pi \pi$ and $\mu-e$ conversion in nuclei.  The previous processes arise in the SM at the loop level and receive strong suppressions due to the GIM mechanism or the smallness of neutrino masses, this makes these processes very sensitive to small new physics contributions.   Charged scalars will also contribute at tree-level to semi-leptonic pseudoscalar meson decays ($M \rightarrow \ell \nu$, $B \rightarrow D^{(*)} \ell \nu$) and leptonic $\tau$ decays ($\tau \rightarrow \ell \bar \nu_{\ell} \nu_{\tau}$),  possibly causing observable violations of lepton universality.  Neutral and charged scalars will contribute at the loop level in processes like $\bar B \rightarrow X_s \gamma$ and $\ell_1 \rightarrow \ell_2 \gamma$ and will in general dominate over the SM contribution which appears at the same level.     The discovery of additional scalars at the LHC and characteristic decay signatures of the non-standard scalars in the  BGL 2HDMs have been analyzed in Ref.~\cite{Bhattacharyya:2013rya,Bhattacharyya:2014nja}.   The main results of these analyses is that within BGL 2HDMs additional charged and neutral scalars can be as light as $150$~GeV while being compatible with present $125$~GeV Higgs, flavor, electroweak precision and collider data~\cite{Botella:2014ska,Bhattacharyya:2013rya,Bhattacharyya:2014nja}.

\section{Conclusions}\label{sec:concl}
In this work we have built a class of invisible axion models with FCNCs at tree level which are controlled by the fermion mixing matrices, therefore extending the work done in Ref.~\cite{Celis:2014iua}.  The scalar sector contains three-Higgs doublets and a complex scalar gauge singlet field.   A flavored Peccei-Quinn symmetry provides a solution to the strong CP problem via the Peccei-Quinn mechanism, giving rise to an invisible axion which could account for the cold dark matter in the Universe.   The main features of such three-Higgs flavored Peccei-Quinn (3HFPQ) class of models are summarized in Table~\ref{3HFPtable}, making the relevant comparisons with the KSVZ and DFSZ axion models.     Experimental limits on the axion have been analyzed taking into account familon searches in rare kaon and muon decays, astrophysical considerations and axion-photon conversion experiments.     
\begin{table}[h]
\begin{center}
\begin{tabular}{r||c|c|c}   \rowcolor{RGray}
Models&KSVZ&DFSZ&3HFPQ\\  \rowcolor{Gray}
\hline
BSM fields&$Q$+$S$&$\Phi_2$+$S$&$\Phi_{2}$+$\Phi_3$+$S$\\ 
\multirow{2}{*}{PQ fields}&\multirow{2}{*}{$Q$, $S$ }&$q,\, l,\, \Phi_{1,2},\, S$ &$q,\, l,\, \Phi_{1,2,3},\, S$ \\ 
&&(flavor blind)&(flavor sensitive)\\     \rowcolor{Gray}
$C_{a\gamma}/C_{ag}$&$6(X^{em}_Q)^2$&$2/3, 8/3$&$[-34/3,44/3]$\\
Tree-level CtM&No&Yes&Yes\\     \rowcolor{Gray}
Tree-level FCAI& No&No& Yes\\
$N_{\mbox{\scriptsize{DW}}}$&1&$3, 6$&$1,\,2,\,\cdots,\,8$\\
\hline
\end{tabular}
\caption{ \it \small  Comparison of the class of models constructed in this work with the usual invisible axion model benchmarks.    The different values for $C_{a\gamma}/C_{ag}$ and $N_{\mbox{\scriptsize{DW}}}$ in the DFSZ and the 3HFPQ models correspond to different implementations of the PQ symmetry.   We use the notation: CtM=Coupling to Matter; FCAI=Flavor Changing Axion Interaction. \label{3HFPtable} }
\end{center}
\end{table}

The most important findings of our analysis are:
\begin{itemize}

\item Models with tree-level FCNCs in the down-quark or charged lepton sectors receive important constraints on the PQ scale from familon searches in kaon and muon decays.  These bounds are very robust for the class of models considered in this work since the flavor changing axion couplings are completely controlled by elements of the fermion mixing matrices due to the underlying PQ symmetry.

\item  Models with tree-level FCNCs in the down-quark sector for which the top quark is singled out receive the strongest upper bound on the axion mass from white-dwarf cooling arguments in general, though this bounds depend strongly on the vevs of the Higgs doublets.      Bounds from $K^+ \rightarrow \pi^+ a$ are very weak due to the strong CKM suppression.   Fig.~\ref{fig:constraintsO} summarizes all the constraints on this scenario.

\item  Models with tree-level FCNCs in the down-quark sector for which the up (or charm) quark is singled out receive the strongest upper bound on the axion mass from $K^+ \rightarrow \pi^+ a$ decays since in this case the flavor changing couplings are not as suppressed $| V_{us}^* V_{ud}| \sim | V_{cs}^* V_{cd}|  \gg | V_{ts}^* V_{td}|$.  Fig.~\ref{fig:constraintsP} summarizes all the constraints on this scenario.

\item Constraints from $\mu^+ \rightarrow e^+ a \gamma$ are very similar in all the models with FCNCs in the charged lepton sector due to the anarchical structure of the PMNS matrix.  The bounds derived from  $\mu^+ \rightarrow e^+ a \gamma$ are stronger than those obtained from $K^+ \rightarrow \pi^+ a$ in models with tree-level FCNCs in the down-quark sector for which the top quark is singled out.

\item The axion of models without FCNCs in the down-quark and charged lepton sectors do not receive important constraints from flavor observables.     In this case the strongest bounds on the axion can be derived from the axion-photon coupling and white-dwarf cooling arguments.     

\item A large variety of the models considered have $N_{\mbox{\scriptsize{DW}}}=1$, avoiding the domain wall problem.    Allowed values for the model dependent quantity $C_{a\gamma}/C_{ag}$ (see Eq.~\eqref{eq_phtc}) in these models was presented in Fig.~\ref{fig:varationCag}, large deviations on the axion-photon coupling compared with the DFSZ model are obtained in some cases. One interesting aspect is the fact that we are able to mimic the DFSZ axion coupling to photons and have at the same time $N_{\mbox{\scriptsize{DW}}}=1$. A zero $C_{a\gamma}$ can be achieved but only in models with $N_{\mbox{\scriptsize{DW}}}>1$.

\end{itemize}

\section*{Acknowledgments}  
The work of A.C. and J.F. has been supported in part by the Spanish Government and ERDF funds from the EU Commission [Grants FPA2011-23778 and CSD2007-00042 (Consolider Project CPAN)].   J.~F.  also acknowledges VLC-CAMPUS for an ``Atracci\'{o} del Talent"  scholarship. 
H.S. wants to thank to IFIC where a large part of the work was done and funded by the European FEDER, Spanish MINECO, under the grant FPA2011-23596. H.S. work was also supported by the National Research Foundation of Korea (NRF) grant funded by the Korea Government (MEST) (No. 2012R1A2A2A01045722), and also supported by Basic Science Research Program through the National Research Foundation of Korea (NRF) funded by the ministry of Education, Science and Technology (No. 2013R1A1A1062597). Finally, H.S. acknowledges the Portuguese FCT project PTDC/FIS-NUC/0548/2012.


\begin{thebibliography}{99}



\bibitem{Aad:2012tfa}
  G.~Aad {\it et al.}  [ATLAS Collaboration],
  Phys.\ Lett.\ B {\bf 716} (2012) 1
  [arXiv:1207.7214 [hep-ex]].
  
\bibitem{Chatrchyan:2012ufa}
  S.~Chatrchyan {\it et al.}  [CMS Collaboration],
  Phys.\ Lett.\ B {\bf 716} (2012) 30
  [arXiv:1207.7235 [hep-ex]].

\bibitem{Cheng:1987gp}
For comprehensive reviews of the strong CP problem see: H.~-Y.~Cheng,
  Phys.\ Rept.\  {\bf 158} (1988) 1;
  R.~D.~Peccei,
  Lect.\ Notes Phys.\  {\bf 741} (2008) 3
  [hep-ph/0607268];
  J.~E.~Kim and G.~Carosi,
  Rev.\ Mod.\ Phys.\  {\bf 82} (2010) 557
  [arXiv:0807.3125 [hep-ph]].
  
\bibitem{Glashow:1970rp}
  S.~L.~Glashow, R.~Jackiw and S.~S.~Shei,
  Phys.\ Rev.\  {\bf 187} (1969) 1916;
  H.~Fritzsch, M.~Gell-Mann and H.~Leutwyler,
  Phys.\ Lett.\ B {\bf 47} (1973) 365.
  
  
\bibitem{'tHooft:1976up}
  G.~'t Hooft,
  Phys.\ Rev.\ Lett.\  {\bf 37} (1976) 8;
  G.~'t Hooft,
  Phys.\ Rev.\ D {\bf 14} (1976) 3432
   [Erratum-ibid.\ D {\bf 18} (1978) 2199].
  
  
     
  %
\bibitem{Baker:2006ts}
  C.~A.~Baker, D.~D.~Doyle, P.~Geltenbort, K.~Green, M.~G.~D.~van der Grinten, P.~G.~Harris, P.~Iaydjiev and S.~N.~Ivanov {\it et al.},
  Phys.\ Rev.\ Lett.\  {\bf 97} (2006) 131801
  [hep-ex/0602020].

  
\bibitem{Baluni:1978rf}
  V.~Baluni,
  Phys.\ Rev.\ D {\bf 19} (1979) 2227;
  R.~J.~Crewther, P.~Di Vecchia, G.~Veneziano and E.~Witten,
  Phys.\ Lett.\ B {\bf 88} (1979) 123
   [Erratum-ibid.\ B {\bf 91} (1980) 487].
  
\bibitem{Peccei:1977hh}
  R.~D.~Peccei and H.~R.~Quinn,
  Phys.\ Rev.\ Lett.\  {\bf 38} (1977) 1440;
  R.~D.~Peccei and H.~R.~Quinn,
  Phys.\ Rev.\ D {\bf 16} (1977) 1791.
  
  

\bibitem{Weinberg:1977ma}
  S.~Weinberg,
  Phys.\ Rev.\ Lett.\  {\bf 40} (1978) 223;
  F.~Wilczek,
  Phys.\ Rev.\ Lett.\  {\bf 40} (1978) 279.

\bibitem{Glashow:1976nt}
  S.~L.~Glashow and S.~Weinberg,
  Phys.\ Rev.\ D {\bf 15} (1977) 1958;
  E.~A.~Paschos,
  Phys.\ Rev.\ D {\bf 15} (1977) 1966.

\bibitem{Bona:2007vi}
  M.~Bona {\it et al.}  [UTfit Collaboration],
  JHEP {\bf 0803} (2008) 049
  [arXiv:0707.0636 [hep-ph]].



\bibitem{Preskill:1982cy}
  J.~Preskill, M.~B.~Wise and F.~Wilczek,
  Phys.\ Lett.\ B {\bf 120} (1983) 127;
  L.~F.~Abbott and P.~Sikivie,
  Phys.\ Lett.\ B {\bf 120} (1983) 133;
  M.~Dine and W.~Fischler,
  Phys.\ Lett.\ B {\bf 120} (1983) 137;
  M.~S.~Turner and F.~Wilczek,
  Phys.\ Rev.\ Lett.\  {\bf 66} (1991) 5;
  D.~H.~Lyth and E.~D.~Stewart,
  Phys.\ Rev.\ D {\bf 46} (1992) 532.


\bibitem{Minkowski:1977sc}
  P.~Minkowski,
  Phys.\ Lett.\ B {\bf 67} (1977) 421;
  M.~Gell-Mann, P.~Ramond and R.~Slansky,
  Conf.\ Proc.\ C {\bf 790927} (1979) 315
  [arXiv:1306.4669 [hep-th]];
  T.~Yanagida,
  Conf.\ Proc.\ C {\bf 7902131} (1979) 95.
  R.~N.~Mohapatra and G.~Senjanovic,
  Phys.\ Rev.\ Lett.\  {\bf 44} (1980) 912;
  J.~Schechter and J.~W.~F.~Valle,
  Phys.\ Rev.\ D {\bf 22} (1980) 2227.
  
  
\bibitem{Chikashige:1980ui}
  Y.~Chikashige, R.~N.~Mohapatra and R.~D.~Peccei,
  Phys.\ Lett.\ B {\bf 98} (1981) 265;
  G.~B.~Gelmini and M.~Roncadelli,
  Phys.\ Lett.\ B {\bf 99} (1981) 411;
  J.~E.~Kim,
  Phys.\ Lett.\ B {\bf 107} (1981) 69;
  P.~Langacker, R.~D.~Peccei and T.~Yanagida,
  Mod.\ Phys.\ Lett.\ A {\bf 1} (1986) 541.
  
  
\bibitem{Zhitnitsky:1980tq}
  A.~R.~Zhitnitsky,
  Sov.\ J.\ Nucl.\ Phys.\  {\bf 31} (1980) 260
   [Yad.\ Fiz.\  {\bf 31} (1980) 497];
  M.~Dine, W.~Fischler and M.~Srednicki,
  Phys.\ Lett.\ B {\bf 104} (1981) 199.
  
  

\bibitem{Kim:1979if}
  J.~E.~Kim,
  Phys.\ Rev.\ Lett.\  {\bf 43} (1979) 103;
  M.~A.~Shifman, A.~I.~Vainshtein and V.~I.~Zakharov,
  Nucl.\ Phys.\ B {\bf 166} (1980) 493.




  


\bibitem{Celis:2014iua}
  A.~Celis, J.~Fuentes-Martin and H.~Serodio,
  arXiv:1410.6217 [hep-ph].


\bibitem{Branco:1996bq}
  G.~C.~Branco, W.~Grimus and L.~Lavoura,
  Phys.\ Lett.\ B {\bf 380} (1996) 119
  [hep-ph/9601383].





  
\bibitem{Cabibbo:1963yz}
  N.~Cabibbo,
  Phys.\ Rev.\ Lett.\  {\bf 10} (1963) 531;
  M.~Kobayashi and T.~Maskawa,
  Prog.\ Theor.\ Phys.\  {\bf 49} (1973) 652.
    
    
  
  
\bibitem{Pontecorvo:1957qd}
  B.~Pontecorvo,
  Sov.\ Phys.\ JETP {\bf 7} (1958) 172
   [Zh.\ Eksp.\ Teor.\ Fiz.\  {\bf 34} (1957) 247];
  Z.~Maki, M.~Nakagawa and S.~Sakata,
  Prog.\ Theor.\ Phys.\  {\bf 28} (1962) 870;
  B.~Pontecorvo,
  Sov.\ Phys.\ JETP {\bf 26} (1968) 984
   [Zh.\ Eksp.\ Teor.\ Fiz.\  {\bf 53} (1967) 1717].






\bibitem{Zeldovich:1974uw}
  Y.~.B.~Zeldovich, I.~Y.~Kobzarev and L.~B.~Okun,
  Zh.\ Eksp.\ Teor.\ Fiz.\  {\bf 67} (1974) 3
   [Sov.\ Phys.\ JETP {\bf 40} (1974) 1].
  
  

\bibitem{Sikivie:1982qv}
  P.~Sikivie [ADMX Collaboration],
  Phys.\ Rev.\ Lett.\  {\bf 48} (1982) 1156;
  G.~Lazarides and Q.~Shafi,
  Phys.\ Lett.\ B {\bf 115} (1982) 21.

  
 \bibitem{Vilenkin:1984ib}
  A.~Vilenkin,
  Phys.\ Rept.\  {\bf 121} (1985) 263.
  
    



  
\bibitem{Wilczek:1982rv}
  F.~Wilczek,
  Phys.\ Rev.\ Lett.\  {\bf 49} (1982) 1549;
  Z.~G.~Berezhiani and M.~Y.~Khlopov,
  Z.\ Phys.\ C {\bf 49} (1991) 73.
 
  
  
\bibitem{Davidson:1983tp}
  A.~Davidson and M.~A.~H.~Vozmediano,
  Phys.\ Lett.\ B {\bf 141} (1984) 177;
  A.~Davidson and M.~A.~H.~Vozmediano,
  Nucl.\ Phys.\ B {\bf 248} (1984) 647;
  C.~Q.~Geng and J.~N.~Ng,
  Phys.\ Rev.\ D {\bf 39} (1989) 1449.
  
     

  
  
\bibitem{Hindmarsh:1997ac}
  M.~Hindmarsh and P.~Moulatsiotis,
  Phys.\ Rev.\ D {\bf 56} (1997) 8074
  [hep-ph/9708281].
  
\bibitem{Hindmarsh:1998ph}
  M.~Hindmarsh and P.~Moulatsiotis,
  Phys.\ Rev.\ D {\bf 59} (1999) 055015
  [hep-ph/9807363].
  
\bibitem{Ahn:2014gva}
  Y.~H.~Ahn,
  arXiv:1410.1634 [hep-ph].

\bibitem{Berezhiani:1990wn}
  Z.~G.~Berezhiani and M.~Y.~Khlopov,
  Sov.\ J.\ Nucl.\ Phys.\  {\bf 51} (1990) 739
   [Yad.\ Fiz.\  {\bf 51} (1990) 1157];
  Z.~G.~Berezhiani and M.~Y.~Khlopov,
  Sov.\ J.\ Nucl.\ Phys.\  {\bf 51} (1990) 935
   [Yad.\ Fiz.\  {\bf 51} (1990) 1479];

\bibitem{Pich:2009sp}
  A.~Pich and P.~Tuzon,
  Phys.\ Rev.\ D {\bf 80} (2009) 091702
  [arXiv:0908.1554 [hep-ph]].
  
  
\bibitem{Bae:2010ai}
  K.~J.~Bae,
  Phys.\ Rev.\ D {\bf 82} (2010) 055004
  [arXiv:1003.5869 [hep-ph]];
  H.~Serodio,
  Phys.\ Lett.\ B {\bf 700} (2011) 133
  [arXiv:1104.2545 [hep-ph]];
  G.~Cree and H.~E.~Logan,
  Phys.\ Rev.\ D {\bf 84} (2011) 055021
  [arXiv:1106.4039 [hep-ph]];
  I.~de Medeiros Varzielas,
  Phys.\ Lett.\ B {\bf 701} (2011) 597
  [arXiv:1104.2601 [hep-ph]];
  A.~Celis, J.~Fuentes-Martin and H.~Serodio,
  Phys.\ Lett.\ B {\bf 737} (2014) 185
  [arXiv:1407.0971 [hep-ph]].

  
\bibitem{Buras:2000dm}
  A.~J.~Buras, P.~Gambino, M.~Gorbahn, S.~Jager and L.~Silvestrini,
  Phys.\ Lett.\ B {\bf 500} (2001) 161
  [hep-ph/0007085];
  G.~D'Ambrosio, G.~F.~Giudice, G.~Isidori and A.~Strumia,
  Nucl.\ Phys.\ B {\bf 645} (2002) 155
  [hep-ph/0207036].
  


\bibitem{Botella:2009pq}
  F.~J.~Botella, G.~C.~Branco and M.~N.~Rebelo,
  Phys.\ Lett.\ B {\bf 687} (2010) 194
  [arXiv:0911.1753 [hep-ph]].
  
  
    
\bibitem{Buras:2010mh}
  A.~J.~Buras, M.~V.~Carlucci, S.~Gori and G.~Isidori,
  JHEP {\bf 1010} (2010) 009
  [arXiv:1005.5310 [hep-ph]].
  
  
   
\bibitem{Ferreira:2010ir}
  P.~M.~Ferreira and J.~P.~Silva,
  Phys.\ Rev.\ D {\bf 83} (2011) 065026
  [arXiv:1012.2874 [hep-ph]].

  
  
\bibitem{Serodio:2013gka}
  H.~Serodio,
  Phys.\ Rev.\ D {\bf 88} (2013) 5,  056015
  [arXiv:1307.4773 [hep-ph]].
  
  
\bibitem{Botella:2014ska}
  F.~J.~Botella, G.~C.~Branco, A.~Carmona, M.~Nebot, L.~Pedro and M.~N.~Rebelo,
  JHEP {\bf 1407} (2014) 078
  [arXiv:1401.6147 [hep-ph]].
  

\bibitem{Bhattacharyya:2013rya}
  G.~Bhattacharyya, D.~Das, P.~B.~Pal and M.~N.~Rebelo,
  JHEP {\bf 1310} (2013) 081
  [arXiv:1308.4297 [hep-ph]].
  

\bibitem{Bhattacharyya:2014nja}
  G.~Bhattacharyya, D.~Das and A.~Kundu,
  Phys.\ Rev.\ D {\bf 89} (2014) 095029
  [arXiv:1402.0364 [hep-ph]].

  
  
  
\bibitem{Botella:2012ab}
  F.~J.~Botella, G.~C.~Branco and M.~N.~Rebelo,
  Phys.\ Lett.\ B {\bf 722} (2013) 76
  [arXiv:1210.8163 [hep-ph]].
  
  
  
  
\bibitem{Krauss:1986bq}
  L.~M.~Krauss and M.~B.~Wise,
  Phys.\ Lett.\ B {\bf 176} (1986) 483.
  

  
\bibitem{Bardeen:1986yb}
  W.~A.~Bardeen, R.~D.~Peccei and T.~Yanagida,
  Nucl.\ Phys.\ B {\bf 279} (1987) 401.
  


  
\bibitem{Feng:1997tn}
  J.~L.~Feng, T.~Moroi, H.~Murayama and E.~Schnapka,
  Phys.\ Rev.\ D {\bf 57} (1998) 5875.

  


\bibitem{Botella:2011ne}
  F.~J.~Botella, G.~C.~Branco, M.~Nebot and M.~N.~Rebelo,
  JHEP {\bf 1110} (2011) 037
  [arXiv:1102.0520 [hep-ph]].
  
  
  
\bibitem{Adler:1969gk}
  S.~L.~Adler,
  Phys.\ Rev.\  {\bf 177} (1969) 2426;
  J.~S.~Bell and R.~Jackiw,
  Nuovo Cim.\ A {\bf 60} (1969) 47;
  W.~A.~Bardeen,
  Phys.\ Rev.\  {\bf 184} (1969) 1848.


\bibitem{Srednicki:1985xd}
  M.~Srednicki,
  Nucl.\ Phys.\ B {\bf 260} (1985) 689;
  D.~B.~Kaplan,
  Nucl.\ Phys.\ B {\bf 260} (1985) 215.

\bibitem{Beringer:1900zz}
  J.~Beringer {\it et al.}  [Particle Data Group Collaboration],
  Phys.\ Rev.\ D {\bf 86} (2012) 010001.


\bibitem{Barr:1986hs}
  S.~M.~Barr, K.~Choi and J.~E.~Kim,
  Nucl.\ Phys.\ B {\bf 283} (1987) 591.
  
  
  
\bibitem{Lyth:1989pb}
  D.~H.~Lyth,
  Phys.\ Lett.\ B {\bf 236} (1990) 408.
  
  

  
\bibitem{Holdom:1982ew}
  B.~Holdom,
  Phys.\ Rev.\ D {\bf 27} (1983) 332.
  



 
\bibitem{Abbott:1989jw}
  L.~F.~Abbott and M.~B.~Wise,
  Nucl.\ Phys.\ B {\bf 325} (1989) 687.

\bibitem{Banks:1989zw}
  T.~Banks,
  Physicalia Mag.\  {\bf 12} (1990) 19.


\bibitem{Hiramatsu:2012sc}
  T.~Hiramatsu, M.~Kawasaki, K.~Saikawa and T.~Sekiguchi,
  JCAP {\bf 1301} (2013) 001
  [arXiv:1207.3166 [hep-ph]].
  
  

  
  

   

  

  



\bibitem{Ghigna:1992iv}
  S.~Ghigna, M.~Lusignoli and M.~Roncadelli,
  Phys.\ Lett.\ B {\bf 283} (1992) 278.

\bibitem{Holman:1992us}
  R.~Holman, S.~D.~H.~Hsu, T.~W.~Kephart, E.~W.~Kolb, R.~Watkins and L.~M.~Widrow,
  Phys.\ Lett.\ B {\bf 282} (1992) 132
  [hep-ph/9203206].


\bibitem{Kamionkowski:1992mf}
  M.~Kamionkowski and J.~March-Russell,
  Phys.\ Lett.\ B {\bf 282} (1992) 137
  [hep-th/9202003].


\bibitem{Georgi:1981pu}
  H.~M.~Georgi, L.~J.~Hall and M.~B.~Wise,
  Nucl.\ Phys.\ B {\bf 192} (1981) 409.

\bibitem{Barr:1992qq}
  S.~M.~Barr and D.~Seckel,
  Phys.\ Rev.\ D {\bf 46} (1992) 539.

\bibitem{Kallosh:1995hi}
  R.~Kallosh, A.~D.~Linde, D.~A.~Linde and L.~Susskind,
  Phys.\ Rev.\ D {\bf 52} (1995) 912
  [hep-th/9502069].

\bibitem{Dobrescu:1996jp}
  B.~A.~Dobrescu,
  Phys.\ Rev.\ D {\bf 55} (1997) 5826
  [hep-ph/9609221].


\bibitem{Choi:2003wr}
  K.~w.~Choi,
  Phys.\ Rev.\ Lett.\  {\bf 92} (2004) 101602
  [hep-ph/0308024].


\bibitem{Dvali:2013cpa}
  G.~Dvali, S.~Folkerts and A.~Franca,
  Phys.\ Rev.\ D {\bf 89} (2014) 105025
  [arXiv:1312.7273 [hep-th]].


\bibitem{Banks:1991xj}
  T.~Banks and M.~Dine,
  Phys.\ Rev.\ D {\bf 45} (1992) 1424
  [hep-th/9109045];
  T.~Banks and M.~Dine,
  Phys.\ Rev.\ D {\bf 50} (1994) 7454
  [hep-th/9406132].


\bibitem{Babu:2002ic}
  K.~S.~Babu, I.~Gogoladze and K.~Wang,
  Phys.\ Lett.\ B {\bf 560} (2003) 214
  [hep-ph/0212339].

\bibitem{Dias:2002gg}
  A.~G.~Dias, V.~Pleitez and M.~D.~Tonasse,
  Phys.\ Rev.\ D {\bf 67} (2003) 095008
  [hep-ph/0211107],
  A.~G.~Dias, V.~Pleitez and M.~D.~Tonasse,
  Phys.\ Rev.\ D {\bf 69} (2004) 015007
  [hep-ph/0210172],
  A.~G.~Dias and V.~Pleitez,
  Phys.\ Rev.\ D {\bf 69} (2004) 077702
  [hep-ph/0308037].
  
  

    
\bibitem{Dias:2014osa}
  A.~G.~Dias, A.~C.~B.~Machado, C.~C.~Nishi, A.~Ringwald and P.~Vaudrevange,
  JHEP {\bf 1406} (2014) 037
  [arXiv:1403.5760 [hep-ph]].
  
  
  


\bibitem{Green:1984sg}
  M.~B.~Green and J.~H.~Schwarz,
  Phys.\ Lett.\ B {\bf 149} (1984) 117,
  M.~B.~Green and J.~H.~Schwarz,
  Nucl.\ Phys.\ B {\bf 255} (1985) 93,
  M.~B.~Green, J.~H.~Schwarz and P.~C.~West,
  Nucl.\ Phys.\ B {\bf 254} (1985) 327.
  
  
\bibitem{Kac:1967jr}
  V.~G.~Kac,
  Funct.\ Anal.\ Appl.\  {\bf 1} (1967) 328;
  R.~V.~Moody,
  Bull.\ Am.\ Math.\ Soc.\  {\bf 73} (1967) 217.


\bibitem{Ibanez:1991hv}
  L.~E.~Ibanez and G.~G.~Ross,
  Phys.\ Lett.\ B {\bf 260} (1991) 291,
  L.~E.~Ibanez and G.~G.~Ross,
  Nucl.\ Phys.\ B {\bf 368} (1992) 3,
  L.~E.~Ibanez,
  Nucl.\ Phys.\ B {\bf 398} (1993) 301
  [hep-ph/9210211].


  

  

\bibitem{Raffelt:2006cw}
G.~G.~Raffelt,
  Lect.\ Notes Phys.\  {\bf 741} (2008) 51
  [hep-ph/0611350].
  





  
\bibitem{Friedland:2012hj}
  A.~Friedland, M.~Giannotti and M.~Wise,
  Phys.\ Rev.\ Lett.\  {\bf 110} (2013) 061101
  [arXiv:1210.1271 [hep-ph]].
  
    
    
  
\bibitem{Ayala:2014pea}
  A.~Ayala, I.~Dominguez, M.~Giannotti, A.~Mirizzi and O.~Straniero,
  arXiv:1406.6053 [astro-ph.SR].
  

  
  
  
  

  
  
  
  
  
  
\bibitem{Zioutas:2004hi}
  K.~Zioutas {\it et al.}  [CAST Collaboration],
  Phys.\ Rev.\ Lett.\  {\bf 94} (2005) 121301
  [hep-ex/0411033];
  S.~Andriamonje {\it et al.}  [CAST Collaboration],
  JCAP {\bf 0704} (2007) 010
  [hep-ex/0702006].


\bibitem{Irastorza:2011gs}
  I.~G.~Irastorza {\it et al.},
  JCAP {\bf 1106} (2011) 013
  [arXiv:1103.5334 [hep-ex]];  I. G. Irastorza. Tech. Rep SPSCI-242, CERN, Geneva (2013). CERN-SPSC-2013-022.  
  


  
\bibitem{DePanfilis:1987dk}
  S.~De Panfilis {\it et al.},
  Phys.\ Rev.\ Lett.\  {\bf 59} (1987) 839;
  W.~Wuensch {\it et al.},
  Phys.\ Rev.\ D {\bf 40} (1989) 3153;
  C.~Hagmann, P.~Sikivie, N.~S.~Sullivan and D.~B.~Tanner,
  Phys.\ Rev.\ D {\bf 42} (1990) 1297.
  
  
  
\bibitem{Carosi:2013rla}
  G.~Carosi {\it et al.},
  arXiv:1309.7035 [hep-ph].
  

\bibitem{Asztalos:2003px}
  S.~J.~Asztalos {\it et al.}  [ADMX Collaboration],
  Phys.\ Rev.\ D {\bf 69} (2004) 011101
  [astro-ph/0310042];
  S.~J.~Asztalos {\it et al.}  [ADMX Collaboration],
  Phys.\ Rev.\ Lett.\  {\bf 104} (2010) 041301
  [arXiv:0910.5914 [astro-ph.CO]].





  
\bibitem{Raffelt:1994ry}
  G.~Raffelt and A.~Weiss,
  Phys.\ Rev.\ D {\bf 51} (1995) 1495
  [hep-ph/9410205].
  
  
\bibitem{Bertolami:2014wua}
  M.~M.~Miller Bertolami, B.~E.~Melendez, L.~G.~Althaus and J.~Isern,
  arXiv:1406.7712 [hep-ph].

  

  
\bibitem{Keil:1996ju}
  W.~Keil, H.~T.~Janka, D.~N.~Schramm, G.~Sigl, M.~S.~Turner and J.~R.~Ellis,
  Phys.\ Rev.\ D {\bf 56} (1997) 2419
  [astro-ph/9612222].
  
  
   
  
  
  
\bibitem{Sikivie:2014lha}
  P.~Sikivie,
  arXiv:1409.2806 [hep-ph].
  
\bibitem{Graham:2011qk}
  P.~W.~Graham and S.~Rajendran,
  Phys.\ Rev.\ D {\bf 84} (2011) 055013
  [arXiv:1101.2691 [hep-ph]].
  

\bibitem{Stadnik:2013raa}
  Y.~V.~Stadnik and V.~V.~Flambaum,
  Phys.\ Rev.\ D {\bf 89} (2014) 4,  043522
  [arXiv:1312.6667 [hep-ph]].



  
\bibitem{Roberts:2014cga}
  B.~M.~Roberts, Y.~V.~Stadnik, V.~A.~Dzuba, V.~V.~Flambaum, N.~Leefer and D.~Budker,
  arXiv:1409.2564 [hep-ph].
  
\bibitem{Budker:2013hfa}
  D.~Budker, P.~W.~Graham, M.~Ledbetter, S.~Rajendran and A.~Sushkov,
  Phys.\ Rev.\ X {\bf 4} (2014) 021030
  [arXiv:1306.6089 [hep-ph]].
  
  
  
  
\bibitem{Reiss:1982sq}
  D.~B.~Reiss,
  Phys.\ Lett.\ B {\bf 115} (1982) 217.
  
  
  


  



\bibitem{Jodidio:1986mz}
  A.~Jodidio {\it et al.},
  Phys.\ Rev.\ D {\bf 34} (1986) 1967
   [Erratum-ibid.\ D {\bf 37} (1988) 237].
  
\bibitem{Goldman:1987hy}
  J.~T.~Goldman, A.~L.~Hallin, C.~M.~Hoffman, L.~E.~Piilonen, D.~Preston, R.~D.~Bolton, M.~D.~Cooper and J.~S.~Frank {\it et al.},
  Phys.\ Rev.\ D {\bf 36} (1987) 1543.
  
  
\bibitem{Bolton:1988af}
  R.~D.~Bolton {\it et al.},
  Phys.\ Rev.\ D {\bf 38} (1988) 2077.

  
  
\bibitem{Adler:2002hy}
  S.~S.~Adler {\it et al.}  [E787 Collaboration],
  Phys.\ Lett.\ B {\bf 537} (2002) 211
  [hep-ex/0201037].
  
  
\bibitem{Leutwyler:1984je}
  H.~Leutwyler and M.~Roos,
  Z.\ Phys.\ C {\bf 25} (1984) 91.
  
\bibitem{Hirsch:2009ee}
  M.~Hirsch, A.~Vicente, J.~Meyer and W.~Porod,
  Phys.\ Rev.\ D {\bf 79} (2009) 055023
   [Erratum-ibid.\ D {\bf 79} (2009) 079901]
  [arXiv:0902.0525 [hep-ph]].
  
  
\bibitem{Fantechi:2014hqa}
  R.~Fantechi [NA62 Collaboration],
  arXiv:1407.8213 [physics.ins-det].
  

  
  
  
\bibitem{Allison:2014hna}
  K.~Allison, C.~T.~Hill and G.~G.~Ross,
  arXiv:1409.4029 [hep-ph].
  

  

\end{thebibliography}
\end{document}